\documentclass[11pt]{article}

\usepackage{graphicx}%
\usepackage{multirow}%
\usepackage{amsmath,amssymb,amsfonts}%
\usepackage{amsthm}%
\usepackage{mathrsfs}%
\usepackage[title]{appendix}%
\usepackage{xcolor}%
\usepackage{textcomp}%
\usepackage{booktabs}%
\usepackage{algorithm}%
\usepackage{algorithmicx}%
\usepackage{algpseudocode}%
\usepackage{listings}%
\usepackage{subfig}
\usepackage{quantikz}

\usepackage{physics}
\usepackage{tcolorbox}
\usepackage{blochsphere}
\usepackage{tikz}
\usepackage{tikz-3dplot}
\usetikzlibrary{angles, quotes, intersections, 3d, decorations.pathmorphing}

\usepackage{braket}

\usepackage{pgfplots}
\usepackage{pgfplotstable}

\usepackage{siunitx}

\usepackage[ letterpaper, margin=2cm ]{geometry}
\usepackage{authblk}

\usepackage[ backend=biber, style=ieee, doi=true, url=true, eprint=true ]{biblatex}
\pgfplotsset{compat=1.18}
\tikzset{>=Latex}

\usepackage[ colorlinks=true, citecolor=blue, linkcolor=blue, urlcolor=blue ]{hyperref}

\begin{document}
\title{Building an analogue simulator of a photonic quantum computer with transparent tape, maple syrup, and cat lasers, and implementing first quantum algorithms in the classroom}

\author{Ghislain Lefebvre} 
\affil{AlgoLab, Institut quantique, Université de Sherbrooke, Sherbrooke, Canada}

\maketitle


\begin{abstract}
This work presents the implementation of single-qubit gates, including $R_x$ and $R_z$ gates realized using transparent adhesive tape, and $R_y$ gates obtained with optically active maple and agave solutions. These gates form the native gate set of a simple photonic system and are subsequently used to construct a Hadamard gate. Two forms of two-qubit gates are introduced using a combination of a calcite crystal and transparent tape. The setups employ both the polarization and the path degree of freedom of a photon as qubits, illustrating how readily accessible materials can be used to manipulate and transform the quantum information they convey.

Calculations are performed to determine the birefringence of two different types of tape and to quantify the specific rotations introduced by multiple layers of transparent tape. Finally, simple algorithms and exercises are proposed for students.

These experimental setups are designed to facilitate hands-on manipulation of quantum effects while lowering the barrier to accessing quantum systems, with total costs kept below \$50 CAD for the single-qubit gates and below \$100 CAD for all experiments combined. The configurations presented serve as an analogue simulator of a photonic quantum computer: although the laser beams used can be described classically, all theoretical considerations and experimental procedures remain valid for \textit{quantized} systems employing single-photon emitters and single-photon detectors.
\end{abstract}

\section{Introduction}
\label{sec1}

Quantum computing is often perceived as a highly sophisticated field requiring expensive equipment \cite{QuantumInsider2023}. While this is true for hardware capable of delivering useful results, an exploration of the fundamentals of quantum computing through a prototype of a photonic quantum computer can be carried out at a much more reasonable cost using simple theory and materials \cite{RefYouTubeYoganathan}. Although every theoretical element and experimental procedure is fully quantum when using single photons, one must keep in mind that using continuous light in experiments transitions to the classical regime. For this reason, we introduce the system as an analogue simulator. For the experiment to operate in the quantum regime, or, in other words, to be truly \textit{quantized}, one would need a single‑photon emitter and a single‑photon detector \cite{Couteau2023}. Aside from this limitation, the remainder of the explanation applies perfectly to the quantum realm and this is why the theory presented below describes experiments from the quantum perspective of single photons.

The purpose of this work is to assemble various scientific concepts and present them coherently as an introduction to quantum computing for a scientifically curious audience, regardless of their access to specialized quantum laboratory equipment. The required mathematical background is limited to trigonometry. On the physics side, an overview of light polarization is presented and used throughout the article, while more fundamental quantum concepts are provided in appendix~\ref{AppendixFoundamentals} and supplementary classroom exercises are in appendix~\ref{AppendixProblemSets}. 

\section{Encoding information in the polarization of a photon} 
\label{SectionEncoding}
Using the polarization of a single photon as a quantum bit, or qubit, is a straightforward way to begin exploring quantum computing \cite{Barz2015}. This paper introduces the physical characteristics of such qubits and relates them to their quantum representation, first through the Bloch sphere (appendix~\ref{AppendixBlochSphere}) and later using simple quantum circuits. Since each point on the Bloch sphere can be used to represent encoded information, it is important to understand how these points correspond to specific polarization states of a single photon (appendix \ref{AppendixLinearPolarization}).
\subsection{Linear polarization on the x–z plane of the Bloch sphere}
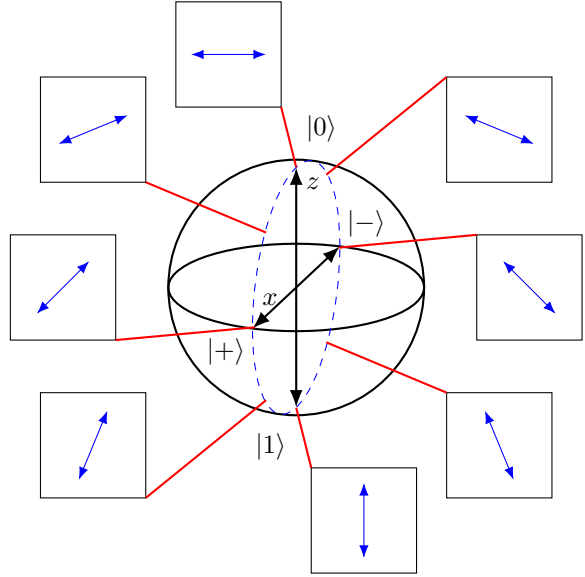
\begin{figure}
    
    \centering
\begin{center}
\tdplotsetmaincoords{70}{110} 
    \begin{tikzpicture}
    \begin{scope}[tdplot_main_coords, scale=1.7]
          \draw[thick] (0,0,0) circle (1); 
          \draw[thick, tdplot_screen_coords] (0,0) circle (1); 
        
          \draw[->, thick] (0,0,0) -- (0,0,1) node[anchor=south west, yshift=6pt] {$\ket{0}$} node[anchor=north west, yshift=0pt] {$z$};
          \draw[->, thick] (0,0,0) -- (0,0,-1) node[anchor=north east, yshift=-6pt] {$\ket{1}$};
          \draw[->, thick] (0,0,0) -- (1,0,0) node[anchor=north, xshift=-10]{$\ket{+}$} node[anchor=south west, yshift=5]{$x$};
          \draw[->, thick] (0,0,0) -- (-1,0,0) node[anchor=south, xshift=10]{$\ket{-}$};

          \def\r{1}
          \def\startAngle{180}
          \def\endAngle{-180}
        
          \draw[dashed, blue,tdplot_main_coords,domain=\startAngle:\endAngle,smooth,variable=\t] 
            plot ({\r*cos(\t)}, 0, {\r*sin(\t)});
    \end{scope}
    \begin{scope}[shift={(-3.8,-4.4)}]
        \coordinate (squareLL) at (2.2,6.8); 
  \coordinate (squareUR) at ($(squareLL)+(1.4,1.4)$); 
  \filldraw[fill=white, draw=black] (squareLL) rectangle (squareUR);
  \draw[<->, blue] ($(squareLL)+(0.2,0.7)$) -- ($(squareUR)-(0.2,0.7)$);
  \draw[thick, red] ($(squareLL)+(1.4,0)$) -- (3.8,6.0); 

  \coordinate (squareLL) at (0.4,5.8); 
  \coordinate (squareUR) at ($(squareLL)+(1.4,1.4)$); 
  \filldraw[fill=white, draw=black] (squareLL) rectangle (squareUR);
  \draw[<->, blue] ($(squareLL)+(0.24,0.51)$) -- ($(squareUR)-(0.24,0.51)$);
  \draw[thick, red] ($(squareLL)+(1.4,0)$) -- (3.4,5.13); 

  \coordinate (squareLL) at (0,3.7); 
  \coordinate (squareUR) at ($(squareLL)+(1.4,1.4)$); 
  \filldraw[fill=white, draw=black] (squareLL) rectangle (squareUR);
  \draw[<->, blue] ($(squareLL)+(0.35,0.35)$) -- ($(squareUR)-(0.35,0.35)$);
  \draw[thick, red] ($(squareLL)+(1.4,0)$) -- (3.25,3.87); 

  \coordinate (squareLL) at (0.4,1.6); 
  \coordinate (squareUR) at ($(squareLL)+(1.4,1.4)$); 
  \filldraw[fill=white, draw=black] (squareLL) rectangle (squareUR);
  \draw[<->, blue] ($(squareLL)+(0.51,0.24)$) -- ($(squareUR)-(0.51,0.24)$);
  \draw[thick, red] ($(squareLL)+(1.4,0)$) -- (3.4,2.9); 

  \coordinate (squareLL) at (4,0.6); 
  \coordinate (squareUR) at ($(squareLL)+(1.4,1.4)$); 
  \filldraw[fill=white, draw=black] (squareLL) rectangle (squareUR);
  \draw[<->, blue] ($(squareLL)+(0.7,0.2)$) -- ($(squareUR)-(0.7,0.2)$);
  \draw[thick, red] ($(squareLL)+(0,1.4)$) -- (3.8,2.8); 

  \coordinate (squareLL) at (5.8,1.6); 
  \coordinate (squareUR) at ($(squareLL)+(1.4,1.4)$); 
  \filldraw[fill=white, draw=black] (squareLL) rectangle (squareUR);
  \draw[<->, blue] ($(squareLL)+(0.89,0.24)$) -- ($(squareUR)-(0.89,0.24)$);
  \draw[thick, red] ($(squareLL)+(0,1.4)$) -- (4.2,3.67); 

  \coordinate (squareLL) at (6.2,3.7); 
  \coordinate (squareUR) at ($(squareLL)+(1.4,1.4)$); 
  \filldraw[fill=white, draw=black] (squareLL) rectangle (squareUR);
  \draw[<->, blue] ($(squareLL)+(1.05,0.35)$) -- ($(squareUR)-(1.05,0.35)$);
  \draw[thick, red] ($(squareLL)+(0,1.4)$) -- (4.37,4.93); 

  \coordinate (squareLL) at (5.8,5.8); 
  \coordinate (squareUR) at ($(squareLL)+(1.4,1.4)$); 
  \filldraw[fill=white, draw=black] (squareLL) rectangle (squareUR);
  \draw[<->, blue] ($(squareLL)+(0.24,0.89)$) -- ($(squareUR)-(0.24,0.89)$);
  \draw[thick, red] ($(squareLL)+(0,1.4)$) -- (4.2,5.9); 
    \end{scope}

\end{tikzpicture}

\end{center}
\caption{Linear polarization associated to various points in the x-z plane of the Bloch sphere.}
    \label{Blochxz}

\begin{tikzpicture}[remember picture, overlay]


\end{tikzpicture}
\end{figure}

By convention, the horizontal linear polarization of a photon corresponds to the $|0\rangle$ state and lies at the top of the Bloch sphere ($z$-axis). If one starts from this point and follows the dashed blue line of figure~\ref{Blochxz} toward the $|+\rangle$ state ($x$-axis), which corresponds to diagonal linear polarization, the polarization rotates counterclockwise from horizontal to diagonal and then to vertical at the bottom of the sphere, corresponding to the $|1\rangle$ state. The polarization remains linear at all points along the dashed blue line, even as one continues toward the $|-\rangle$ state, with anti-diagonal polarization, and finally returns to the top \cite{Shen2020_SU2_Poincare}.

The general equation for the state vector \cite{nielsen_chuang_2000} on the dashed blue line of figure~\ref{Blochxz} is
\begin{equation}
\label{GeneralEquationStateVectorxz}
|\psi\rangle = \cos\frac{\theta}{2}|0\rangle + \sin\frac{\theta}{2}|1\rangle\
\end{equation}
where $\theta$ denotes the angle between the $z$-axis and the state vector.

\subsection{Elliptical polarization on the x–y plane of the Bloch sphere}
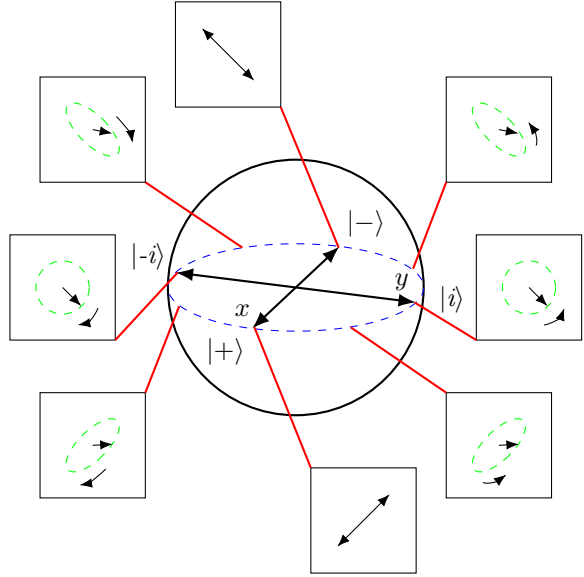
\begin{figure}
    \centering
\begin{center}
\tdplotsetmaincoords{70}{110} 
    \begin{tikzpicture}
    \begin{scope}[tdplot_main_coords, scale=1.7]

  \draw[thick, tdplot_screen_coords] (0,0) circle (1); 

  \draw[white, ->, thick] (0,0,0) -- (0,0,1) node[anchor=south, yshift=6pt] {$\ket{0}$};
  \draw[white, ->, thick] (0,0,0) -- (0,0,-1) node[anchor=north, yshift=-6pt] {$\ket{1}$};\draw[->, thick] (0,0,0) -- (1,0,0) node[anchor=north, xshift=-10]{$\ket{+}$} node[anchor=south east, xshift=3, yshift=1]{$x$};
  \draw[->, thick] (0,0,0) -- (-1,0,0) node[anchor=south, xshift=10]{$\ket{-}$};
  \draw[->, thick] (0,0,0) -- (0,1,0) node[anchor=west, xshift=5pt, yshift=1] {$\ket{\textit{i}}$} node[anchor=south east, xshift=1, yshift=1]{$y$};
  \draw[->, thick] (0,0,0) -- (0,-1,0) node[anchor=east, xshift=0pt, yshift=6] {$\ket{\textit{-i}}$};
  
  \end{scope}

  \begin{scope}[tdplot_main_coords, scale=1.7]
    \draw[dashed, blue] (0,1) arc (90:-270:1);
  \end{scope}

 \begin{scope}[shift={(-3.8,-4.4)}]

    \coordinate (squareLL) at (4,0.6); 
  \coordinate (squareUR) at ($(squareLL)+(1.4,1.4)$); 
  \filldraw[fill=white, draw=black] (squareLL) rectangle (squareUR);
  \draw[<->] ($(squareLL)+(0.35,0.35)$) -- ($(squareUR)-(0.35,0.35)$);
  \draw[thick, red] ($(squareLL)+(0,1.4)$) -- (3.25,3.87); 

  \def\A{135}
   \coordinate (squareLL) at (5.8,1.6); 
   \coordinate (squareUR) at ($(squareLL)+(1.4,1.4)$); 
  \coordinate (startPT) at ($(squareLL) + ({0.7},0.7)$);
  \coordinate (startPTellipse) at ($(startPT)+({cos(\A/2)/2-0.12},{sin(-\A/2)/2+0.28})$);
 \filldraw[fill=white, draw=black] (squareLL) rectangle (squareUR);
   \draw[rotate=(-45), dashed, green] (startPT) ellipse [x radius=cos(\A/2)/2, y radius=sin(-\A/2)/2];
  \begin{scope}[rotate=-45]
    
    \draw[->] 
      ($(startPT) + (0.2,-0.5)$) arc[start angle=-80, end angle=-20, x radius=cos(\A/2)/2, y radius=sin(\A/2)/2];
    \draw[->] (startPT) -- ($(startPT) + (0.18,0.2)$);
  \end{scope}
  \draw[thick, red] ($(squareLL)+(0,1.4)$) -- (4.53,3.87); 

  \coordinate (squareLL) at (6.2,3.7); 
  \coordinate (squareUR) at ($(squareLL)+(1.4,1.4)$);
  \coordinate (startPT) at ($(squareLL)+(0.7,0.7)$);
  \filldraw[fill=white, draw=black] (squareLL) rectangle (squareUR);
  \draw[rotate=0, dashed, green] (startPT) ellipse [x radius=0.707/2, y radius=0.707/2];
   \draw[->] (startPT) -- ($(startPT) + (0.25,-0.25)$);
  \begin{scope} [rotate=0]
    \draw[->] 
      ($(startPT) + (0.2,-0.5)$) arc[start angle=-80, end angle=-20, x radius=0.707/2, y radius=0.707/2];
  \end{scope}
  \draw[thick, red] ($(squareLL)+(0,0)$) -- (5.38,4.2); 

    \coordinate (squareLL) at (5.8,5.8); 
  \coordinate (squareUR) at ($(squareLL)+(1.4,1.4)$); 
  \coordinate (startPT) at ($(squareLL) + ({0.7},0.7)$);
  \coordinate (startPTellipse) at ($(startPT)+({cos(45/2)/2-0.04},{sin(-45/2)/2+0.01})$);
 \filldraw[fill=white, draw=black] (squareLL) rectangle (squareUR);
   \draw[rotate=(45), dashed, green] (startPT) ellipse [x radius=cos(135/2)/2, y radius=sin(-135/2)/2];
  \begin{scope}[rotate=45]
    \draw[->] 
      ($(startPT) + (0.2,-0.5)$) arc[start angle=-80, end angle=-20, x radius=cos(135/2)/2, y radius=sin(135/2)/2];
    \draw[->] (startPT) -- ($(startPT) + (0.15,-0.22)$);
  \end{scope}
  \draw[thick, red] ($(squareLL)+(0,0)$) -- (5.36,4.65); 

 \coordinate (squareLL) at (2.2,6.8); 
  \coordinate (squareUR) at ($(squareLL)+(1.4,1.4)$); 
  \filldraw[fill=white, draw=black] (squareLL) rectangle (squareUR);
  \draw[<->] ($(squareLL)+(1.05,0.35)$) -- ($(squareUR)-(1.05,0.35)$);
  \draw[thick, red] ($(squareLL)+(1.4,0)$) -- (4.37,4.93); 

  \coordinate (squareLL) at (0.4,5.8); 
  \coordinate (squareUR) at ($(squareLL)+(1.4,1.4)$); 
  \coordinate (startPT) at ($(squareLL) + ({0.7},0.7)$);
  \coordinate (startPTellipse) at ($(startPT)+({cos(45/2)/2-0.04},{sin(-45/2)/2+0.01})$);
 \filldraw[fill=white, draw=black] (squareLL) rectangle (squareUR);
   \draw[rotate=(45), dashed, green] (startPT) ellipse [x radius=cos(135/2)/2, y radius=sin(-135/2)/2];
  \begin{scope}[rotate=45]
    \draw[<-] 
      ($(startPT) + (0.25,-0.5)$) arc[start angle=-60, end angle=0, x radius=cos(135/2)/2, y radius=sin(135/2)/2];
    \draw[->] (startPT) -- ($(startPT) + (0.15,-0.22)$); 
  \end{scope}
  \draw[thick, red] ($(squareLL)+(1.4,0)$) -- (3.09,4.93); 
  
 \coordinate (squareLL) at (0,3.7); 
  \coordinate (squareUR) at ($(squareLL)+(1.4,1.4)$);
  \coordinate (startPT) at ($(squareLL) + ({0.7},0.7)$);
  \coordinate (startPT) at ($(squareLL)+(0.7,0.7)$);
  \filldraw[fill=white, draw=black] (squareLL) rectangle (squareUR);
   \draw[rotate=0, dashed, green] (startPT) ellipse [x radius=0.707/2, y radius=0.707/2];
   \draw[->] (startPT) -- ($(startPT) + (0.25,-0.25)$);
  \begin{scope} [rotate=0]
    \draw[<-] 
      ($(startPT) + (0.2,-0.5)$) arc[start angle=-80, end angle=-20, x radius=0.707/2, y radius=0.707/2];
  \end{scope}
  \draw[thick, red] ($(squareLL)+(1.4,0)$) -- (2.23,4.6); 

  \coordinate (squareLL) at (0.4,1.6); 
   \coordinate (squareUR) at ($(squareLL)+(1.4,1.4)$); 
  \coordinate (startPT) at ($(squareLL) + ({0.7},0.7)$);
  \coordinate (startPTellipse) at ($(startPT)+({cos(315/2)/2-0.04},{sin(-315/2)/2+0.01})$);
 \filldraw[fill=white, draw=black] (squareLL) rectangle (squareUR);
   \draw[rotate=(-45), dashed, green] (startPT) ellipse [x radius=cos(135/2)/2, y radius=sin(-135/2)/2];
  \begin{scope}[rotate=-45]
    \draw[<-] 
      ($(startPT) + (0.25,-0.5)$) arc[start angle=-60, end angle=0, x radius=cos(135/2)/2, y radius=sin(135/2)/2];
    \draw[->] (startPT) -- ($(startPT) + (0.18,0.2)$);
  \end{scope}
  \draw[thick, red] ($(squareLL)+(1.4,1.4)$) -- (2.25,4.15); 
\end{scope}

\end{tikzpicture}
\end{center}
\caption{Polarization of various points on the x-y plane of the Bloch sphere.}
    \label{Blochxy}
\end{figure}

Figure~\ref{Blochxy} illustrates the evolution of the polarization of a photon starting in the diagonal $|+\rangle$ state and moving counterclockwise along the equator of the Bloch sphere. Along this trajectory, the polarization becomes elliptical until it reaches the circularly polarized $|i\rangle$ state ($y$-axis). It then becomes elliptical once more as it continues toward the anti-diagonal $|-\rangle$ state. One may note that states located on the hemisphere corresponding to the positive $y$-axis exhibit a counterclockwise rotation of the polarization, whereas states on the negative $y$-axis display clockwise rotation.
The equation for a state vector \cite{nielsen_chuang_2000} on the dashed blue line of figure~\ref{Blochxy} is
\begin{equation}
\label{GeneralEquationStateVector}
|\psi\rangle =\frac{1}{\sqrt{2}}|0\rangle + \frac{1}{\sqrt{2}}e^{i\phi}|1\rangle\
\end{equation}
where $\phi$ is the angle between the position of the state vector on the equator relative to the $|+\rangle$ state. When the Bloch vector lies along the positive $y$‑axis, the state is denoted $|i\rangle$ ($\phi = \frac{\pi}{2}$), and when it lies along the negative $y$‑axis, the state is $|-i\rangle$ ($\phi = -\frac{\pi}{2}$).

\subsection{Physical representation of the phase shift of a photon}
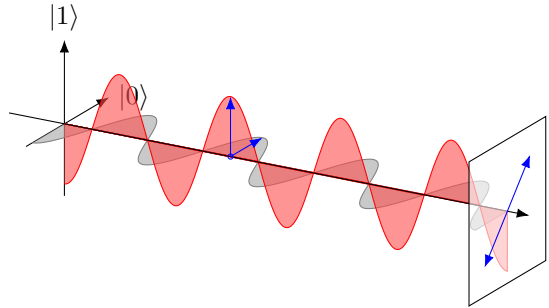
\begin{figure}
    \centering
\begin{center}

   \tdplotsetmaincoords{70}{30}
\begin{tikzpicture}[scale=1.7,tdplot_main_coords]

\draw[->] (0,-0.6,0) -- (0,0.7,0) node[anchor=west]{$|0\rangle$};
\draw[->] (0,0,-0.6) -- (0,0,0.7) node[anchor=south]{$|1\rangle$};

\foreach \i in {1,...,4}{
  \draw[draw=gray, fill=gray!70, variable=\x, domain={\i - 0.75}:{\i - 0.25}, samples=100,fill opacity=0.6] 
      (0,0,0) plot (\x,{-0.5*cos(2*\x*pi r)},0) -- 
      (0.25,0,0) -- (0,0,0);
}
\foreach \i in {1,...,4}{
  \draw[draw=red, fill=red!70, variable=\x, domain={\i - 0.75}:{\i - 0.25}, samples=100,fill opacity=0.6] 
      (0,0,0) plot (\x,0,{-0.5*cos(2*\x*pi r)}) -- 
      (0.25,0,0) -- (0,0,0);
}

\draw[draw=gray, fill=gray!70, variable=\x, domain=0:0.25,samples=50,fill  opacity=0.6] 
    (0,0,0) plot (\x,{-0.5*cos(2*\x*pi r)},0) -- 
    (0.25,0,0) -- (0,0,0);

\draw[draw=red, fill=red!70, variable=\x, domain=0:0.25,samples=50,fill  opacity=0.6] 
    (0,0,0) plot (\x,0,{-0.5*cos(2*\x*pi r)}) -- 
    (0.25,0,0) -- (0,0,0);

\foreach \i in {1,...,3}{
  \draw[draw=gray, fill=gray!70, variable=\x, domain={\i - 0.25}:{\i + 0.25}, samples=100,fill opacity=0.6] 
    (0,0,0) plot (\x,{-0.5*cos(2*\x*pi r)},0) -- 
    (0.25,0,0) -- (0,0,0);
}
\foreach \i in {1,...,3}{
  \draw[draw=red, fill=red!70, variable=\x, domain={\i - 0.25}:{\i + 0.25}, samples=100,fill opacity=0.6] 
    (0,0,0) plot (\x,0,{-0.5*cos(2*\x*pi r)}) -- 
    (0.25,0,0) -- (0,0,0);
}

\draw[draw=gray, fill=gray!70, variable=\x, domain=3.75:4,samples=50,fill  opacity=0.6] 
    (0,0,0) plot (\x,{-0.5*cos(2*\x*pi r)},0) -- 
    (4,0,0) -- (0,0,0);

\draw[draw=red, fill=red!70, variable=\x, domain=3.75:4,samples=50,fill  opacity=0.6] 
    (0,0,0) plot (\x,0,{-0.5*cos(2*\x*pi r)}) -- 
    (4,0,0) -- (0,0,0);

\draw (-0.5,0,0) -- (4,0,0) node[anchor=north east]{};
\draw[->, blue] (1.5,0,0) -- (1.5,0.5,0) node[anchor=north east]{};
\draw[blue] (1.5,0,0) circle (0.02 cm);
\draw[->, blue] (1.5,0,0) -- (1.5,0,0.5) node[anchor=north east]{};


\draw[fill=white,fill opacity=0.6] (4,-0.6,-0.6) -- (4,-0.6,0.6) -- (4,0.6,0.6) -- (4,0.6,-0.6) -- cycle;
\draw[->] (4,0,0) -- (4.2,0,0);
\draw[<->,blue] (4,-0.5/1.4,-0.5/1.4) -- (4,0.5/1.4,0.5/1.4); 

\end{tikzpicture}
    \caption{Decomposition of diagonal polarization into its horizontal and vertical components (blue arrows). The positive horizontal component coincides with the positive vertical component. The two waves are said to be in phase.}
    \label{SansPhaseShift}
\end{center}
\end{figure}

Figure~\ref{SansPhaseShift} shows the diagonal polarization of a photon in the $|+\rangle$ state and its horizontal and vertical components. Since the polarization vector $|+\rangle$ has a magnitude of 1, its horizontal and vertical components are respectively $\frac{1}{\sqrt{2}}|0\rangle$ and $\frac{1}{\sqrt{2}}|1\rangle$. In this instance, no phase is introduced since the positive vertical crests align with the positive horizontal crests (blue arrows). 

\begin{figure}
    \centering
\begin{center}
    
   \tdplotsetmaincoords{70}{30}
\begin{tikzpicture}[scale=1.7,tdplot_main_coords]

\draw[->] (0,-0.6,0) -- (0,0.7,0) node[anchor=west]{$|0\rangle$};
\draw[->] (0,0,-0.6) -- (0,0,0.7) node[anchor=south]{$|1\rangle$};

\foreach \i in {1,...,4}{
  \draw[draw=gray, fill=gray!70, variable=\x, domain={\i - 0.75}:{\i - 0.25}, samples=100,fill opacity=0.6] 
      (0,0,0) plot (\x,{-0.5*cos(2*\x*pi r)},0) -- 
      (0.25,0,0) -- (0,0,0);
}
\foreach \i in {1,...,4}{
  \draw[draw=red, fill=red!70, variable=\x, domain={\i - 0.75}:{\i - 0.25}, samples=100,fill opacity=0.6] 
      (0,0,0) plot (\x,0,{0.5*cos(2*\x*pi r)}) -- 
      (0.25,0,0) -- (0,0,0);
}

\draw[draw=gray, fill=gray!70, variable=\x, domain=0:0.25,samples=50,fill  opacity=0.6] 
    (0,0,0) plot (\x,{-0.5*cos(2*\x*pi r)},0) -- 
    (0.25,0,0) -- (0,0,0);

\draw[draw=red, fill=red!70, variable=\x, domain=0:0.25,samples=50,fill  opacity=0.6] 
    (0,0,0) plot (\x,0,{0.5*cos(2*\x*pi r)}) -- 
    (0.25,0,0) -- (0,0,0);

\foreach \i in {1,...,3}{
  \draw[draw=gray, fill=gray!70, variable=\x, domain={\i - 0.25}:{\i + 0.25}, samples=100,fill opacity=0.6] 
    (0,0,0) plot (\x,{-0.5*cos(2*\x*pi r)},0) -- 
    (0.25,0,0) -- (0,0,0);
}
\foreach \i in {1,...,3}{
  \draw[draw=red, fill=red!70, variable=\x, domain={\i - 0.25}:{\i + 0.25}, samples=100,fill opacity=0.6] 
    (0,0,0) plot (\x,0,{0.5*cos(2*\x*pi r)}) -- 
    (0.25,0,0) -- (0,0,0);
}

\draw[draw=gray, fill=gray!70, variable=\x, domain=3.75:4,samples=50,fill  opacity=0.6] 
    (0,0,0) plot (\x,{-0.5*cos(2*\x*pi r)},0) -- 
    (4,0,0) -- (0,0,0);

\draw[draw=red, fill=red!70, variable=\x, domain=3.75:4,samples=50,fill  opacity=0.6] 
    (0,0,0) plot (\x,0,{0.5*cos(2*\x*pi r)}) -- 
    (4,0,0) -- (0,0,0);

\draw(-0.5,0,0) -- (4,0,0) node[anchor=north east]{};
\draw[->, blue] (2,0,0) -- (2,0,0.5) node[anchor=north east]{};
\draw[blue] (2,0,0) circle (0.02 cm);
\draw[->, blue] (2.5,0,0) -- (2.5,0.5,0) node[anchor=north east]{};
\draw[blue] (2.5,0,0) circle (0.02 cm);

\draw[<->] (2.5,0,0.55) -- (2,0,0.55) node[midway,anchor=south]{$\frac{\lambda}{2}$};

\draw[fill=white,fill opacity=0.6] (4,-0.6,-0.6) -- (4,-0.6,0.6) -- (4,0.6,0.6) -- (4,0.6,-0.6) -- cycle;
\draw[->] (4,0,0) -- (4.2,0,0);
\draw[<->,blue] (4,-0.5/1.4,0.5/1.4) -- (4,0.5/1.4,-0.5/1.4); 

\end{tikzpicture}
    \caption{Decomposition of anti-diagonal polarization into its horizontal and vertical components. The positive horizontal component comes $\frac{\lambda}{2}$ before the positive vertical component. The two waves are said to have a phase of $\phi=\pi$ between them.}
    \label{PhaseShift}
\end{center}
\end{figure}
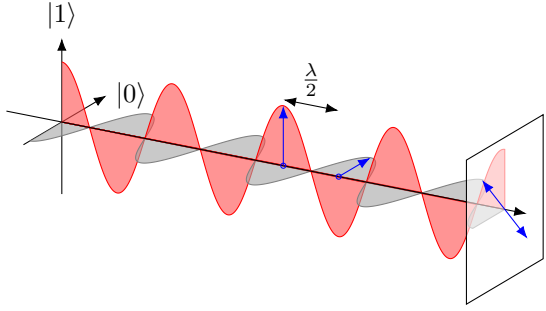

Figure~\ref{PhaseShift} presents the anti-diagonal polarization $|-\rangle$. The only difference in the vector decomposition compared to figure~\ref{SansPhaseShift} is a phase shift of $\frac{\lambda}{2}$ in the positive horizontal component relative to the positive vertical one (blue arrows). On the Bloch sphere (figure~\ref{Blochxy}), this phase difference corresponds to a shift of $\pi$ between the $|+\rangle$ and $|-\rangle$ states.

\begin{figure}
    \centering
\begin{center}
    
   \tdplotsetmaincoords{70}{30}
\begin{tikzpicture}[scale=1.7,tdplot_main_coords]

\draw[->] (0,-0.6,0) -- (0,0.7,0) node[anchor=west]{$|0\rangle$};
\draw[->] (0,0,-0.6) -- (0,0,0.7) node[anchor=south]{$|1\rangle$};

\foreach \i in {1,...,4}{
  \draw[draw=gray, fill=gray!70, variable=\x, domain={\i - 0.75}:{\i - 0.25}, samples=100,fill opacity=0.6] 
      (0,0,0) plot (\x,{0.5*cos(2*\x*pi r)},0) -- 
      (0.25,0,0) -- (0,0,0);
}
\foreach \i in {1,...,4}{
  \draw[draw=red, fill=red!70, variable=\x, domain={\i - 0.5}:{\i}, samples=100,fill opacity=0.6] 
      (0,0,0) plot (\x,0,{0.5*cos(2*(\x+0.25)*pi r)}) -- 
      (0.25,0,0) -- (0,0,0);
}

\draw[draw=gray, fill=gray!70, variable=\x, domain=0:0.25,samples=50,fill  opacity=0.6] 
    (0,0,0) plot (\x,{0.5*cos(2*\x*pi r)},0) -- 
    (0.25,0,0) -- (0,0,0);

\foreach \i in {1,...,3}{
  \draw[draw=gray, fill=gray!70, variable=\x, domain={\i - 0.25}:{\i + 0.25}, samples=100,fill opacity=0.6] 
    (0,0,0) plot (\x,{0.5*cos(2*\x*pi r)},0) -- 
    (0.25,0,0) -- (0,0,0);
}
\foreach \i in {0,...,3}{
  \draw[draw=red, fill=red!70, variable=\x, domain={\i}:{\i + 0.5}, samples=100,fill opacity=0.6] 
    (0,0,0) plot (\x,0,{0.5*cos(2*(\x+0.25)*pi r)}) -- 
    (0.25,0,0) -- (0,0,0);
}

\draw[draw=gray, fill=gray!70, variable=\x, domain=3.75:4,samples=50,fill  opacity=0.6] 
    (0,0,0) plot (\x,{0.5*cos(2*\x*pi r)},0) -- 
    (4,0,0) -- (0,0,0);

\draw(-0.5,0,0) -- (4,0,0) node[anchor=north east]{};
\draw[->, blue] (1.75,0,0) -- (1.75,0,0.5) node[anchor=north east]{};
\draw[blue] (1.75,0,0) circle (0.02 cm);
\draw[->, blue] (2,0,0) -- (2,0.5,0) node[anchor=north east]{};
\draw[green] (2,0,0) circle (0.02 cm);

\draw[<->] (2,0,0.55) -- (1.75,0,0.55) node[midway,anchor=south]{$\frac{\lambda}{4}$};


\draw[fill=white,fill opacity=0.6] (4,-0.6,-0.6) -- (4,-0.6,0.6) -- (4,0.6,0.6) -- (4,0.6,-0.6) -- cycle;
\draw[->] (4,0,0) -- (4.2,0,0);
\draw[->,blue] (4,0,0) -- (4,0.5/1.4,-0.5/1.4);
          \def\r{0.5}
          \def\startAngle{180}
          \def\endAngle{-180}
        
          \draw[dashed, blue,tdplot_main_coords,domain=\startAngle:\endAngle,smooth,variable=\t] 
            plot (4,{\r*cos(\t)},{\r*sin(\t)});

  \begin{scope} [canvas is yz plane at x=4]
    \draw[<-] (0.53,-0.33) arc (-35:-60:0.6);
  \end{scope}
    
\end{tikzpicture}
    \caption{Decomposition of counterclockwise circular polarization into its horizontal and vertical components. The positive horizontal component comes $\frac{\lambda}{4}$ before the positive vertical component. The two waves are said to have a phase of $\phi=\frac{\pi}{2}$ between the two.}
    \label{PhaseShiftRamdom}
\end{center}
\end{figure}
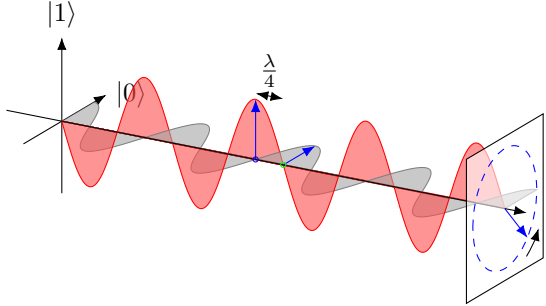

Figure~\ref{PhaseShiftRamdom} shows a circularly polarized photon. In this case, the phase shift between a positive horizontal crest and a positive vertical crest is $\frac{\lambda}{4}$ (that is, $\frac{\pi}{2}$ on the Bloch sphere). For circular polarization, the magnitude of the electric field (the length of the polarization vector) remains constant while the vector rotates uniformly in a circle. Any other phase shift would yield an elliptical polarization. Note: The sense of indicated rotation is always defined for a photon propagating toward the observer. However, it is crucial never to look directly into an incoming laser beam, as this may cause severe eye damage.

\subsection{Elliptical polarization on the y–z plane of the Bloch sphere}
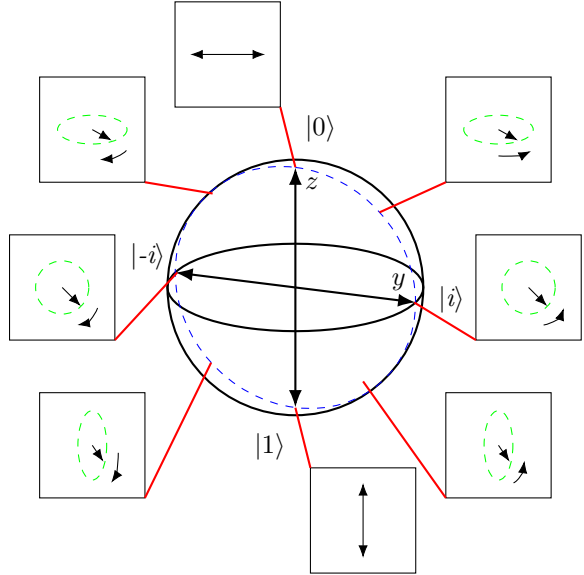
\begin{figure}
    \centering
\begin{center}
\tdplotsetmaincoords{70}{110} 
    \begin{tikzpicture}
    \begin{scope}[tdplot_main_coords, scale=1.7]

  \draw[thick] (0,0,0) circle (1); 
  \draw[thick, tdplot_screen_coords] (0,0) circle (1); 

  \draw[->, thick] (0,0,0) -- (0,0,1) node[anchor=south west, yshift=6pt] {$\ket{0}$} node[anchor=north west, yshift=0pt] {$z$};
  \draw[->, thick] (0,0,0) -- (0,0,-1) node[anchor=north east, yshift=-6pt] {$\ket{1}$};
  \draw[->, thick] (0,0,0) -- (0,1,0) node[anchor=west, xshift=5pt, yshift=1] {$\ket{\textit{i}}$} node[anchor=south east, xshift=0, yshift=1]{$y$};
  \draw[->, thick] (0,0,0) -- (0,-1,0) node[anchor=east, xshift=0pt, yshift=6] {$\ket{\textit{-i}}$};
   
  \end{scope}

  \begin{scope}[tdplot_main_coords, canvas is yz plane at x=0, scale=1.7]
    \draw[dashed, blue] (0,1) arc (90:-270:1);
  \end{scope}

 \begin{scope}[shift={(-3.8,-4.4)}]

  \coordinate (squareLL) at (2.2,6.8); 
  \coordinate (squareUR) at ($(squareLL)+(1.4,1.4)$); 
  \filldraw[fill=white, draw=black] (squareLL) rectangle (squareUR);
  \draw[<->] ($(squareLL)+(0.2,0.7)$) -- ($(squareUR)-(0.2,0.7)$);
  \draw[thick, red] ($(squareLL)+(1.4,0)$) -- (3.8,6.0); 

  \coordinate (squareLL) at (0.4,5.8); 
  \coordinate (squareUR) at ($(squareLL)+(1.4,1.4)$); 
  \coordinate (startPT) at ($(squareLL) + ({0.7},0.7)$);
  \coordinate (startPTellipse) at ($(startPT)+({cos(45/2)/2-0.04},{sin(-45/2)/2+0.01})$);
 \filldraw[fill=white, draw=black] (squareLL) rectangle (squareUR);
   \draw[rotate=0, dashed, green] (startPT) ellipse [x radius=cos(45/2)/2, y radius=sin(-45/2)/2];
  \begin{scope} [rotate=0]
    \draw[<-] 
      ($(startPT) + (0.1,-0.4)$) arc[start angle=-80, end angle=-20, x radius=cos(45/2)/2, y radius=sin(45/2)/2];
    \draw[->] (startPT) -- ($(startPT) + (0.26,-0.145)$); 
  \end{scope}
  \draw[thick, red] ($(squareLL)+(1.4,0)$) -- (2.7,5.65); 

  \coordinate (squareLL) at (0,3.7); 
  \coordinate (squareUR) at ($(squareLL)+(1.4,1.4)$);
  \coordinate (startPT) at ($(squareLL) + ({0.7},0.7)$);
  \coordinate (startPT) at ($(squareLL)+(0.7,0.7)$);
  \filldraw[fill=white, draw=black] (squareLL) rectangle (squareUR);
   \draw[rotate=0, dashed, green] (startPT) ellipse [x radius=0.707/2, y radius=0.707/2];
   \draw[->] (startPT) -- ($(startPT) + (0.25,-0.25)$);
  \begin{scope} [rotate=0]
    \draw[<-] 
      ($(startPT) + (0.2,-0.5)$) arc[start angle=-80, end angle=-20, x radius=0.707/2, y radius=0.707/2];
  \end{scope}
  \draw[thick, red] ($(squareLL)+(1.4,0)$) -- (2.23,4.6); 

  \coordinate (squareLL) at (0.4,1.6); 
   \coordinate (squareUR) at ($(squareLL)+(1.4,1.4)$); 
  \coordinate (startPT) at ($(squareLL) + ({0.7},0.7)$);
  \coordinate (startPTellipse) at ($(startPT)+({cos(135/2)/2-0.12},{sin(-135/2)/2+0.28})$);
 \filldraw[fill=white, draw=black] (squareLL) rectangle (squareUR);
   \draw[rotate=(0), dashed, green] (startPT) ellipse [x radius=cos(135/2)/2, y radius=sin(-135/2)/2];
  \begin{scope}[rotate=0]
    \draw[<-] 
      ($(startPT) + (0.25,-0.5)$) arc[start angle=-60, end angle=0, x radius=cos(135/2)/2, y radius=sin(135/2)/2];
    \draw[->] (startPT) -- ($(startPT) + (0.15,-0.22)$);
  \end{scope}
  \draw[thick, red] ($(squareLL)+(1.4,0)$) -- (2.68,3.4); 

 \coordinate (squareLL) at (4,0.6); 
  \coordinate (squareUR) at ($(squareLL)+(1.4,1.4)$); 
  \filldraw[fill=white, draw=black] (squareLL) rectangle (squareUR);
  \draw[<->] ($(squareLL)+(0.7,0.2)$) -- ($(squareUR)-(0.7,0.2)$);
  \draw[thick, red] ($(squareLL)+(0,1.4)$) -- (3.8,2.8); 

  \coordinate (squareLL) at (5.8,1.6); 
   \coordinate (squareUR) at ($(squareLL)+(1.4,1.4)$); 
  \coordinate (startPT) at ($(squareLL) + ({0.7},0.7)$);
  \coordinate (startPTellipse) at ($(startPT)+({cos(135/2)/2-0.12},{sin(-135/2)/2+0.28})$);
 \filldraw[fill=white, draw=black] (squareLL) rectangle (squareUR);
   \draw[rotate=(0), dashed, green] (startPT) ellipse [x radius=cos(135/2)/2, y radius=sin(-135/2)/2];
  \begin{scope}[rotate=0]
    \draw[->] 
      ($(startPT) + (0.2,-0.5)$) arc[start angle=-80, end angle=-20, x radius=cos(135/2)/2, y radius=sin(135/2)/2];
    \draw[->] (startPT) -- ($(startPT) + (0.15,-0.22)$);
  \end{scope}
  \draw[thick, red] ($(squareLL)+(0,0)$) -- (4.7,3.15); 

  \coordinate (squareLL) at (6.2,3.7); 
  \coordinate (squareUR) at ($(squareLL)+(1.4,1.4)$);
  \coordinate (startPT) at ($(squareLL) + ({0.7},0.7)$);
  \coordinate (startPT) at ($(squareLL)+(0.7,0.7)$);
  \filldraw[fill=white, draw=black] (squareLL) rectangle (squareUR);
   \draw[rotate=0, dashed, green] (startPT) ellipse [x radius=0.707/2, y radius=0.707/2];
   \draw[->] (startPT) -- ($(startPT) + (0.25,-0.25)$);
  \begin{scope} [rotate=0]
    \draw[->] 
      ($(startPT) + (0.2,-0.5)$) arc[start angle=-80, end angle=-20, x radius=0.707/2, y radius=0.707/2];
  \end{scope}
  \draw[thick, red] ($(squareLL)+(0,0)$) -- (5.38,4.2); 

  \def\A{315}
  \coordinate (squareLL) at (5.8,5.8); 
  \coordinate (squareUR) at ($(squareLL)+(1.4,1.4)$); 
  \coordinate (startPT) at ($(squareLL) + ({0.7},0.7)$);
  \coordinate (startPTellipse) at ($(startPT)+({cos(\A/2)/2-0.04},{sin(-\A/2)/2+0.01})$);
 \filldraw[fill=white, draw=black] (squareLL) rectangle (squareUR);
   \draw[rotate=0, dashed, green] (startPT) ellipse [x radius=cos(\A/2)/2, y radius=sin(-\A/2)/2];
  \begin{scope} [rotate=0]
    \draw[->] 
      ($(startPT) + (0,-0.35)$) arc[start angle=-100, end angle=-40, x radius=cos(45/2)/2, y radius=sin(45/2)/2];
    \draw[->] (startPT) -- ($(startPT) + (0.26,-0.145)$); 
  \end{scope}
  \draw[thick, red] ($(squareLL)+(0,0)$) -- (4.92,5.4); 
\end{scope}
\end{tikzpicture}
\caption{Polarization of various points on the y-z plane of the Bloch sphere.}
    \label{Blochyz}
\end{center}
\end{figure}

If one starts in the $|0\rangle$ state and moves clockwise along the meridian of the Bloch sphere in the $y$–$z$ plane, following the dashed blue line in figure~\ref{Blochyz}, the polarization evolves from the horizontal linear polarization of the $|0\rangle$ state to an elliptical polarization whose short axis is vertical. As one progresses along the meridian, the vertical axis gradually increases in magnitude until reaching circular polarization, corresponding to the $|i\rangle$ state. Continuing toward the $|1\rangle$ state, the vertical axis becomes the long axis of the ellipse while the horizontal axis becomes the short one.

It is worth noting that the representation of quantum states on the Bloch sphere is analogous to how physicists use the Poincar\'e sphere in photonics \cite{Shen2020_SU2_Poincare}.

Figures~\ref{Blochxz},~\ref{Blochxy}, and~\ref{Blochyz} illustrate how each point on the Bloch sphere corresponds to a specific polarization pattern of a photon. This correspondence shows that polarization provides a natural means of encoding information in a photonic qubit.

\section{Modifying the polarization of a photon}
\label{SectionModifyingPol}
One can change the state of a photonic qubit by applying a quantum logic gate. Before introducing the different gates, it is important to understand how the physical properties of certain materials can be used to modify the polarization of a photon. The following section presents key material behaviours that are required to fabricate such logic gates.
\subsection{Rotation of linearly polarized photons by optically active materials}
Linearly polarized light can be rotated using \emph{optically active} materials. These materials possess a chiral chemical structure that causes a linearly polarized photon to rotate within the plane perpendicular to the photon’s direction of propagation. One well-known optically active substance is sucrose, which rotates the polarization clockwise (equivalently, the upper part of the electric‑field vector tilts to the right). Another is fructose, which rotates polarization counterclockwise \cite{ParraCordova2025}.

\begin{figure}
    \centering
\begin{center}

   \tdplotsetmaincoords{70}{30}
\begin{tikzpicture}[scale=1.7,tdplot_main_coords]

\foreach \i in {1,...,1.25}{
  \draw[draw=gray, fill=gray!70, variable=\x, domain={\i - 0.75}:{\i - 0.25}, samples=100,fill opacity=0.6] 
      (0,0,0) plot (\x,{-0.5*cos(2*\x*pi r)},0) -- 
      (0.25,0,0) -- (0,0,0);
}

\draw[draw=gray, fill=gray!70, variable=\x, domain=0:0.25,samples=50,fill opacity=0.6] 
    (0,0,0) plot (\x,{-0.5*cos(2*\x*pi r)},0) -- 
    (0.25,0,0) -- (0,0,0);

\foreach \i in {1,...,1.25}{
  \draw[draw=gray, fill=gray!70, variable=\x, domain={\i - 0.25}:{\i + 0.25}, samples=100,fill opacity=0.6] 
    (0,0,0) plot (\x,{-0.5*cos(2*\x*pi r)},0) -- 
    (0.25,0,0) -- (0,0,0);
}

\draw(-0.5,0,0) -- (1.25,0,0) node[anchor=north east]{};

\draw[fill=white,fill opacity=0.6] (1.25,-0.6,-0.6) -- (1.25,-0.6,0.6) -- (1.25,0.6,0.6) -- (1.25,0.6,-0.6) -- cycle;
\draw[fill=white,fill opacity=0.6] (1.3,-0.55,0.4) -- (1.3,-0.55,0.6) -- (1.3,0.55,0.6) -- (1.3,0.55,0.4) -- cycle;

\draw[fill=gray,fill opacity=0.6] (1.3,-0.55,-0.55) -- (1.3,-0.55,0.4) -- (1.3,0.55,0.4) -- (1.3,0.55,-0.55) -- cycle;
\draw[] (1.3,0.55,-0.55) -- (2.2,0.55,-0.55);
\draw[fill=gray,fill opacity=0.6] (1.3,-0.55,0.4) -- (1.3,0.55,0.4) -- (2.2,0.55,0.4) -- (2.2,-0.55,0.4) -- cycle;
\draw[fill=gray,fill opacity=0.6] (1.3,-0.55,-0.55) -- (1.3,-0.55,0.4) -- (2.2,-0.55,0.4) -- (2.2,-0.55,-0.55) -- cycle;
\draw[fill=gray,fill opacity=0.6] (2.2,-0.55,-0.55) -- (2.2,-0.55,0.4) -- (2.2,0.55,0.4) -- (2.2,0.55,-0.55) -- cycle;

\draw[] (1.3,0.55,0.4) -- (2.2,0.55,0.4);
\draw[] (1.3,-0.55,0.6) -- (1.3,0.55,0.6) -- (2.2,0.55,0.6) -- (2.2,-0.55,0.6) -- cycle;
\draw[fill=white,fill opacity=0.6] (1.3,-0.55,0.4) -- (1.3,-0.55,0.6) -- (2.2,-0.55,0.6) -- (2.2,-0.55,0.4) -- cycle;
\draw[fill=white,fill opacity=0.6] (2.2,-0.55,0.4) -- (2.2,-0.55,0.6) -- (2.2,0.55,0.6) -- (2.2,0.55,0.4) -- cycle;

\draw[] (1.25,0.6,-0.6) -- (2.25,0.6,-0.6);
\draw[] (1.25,-0.6,0.6) -- (1.25,0.6,0.6) -- (2.25,0.6,0.6) -- (2.25,-0.6,0.6) -- cycle;
\draw[fill=white,fill opacity=0.6] (1.25,-0.6,-0.6) -- (1.25,-0.6,0.6) -- (2.25,-0.6,0.6) -- (2.25,-0.6,-0.6) -- cycle;
\draw[fill=white,fill opacity=0.6] (2.25,-0.6,-0.6) -- (2.25,-0.6,0.6) -- (2.25,0.6,0.6) -- (2.25,0.6,-0.6) -- cycle;

\draw[<->] (1.3,-0.65,-0.65) -- (2.2,-0.65,-0.65) node[rotate=-10, anchor=north west, xshift=-38, yshift=0] {100mm};

\draw[<->, blue] (0,-0.5) -- (0,0.5,0) node[anchor=west]{$|0\rangle$};
\draw[->] (0,0,-0.6) -- (0,0,0.7) node[anchor=south]{$|1\rangle$};

\foreach \i in {2.25,...,4}{
  \draw[draw=gray, fill=gray!70, variable=\x, domain=2.25:4, samples=100,fill opacity=0.6] 
    (2.25,0,0) plot (\x,{-0.5*cos(2*\x*pi r)*cos(-66.5)},{-0.5*cos(2*\x*pi r)*sin(-66.5)}) -- 
    (4,0,0) -- (2.25,0,0);
}

\draw[fill=white,fill opacity=0.6] (4,-0.6,-0.6) -- (4,-0.6,0.6) -- (4,0.6,0.6) -- (4,0.6,-0.6) -- cycle;

\draw[<->,blue] (4,{-0.5*cos(66.5)},{0.5*sin(66.5)}) -- (4,{0.5*cos(66.5)},{-0.5*sin(66.5)}) node[anchor=north east, xshift=0, yshift=4]{$|\psi\rangle$};

\draw[dashed] (4,-0.5,0) -- (4,0.5,0);

  \begin{scope} [canvas is yz plane at x=4]
    \draw[->] (0.3,0) arc (-0.2:-66.5:0.3); 
  \end{scope}

\end{tikzpicture}
    \caption{Rotation of linearly polarized 589nm ray of light from the horizontal plane to a plane tilted at 66.5° clockwise through a sucrose solution of 10g per 100mL concentration and 100mm long container.}
    \label{opticallyActive}
\end{center}
\end{figure}
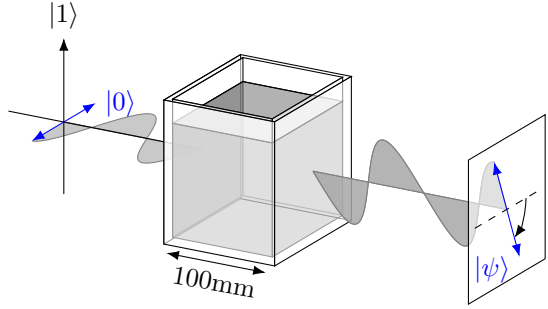

The “strength’’ with which a material rotates the polarization of light is quantified by its specific rotation. For a sucrose solution in water, the specific rotation $\alpha_s$ is
\begin{equation}
[\alpha_s]^{20}_{589} = \frac{+66.5^\circ}{100\,\text{mm}}\
\end{equation}
while, for a fructose solution the specific rotation $\alpha_f$ is
\begin{equation}
[\alpha_f]^{20}_{589} = \frac{-92.4^\circ}{100\,\text{mm}}\
\end{equation}
In both expressions, the upper index 20 indicates that the values are for a solution at $20^\circ\mathrm{C}$, while the lower index 589 indicates the value is for light with a  wavelength of $589\mathrm{nm}$ \cite{BigChemicalEncyclopedia_p271,Miljkovic2009_Isomerization,PASCO_Polarimetry}.
Thus, if a linearly polarized photon of such a wavelength passes through $100\mathrm{mm}$ of a sucrose solution at a concentration of $10\mathrm{g}$ per $100\mathrm{mL}$, its polarization would rotate clockwise by $66.5^\circ$ (figure~\ref{opticallyActive}). Under the same conditions, passage through $100\mathrm{mm}$ of a fructose solution would rotate the polarization by $92.4^\circ$ counterclockwise. 

\subsection{Birefringence}
\label{SectionBirefringence}
Birefringence is the property of certain materials to exhibit two (or more) distinct refractive indices \textit{n} when photons propagate along different axes \cite{raymer_2017,RefSlepkov}. Many materials display birefringence, including anisotropic minerals with crystalline structures, such as hexagonal or tetragonal lattices \cite{klein_hurlbut_1993}. Non‑crystalline materials can also be birefringent, among them several commercially available transparent adhesive tapes \cite{EdwardsLangley1981}.

The optical axis of a birefringent material is the direction along which light experiences no birefringence. It corresponds to the axis along which electrons move more freely. This results in photons propagate faster when their electric field is aligned parallel to this axis; i.e., when their polarization vector is completely parallel to the optical axis. This defines the fast axis. In contrast, photons propagate more slowly when their polarization vector  is perpendicular to the optical axis, which defines the slow axis. In general, we define the refractive index as the ratio of the speed of light in vacuum to its speed in the material. Given the photon speed depends on whether its polarization vector is parallel or perpendicular to the optical axis, we can define two indices of refraction for the material: one for the fast axis $n_f$ and one for the slow axis $n_s$. The birefringence $\Delta n=n_s-n_f$ is defined as the difference between the refractive index of the slow axis and that of the fast axis \cite{EdwardsLangley1981}.

The reason light propagates more easily along certain directions within the material is its electronic structure; i.e., the shape of the electron orbitals. Because electrons in the polymer can move more readily parallel to this elongated structure, the interaction of a photon with the electronic structure produces a faster path when the electric field is parallel to that direction. In transparent polymer films, the optical axis typically aligns with the edges of the tape. This alignment arises from the stretching of the plastic during fabrication, which orients the polymer chains along the direction of the applied tension \cite{RefSlepkov,EdwardsLangley1981}.

\begin{figure}
    \centering
   \tdplotsetmaincoords{70}{30}
\begin{tikzpicture}[scale=1.7,tdplot_main_coords]

\foreach \i in {1,...,2}{
  \draw[draw=gray, fill=gray!70, variable=\x, domain={\i - 0.75}:{\i - 0.25}, samples=100,fill opacity=0.6] 
      (0,0,0) plot (\x,{-0.5*cos(2*\x*pi r)},0) -- 
      (0.25,0,0) -- (0,0,0);
}
\foreach \i in {1,...,2}{
  \draw[draw=red, fill=red!70, variable=\x, domain={\i - 0.75}:{\i - 0.25}, samples=100,fill opacity=0.6] 
      (0,0,0) plot (\x,0,{-0.5*cos(2*\x*pi r)}) -- 
      (0.25,0,0) -- (0,0,0);
}

\draw[draw=gray, fill=gray!70, variable=\x, domain=0:0.25,samples=50,fill opacity=0.6] 
    (0,0,0) plot (\x,{-0.5*cos(2*\x*pi r)},0) -- 
    (0.25,0,0) -- (0,0,0);

\draw[draw=red, fill=red!70, variable=\x, domain=0:0.25,samples=50,fill opacity=0.6] 
    (0,0,0) plot (\x,0,{-0.5*cos(2*\x*pi r)}) -- 
    (0.25,0,0) -- (0,0,0);

\foreach \i in {1,...,1.5}{
  \draw[draw=gray, fill=gray!70, variable=\x, domain={\i - 0.25}:{\i + 0.25}, samples=100,fill opacity=0.6] 
    (0,0,0) plot (\x,{-0.5*cos(2*\x*pi r)},0) -- 
    (0.25,0,0) -- (0,0,0);
}
\foreach \i in {1,...,1.5}{
  \draw[draw=red, fill=red!70, variable=\x, domain={\i - 0.25}:{\i + 0.25}, samples=100, fill opacity=0.6] 
    (0,0,0) plot (\x,0,{-0.5*cos(2*\x*pi r)}) -- 
    (0.25,0,0) -- (0,0,0);
}

\draw[draw=gray, fill=gray!70, variable=\x, domain=1.75:2,samples=50, fill opacity=0.6] 
    (0,0,0) plot (\x,{-0.5*cos(2*\x*pi r)},0) -- 
    (2,0,0) -- (0,0,0);

\draw[draw=red, fill=red!70, variable=\x, domain=1.75:2,samples=50, fill opacity=0.6] 
    (0,0,0) plot (\x,0,{-0.5*cos(2*\x*pi r)}) -- 
    (2,0,0) -- (0,0,0);

\draw (-0.5,0,0) -- (2,0,0) node[anchor=north east]{};

\draw[fill=white,fill opacity=0.6] (2,-1.2,-0.6) -- (2,-1.2,0.6) -- (2.25,-1.2,0.6) -- (2.25,-1.2,-0.6) -- cycle;
\draw[fill=white,fill opacity=0.6] (2,-1.2,0.6) -- (2,1.2,0.6) -- (2.25,1.2,0.6) -- (2.25,-1.2,0.6) -- cycle;
\draw[fill=white,fill opacity=0.6] (2,-1.2,-0.6) -- (2,-1.2,0.6) -- (2,1.2,0.6) -- (2,1.2,-0.6) -- cycle;

\draw[] (2,1.2,-0.6) -- (2.2,1.2,-0.6);

\draw[fill=white,fill opacity=0.6] (2.25,-1.2,-0.6) -- (2.25,-1.2,0.6) -- (2.25,1.2,0.6) -- (2.25,1.2,-0.6) -- cycle;
\draw[<->] (2.25,-0.4,-0.3) -- (2.25,0.4,-0.3) node[rotate=30, anchor=south east, xshift=0, yshift=0] {\tiny {O.A.}};

\draw[->] (0,-0.6,0) -- (0,0.7,0) node[anchor=west]{$|0\rangle$};
\draw[->] (0,0,-0.6) -- (0,0,0.7) node[anchor=south]{$|1\rangle$};
\draw[<->, blue] (0,-0.5/1.4,-0.5/1.4) -- (0,0.5/1.4,0.5/1.4) node [anchor=south west] {$|+\rangle$};

\foreach \i in {2.25,...,4}{
  \draw[draw=gray, fill=gray!70, variable=\x, domain=2.25:4, samples=100,fill opacity=0.6] 
    (2.25,0,0) plot (\x,{-0.5*cos(2*\x*pi r)},0) -- 
    (4,0,0) -- (2.25,0,0);
}
\foreach \i in {2.25,...,4}{
  \draw[draw=red, fill=red!70, variable=\x, domain=2.25:4, samples=100,fill opacity=0.6] 
    (2.25,0,0) plot (\x,0,{-0.5*cos(2*(\x+0.25)*pi r)}) -- 
    (4,0,0) -- (2.25,0,0);
}

\draw[fill=white,fill opacity=0.6] (4,-0.6,-0.6) -- (4,-0.6,0.6) -- (4,0.6,0.6) -- (4,0.6,-0.6) -- cycle;
\draw[->] (4,0,0) -- (4.2,0,0);
\draw[->, blue] (3.25,0,0) -- (3.25,0,0.5) node[anchor=north east]{};
\draw[->, blue] (3.5,0,0) -- (3.5,0.5,0) node[anchor=north east]{};

\draw[->,blue] (4,0,0) -- (4,0.5/1.4,-0.5/1.4);
          \def\r{0.5}
          \def\startAngle{180}
          \def\endAngle{-180}
        
          \draw[dashed, green,tdplot_main_coords,domain=\startAngle:\endAngle,smooth,variable=\t] 
            plot (4,{\r*cos(\t)},{\r*sin(\t)});

  \begin{scope} [canvas is yz plane at x=4]
    \draw[<-] (0.53,-0.33) arc (-35:-60:0.6);
  \end{scope}

\draw[<->] (3.25,0,0) -- (3.5,0,0) node[anchor=north east, xshift=-1, yshift=2] {$\frac{\lambda}{4}$};
    
\end{tikzpicture}
    \caption{A diagonal polarized ray of light enters a so-called quarter-wavelength ($\frac{\lambda}{4}$) plate creating a phase shift of $\phi=\frac{\pi}{2}$ resulting in a counterclockwise circular polarization. The wave plate is made of a birefringent material for which the optical axis (O.A.) is horizontal and for which light travels faster when its electric field is parallel to it.}
    \label{birefringence}
\end{figure}
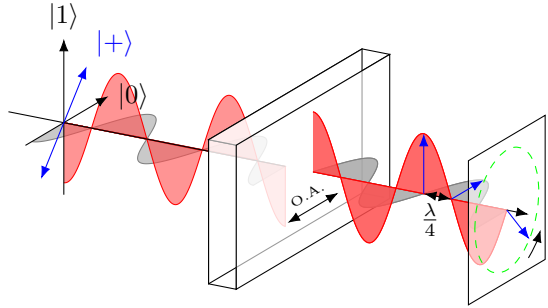

Figure~\ref{birefringence} illustrates a so-called quarter‑wavelength ($\frac{\lambda}{4}$) plate in a configuration where the horizontal component of the polarization vector propagates along the optical axis of the birefringent material, while the vertical component propagates more slowly \cite{RefTagaya}. As a result, the vertical component is one quarter of a wavelength behind the horizontal component. Under these conditions, an incoming diagonally polarized photon is transformed by the birefringent material into a counterclockwise circularly polarized photon. If the phase shift deviates from exactly $\frac{\lambda}{4}$, the output polarization becomes elliptical instead \cite{EdwardsLangley1981}.

\subsection{The case of calcite}
\label{SectionCaseofCalcite}
\begin{figure}
  \centering
  \subfloat[]{
    \includegraphics[width=0.3\textwidth]{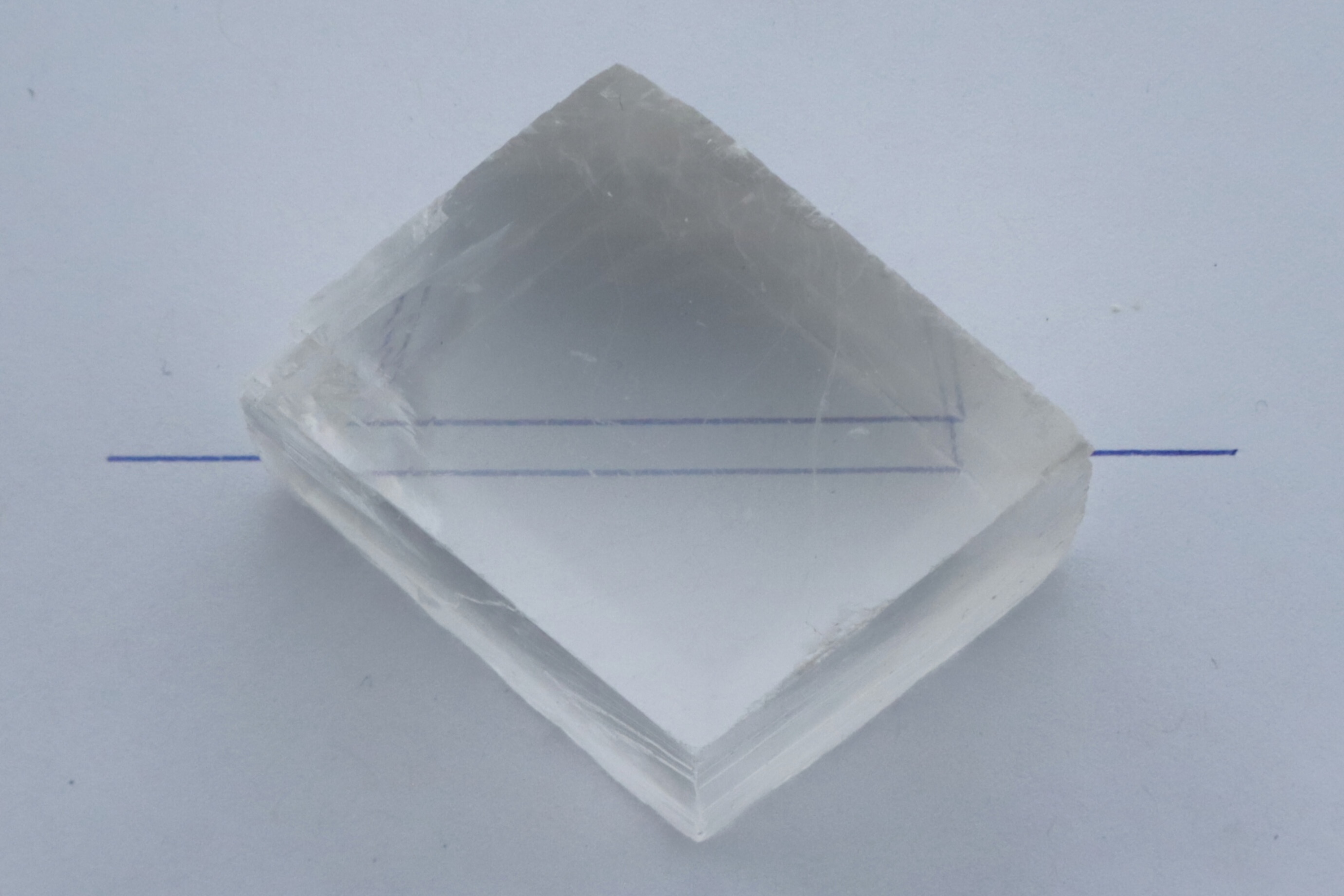}
    \label{PhotoCalcite}
  }
  \hfill
  \subfloat[]{
    \includegraphics[width=0.3\textwidth]{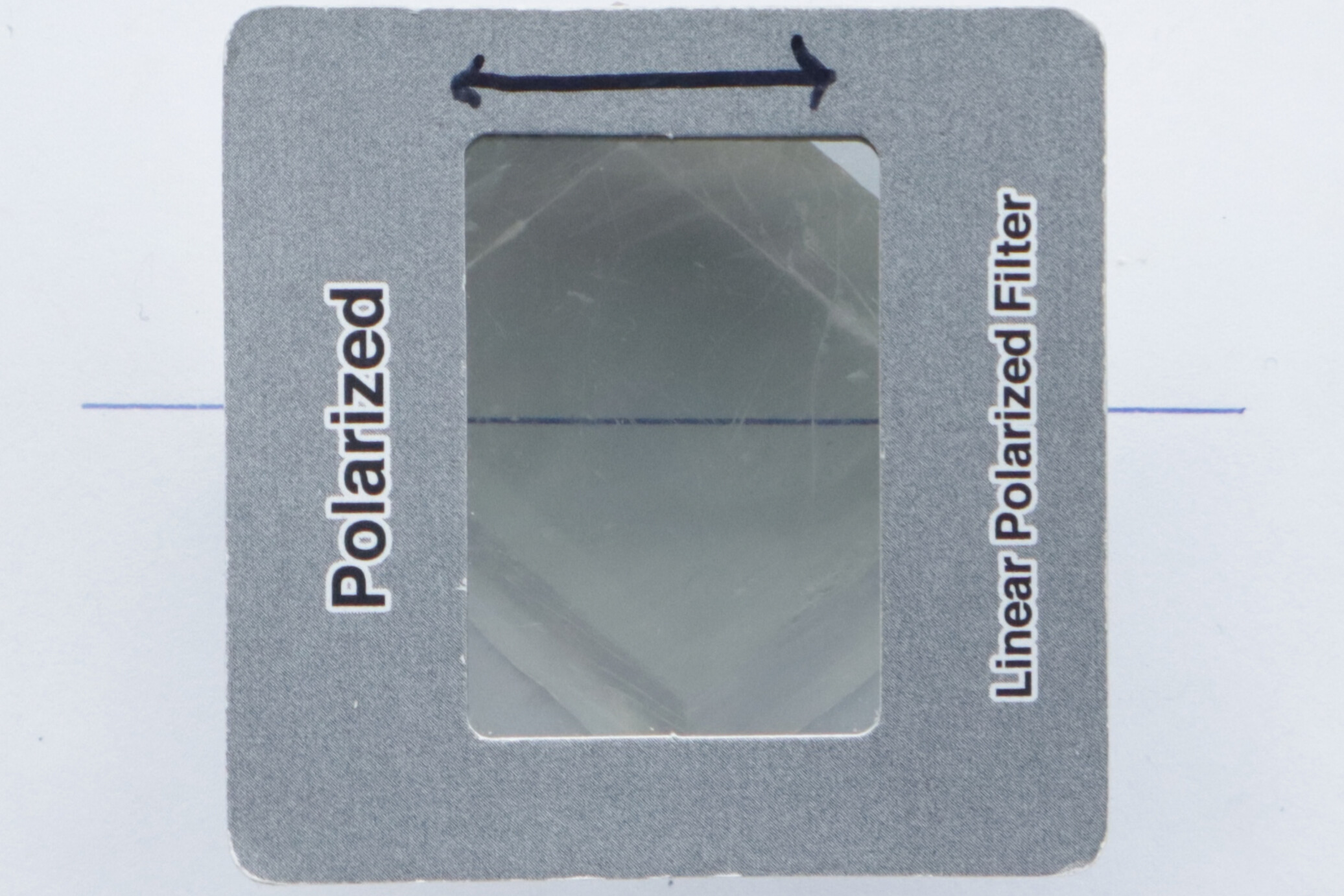}
    \label{PhotoCalciteH}
  }
  \hfill
  \subfloat[]{
    \includegraphics[width=0.3\textwidth]{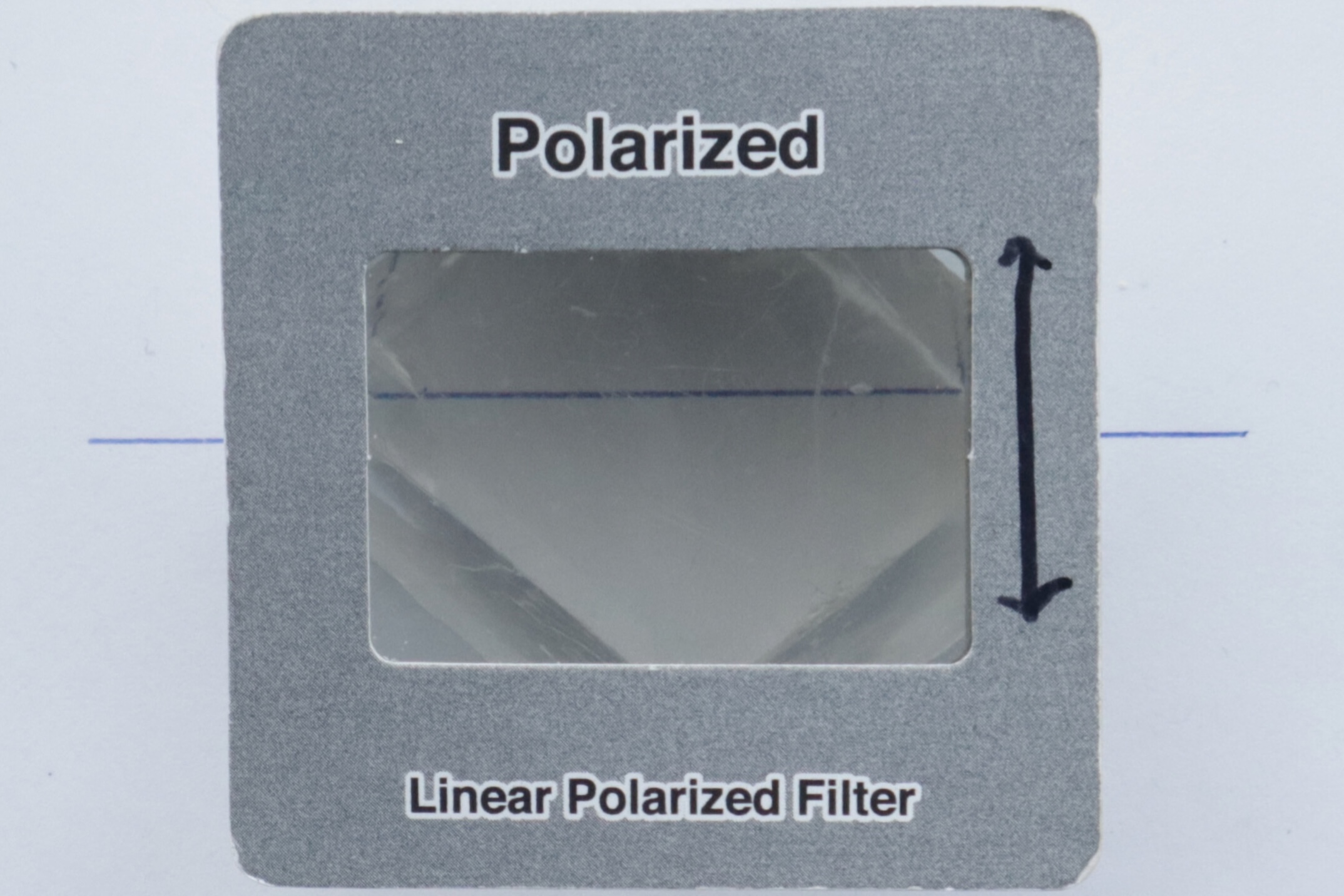}
    \label{PhotoCalciteV}
  }
  \caption{A clear calcite crystal is placed over a horizontal straight line drawn on a white sheet of paper, resulting in two visible lines (a). The lower path is called the ordinary ray and follows Snells' law, whereas the upper path is called the extraordinary ray. For this ray, the angle of refraction is determined by the optical axis, which is not parallel to the surface of the crystal. The ordinary ray is horizontally polarized, as shown by the horizontal polarizer placed on top of the crystal (b) and the extraordinary ray is vertically polarized and displaced upward, as shown by the vertical polarizer placed on top of the crystal (c).}
  \label{PhotosCalcite}
\end{figure}

Calcite is a very common mineral found in calcareous rocks such as limestone, as well as in the shells of many animals with exoskeletons, including mussels, clams, and coral. Its chemical composition is CaCO$_3$, and it crystallizes into a tetragonal lattice. Naturally occurring, highly transparent calcite crystals can be found in many regions worldwide. These crystals are well known for producing double images when placed over simple drawings or printed text (figure~\ref{PhotoCalcite}) \cite{klein_hurlbut_1993,duda_rejl_1991,RefNikon}.

Notice that birefringence manifests differently in transparent tape compared to calcite; i.e., the tape does not produce a double image. This is because, for calcite, the optical axis is not parallel to the naturally occurring faces of the crystal \cite{klein_hurlbut_1993}, but instead lies at an angle of about 44.6$^\circ$ relative to the surface normal \cite{SkalwoldBassett2015}. When a photon enters a calcite crystal perpendicular to one of these faces, the so‑called ordinary ray propagates straight through the crystal without deviation, in accordance with Snell’s law. As can be seen in figure~\ref{PhotoCalciteH}, the ordinary ray consists of horizontally polarized light. In the photos of figure~\ref{PhotosCalcite}, the crystal is positioned such that the optical axis is tilted towards the top of the image with respect to the surface normal, which aligns with the observer’s line of sight. The extraordinary ray, which is vertically polarized, is refracted toward the optical axis and thus appears displaced towards the top of figure~\ref{PhotoCalciteV}.

When a single photon with a linearly polarized electric field enters a calcite crystal at an angle relative to the polarization directions of the ordinary and extraordinary rays, it emerges in a quantum superposition of two spatially separated paths. Upon measurement, if the photon is detected along the ordinary-ray path (the lower path in figure~\ref{PhotoCalciteH}), it is found to be horizontally polarized. Conversely, if it is detected along the extraordinary-ray path (the upper path in figure~\ref{PhotoCalciteV}), it is found to be vertically polarized \cite{klein_hurlbut_1993}.

Light thus offers several ways of encoding quantum information. That is, in addition to polarization, the \textit{propagation path} of a photon can be used to form a qubit, as seen with the possibility of using calcite to create a superposition between two optical paths.

\section{Materials and method for building an analogue simulator of a photonic quantum computer}
In section \ref{SectionModifyingPol}, we discussed the fundamental physical properties needed to modify the polarization state of photons, namely optical activity and birefringence. Here, we discuss exactly what materials to use and how to put them together to create an analogue simulator of a photonic quantum computer.
\subsection{$R_y$ gates: maple syrup and agave syrup}
To fabricate an $R_y$ gate (figure \ref{fig:RyBloch}), it is essential to keep the polarization of the photon linear throughout the process if the polarization was originally linear. For elliptically polarized light, we want to keep the two axes of the ellipse the same relative lengths while rotating the ellipse. Optically active materials such as fructose or sucrose solutions are well suited for this purpose.

One approach is to prepare solutions with the appropriate concentration of these sugars so as to obtain the desired rotation of the polarization, using fructose for a counterclockwise rotation, thus an $R_y$ gate, and sucrose for a clockwise rotation, thus an $R_y^\dagger$ gate. To obtain a targeted rotation angle, we adjust the sugar concentration based on the wavelength and specific container width. Note that chemists attribute a positive sign to a clockwise rotation, which corresponds to the physicists' $R_y^\dagger$ gate, and a negative sign to a counterclockwise rotation, which corresponds to the $R_y$ gate.

For our apparatus, we used common grocery store granulated sugar (a sucrose-rich substance) \cite{PubChemSucrose2026} and common grocery store fructose. For the sake of calibration, we determined their specific rotations for lasers with wavelengths of 405\,nm (blue), 532\,nm (green), and 650\,nm (red). The Drude equation gives an approximation of the optical power of a solution \cite{ParraCordova2025,Soetedjo2014_DrudeSugars,Guclu2004_DrudeGlucose, Mahurin1999} and is expressed as: 
\begin{equation}
[\alpha]_{\lambda} = \frac{A}{\lambda^2-\lambda_0^2}\
\end{equation}

where $\alpha$ is the specific rotation, $A$ is the rotation constant, $\lambda_0$ is the dispersion constant, and $\lambda$ the wavelength of interest. 

The magnitudes of the specific rotations measured with our setup were lower than those predicted by the Drude equation, as shown in figure~\ref{graphOpticalPower}. Such reduced values are expected, since these available sugars are not laboratory-grade and contain impurities. Nonetheless, both substances can be used to fabricate $R_y$ gates with any desired rotation.

Alternatively, one can use agave syrup (a fructose-rich solution) \cite{Saraiva2022AgaveSyrup} or maple syrup (a sucrose-rich solution) \cite{Saraiva2022MapleSyrup}, diluting them to achieve the desired rotation. By selecting the appropriate solution concentration for a given sugar type, container, and wavelength, one can implement any required $R_y$ gate.

\begin{figure}[htbp]
\centering

\begin{minipage}[c]{0.35\linewidth}
\centering
\begin{quantikz}
\lstick{$q_0\;\ket{0}$} & \gate{R_y(\theta)} & \qw
\end{quantikz}
\end{minipage}
\hfill
\begin{minipage}[c]{0.6\linewidth}
\centering

\tdplotsetmaincoords{70}{110}
\begin{tikzpicture}[tdplot_main_coords, scale=2]

  \draw[thick] (0,0,0) circle (1);
  \draw[thick, tdplot_screen_coords] (0,0) circle (1);

  \draw[->] (0,0,0) -- (0,0,1) node[anchor=south, yshift=6pt] {$\ket{0}$};
  \draw[->] (0,0,0) -- (0,0,-1) node[anchor=north, yshift=-6pt] {$\ket{1}$};
  \draw[->] (0,0,0) -- (1,0,0) node[anchor=north, xshift=-10] {$\ket{+}$};
  \draw[->] (0,0,0) -- (-1,0,0) node[anchor=south, xshift=10] {$\ket{-}$};
  \draw[->] (0,0,0) -- (0,1,0) node[anchor=west, xshift=6pt, yshift=2] {$\ket{\textit{i}}$};
  \draw[->] (0,0,0) -- (0,-1,0) node[anchor=east, xshift=-6pt, yshift=7] {$\ket{\textit{-i}}$};

  \draw[->] (0,1.4,0) -- (0,1.9,0) node[anchor=west] {$y$};

  \begin{scope}[canvas is xz plane at y=1.7]
    \draw[blue,->] (-0.16,0) arc (180:-140:0.2);
  \end{scope}

  \draw[->, thick, blue] (0,0,0) -- (0,0,1);

  \def\r{1}
  \draw[<-, thick, blue, domain=-30:90, smooth, variable=\t]
    plot ({\r*cos(\t)},0,{\r*sin(\t)})
    node[anchor=east, xshift=-15, yshift=-25] {$R_y$};

  \draw[dashed, blue, domain=180:-180, smooth, variable=\t]
    plot ({\r*cos(\t)},0,{\r*sin(\t)});

\end{tikzpicture}

\end{minipage}

\caption{Rotation of an $R_y$ gate of a quantum circuit (left) illustrated on the Bloch sphere (right), from the $\ket{0}$ state to a final state through an arbitrary rotation $\theta$.}
\label{fig:RyBloch}

\end{figure}

\begin{figure}
  \centering
\begin{center}

\begin{tikzpicture}
  \begin{axis}[
    width=7.6cm, height=6.2cm,
    xlabel={Wavelength (\si{nm})},
    ylabel={Optical power (\si{^\circ/100\,mm})},
    xmin=375, xmax=675,
    ymin=-300, ymax=250,
    axis lines=left, 
    ytick distance=50,
    scaled x ticks=false,
    scaled y ticks=false,
    yticklabel style={/pgf/number format/fixed, /pgf/number format/fixed zerofill, /pgf/number format/precision=0},
    /pgf/number format/1000 sep={\,},
    /pgf/number format/.cd,
    x tick scale label code/.code={},
    y tick scale label code/.code={},
    tick align=outside,
    tick style={black},
    label style={font=\small},
    tick label style={font=\small},
    legend style={draw=none, font=\tiny{}, at={(1.032,1.05)}, anchor=north east},
    legend cell align=left,
    every axis plot/.append style={line width=1.2pt}
  ]

    \addplot[
      domain=375:675,
      samples=800,
      smooth,
      thick,
      color=blue!75!black
    ] {(2.16e7)/(x^2 - 2.132e4)};
    \addlegendentry{Drude equation for sucrose}

    \addplot[
      only marks,
      mark=o,
      mark size=2.2pt,
      color=blue!75!black
    ] coordinates {
      (405,105)
      (532,55)
      (650,36)
    };
    \addlegendentry{Experimental sucrose}

    \addplot[
      domain=375:675,
      samples=800,
      smooth,
      thick,
      color=purple!70!black
    ] {(-3.8e7)/(x^2 - 2.2e4)}; 
    \addlegendentry{Drude equation for fructose}

    \addplot[
      only marks,
      mark=o,
      mark size=2.2pt,
      color=purple!70!black
    ] coordinates {
      (405,-151)
      (532,-87)
      (650,-58)
    };
    \addlegendentry{Experimental fructose}
    
    \draw[ultra thin, dashed] (375,0) -- (675,0);

  \end{axis}
\end{tikzpicture}
\end{center}
\caption{Experimental result for table white sugar (sucrose) and table fructose against theoretical curves calculated from the Drude equation. The experimental values show a smaller rotation power compared to their laboratory grade counterparts.}
\label{graphOpticalPower}

\end{figure}

\subsection{\texorpdfstring{$R_x$ and $R_z$}{R\_x and R\_z} gates: transparent films}
\label{SectionRxRzTape}
\begin{figure}
    \centering
\begin{center}
\includegraphics[width=0.3\textwidth]{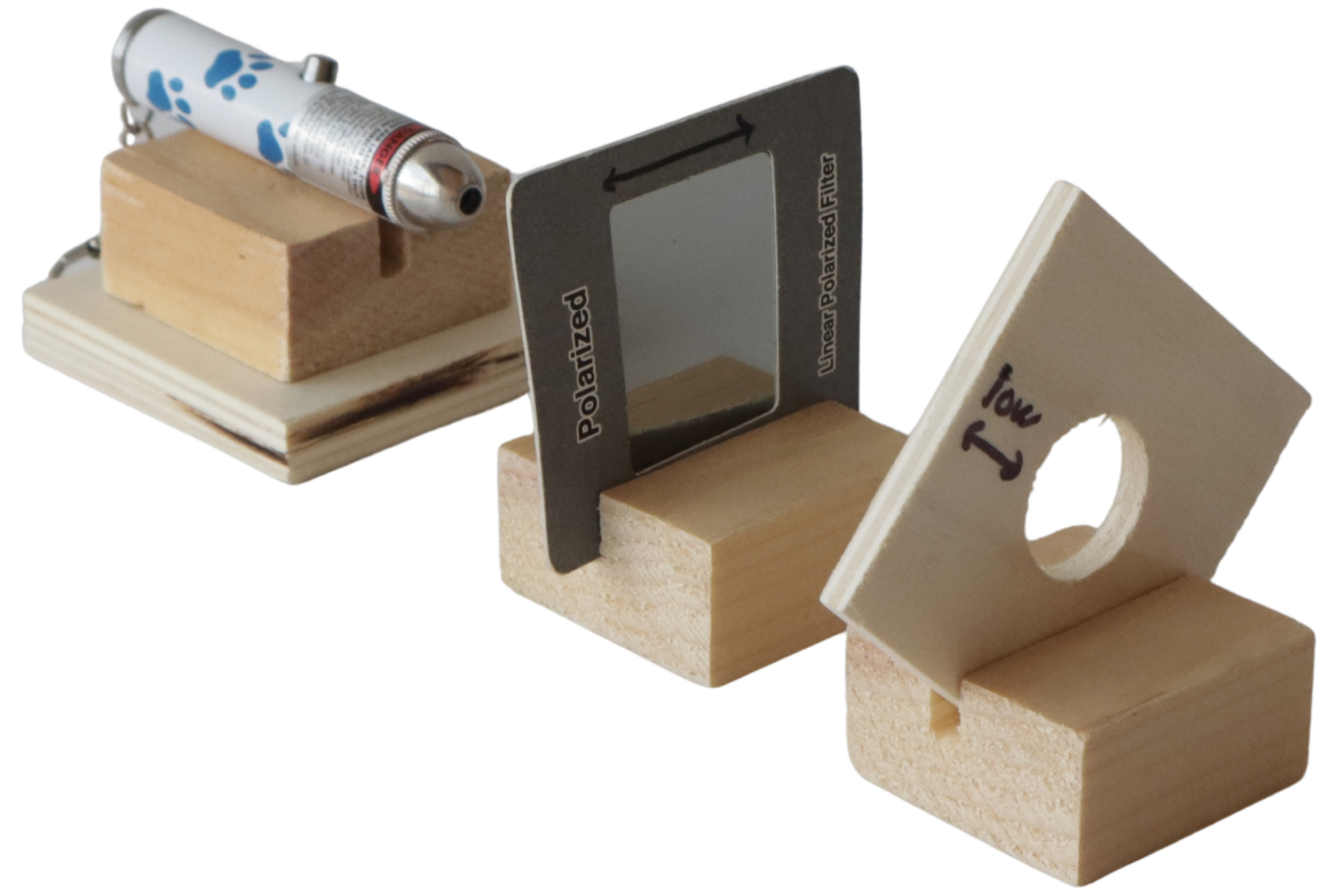}

\end{center}
    \caption{Experimental setup for applying an $R_x$ gate on state $|0\rangle$. The laser first goes through a horizontal polarizer before entering one layer of Office Works tape in the anti-diagonal position.}
    \label{PhotoRxGate}
\end{figure}

When an $R_x$ gate is applied to a qubit initially in the $|0\rangle$ state, it performs a rotation by a chosen angle around the $x$‑axis, as shown in figure~\ref{fig:RxBloch}. Figure~\ref{RxGateTape} illustrates how to orient transparent tape to implement an $R_x$ gate or an $R_x^\dagger$ gate.
Starting from a laser directed onto a horizontally polarizing filter (preparing the $|0\rangle$ state), the emerging light is horizontally polarized, as shown in figure~\ref{PhotoRxGate}. Placing a transparent film so that its optical axis lies in the anti‑diagonal direction creates an $R_x$ gate: a photon whose polarization enters the tape at a $45^\circ$ angle to the optical axis experiences a retardation of the component perpendicular to the optical axis relative to the component parallel to it. This converts the linearly polarized photon into an elliptically polarized one (Section \ref{SectionBirefringence}) \cite{EdwardsLangley1981}.

\begin{figure}[htbp]
\centering

\begin{minipage}[c]{0.35\linewidth}
\centering
\begin{quantikz}
\lstick{$q_0\ \ket{0}$} & \gate{R_x (\theta)} & \qw \\
\end{quantikz}
\end{minipage}
\hfill
\begin{minipage}[c]{0.6\linewidth}
  \centering

\tdplotsetmaincoords{70}{110} 
\begin{center}
\begin{tikzpicture}[tdplot_main_coords, scale=2]

  \draw[thick] (0,0,0) circle (1); 
  \draw[thick, tdplot_screen_coords] (0,0) circle (1); 

  \draw[->] (0,0,0) -- (0,0,1) node[anchor=south, yshift=6pt] {$\ket{0}$};
  \draw[->] (0,0,0) -- (0,0,-1) node[anchor=north, yshift=-6pt] {$\ket{1}$};
  \draw[->] (0,0,0) -- (1,0,0) node[anchor=north, xshift=-10] {$\ket{+}$};
  \draw[->] (0,0,0) -- (-1,0,0) node[anchor=south, xshift=10] {$\ket{-}$};
  \draw[->] (0,0,0) -- (0,1,0) node[anchor=west, xshift=6pt, yshift=2] {$\ket{\textit{i}}$};
  \draw[->] (0,0,0) -- (0,-1,0) node[anchor=east, xshift=-6pt, yshift=7] {$\ket{\textit{-i}}$};
  \draw [->] (2,0,0)  -- (3.2,0,0) node[anchor=north, xshift=-6, yshift=0]{$x$};
  \begin{scope} [canvas is yz plane at x=2.3]
    \draw[blue, <-] (-0.32,-0.15) arc (180:-100:0.2);
  \end{scope}

  \draw[->, thick, blue] (0,0,0) -- (0,0,1);

  \begin{scope}[canvas is yz plane at x=0]
    \draw[dashed, blue] (0,1) arc (90:-270:1);
  \end{scope}

  \begin{scope}[canvas is yz plane at x=0]
    \draw[blue, thick, ->] (0,1) arc (90:215:1) node[anchor=west, xshift=-5, yshift=80]{$R_x$};
  \end{scope}
  
\end{tikzpicture}
\end{center}
\end{minipage}
    \caption{Rotation of an $R_x$ gate of a quantum
circuit (left) illustrated on the Bloch sphere (right), from the $|0\rangle$
state to a final state through an arbitrary rotation $\theta$.}
    \label{fig:RxBloch}
  \end{figure}

  \begin{figure}
    \centering
   \tdplotsetmaincoords{90}{90}
\begin{tikzpicture}[scale=2.7,tdplot_main_coords]

\begin{scope}[shift={(0,-0.5,0)}]
\draw[rotate=-45, fill=gray,fill opacity=0.6] (0,-0.2,-0.6) -- (0,-0.2,0.6) -- (0,-0.15,0.55) -- (0,-0.1,0.59) -- (0,-0.05,0.54) -- (0,0,0.58) -- (0,0.05,0.53) -- (0,0.1,0.57) -- (0,0.15,0.52) -- (0,0.2,0.56) --(0,0.2,-0.64) -- (0,0.15,-0.59) -- (0,0.1,-0.63) -- (0,0.05,-0.58) -- (0,0,-0.62) -- (0,-0.05,-0.57) -- (0,-0.1,-0.61) -- (0,-0.15,-0.56) -- cycle;
\end{scope}

\begin{scope}[shift={(0,-2,0)}]
\draw[rotate=45, fill=gray,fill opacity=0.6] (0,-0.2,-0.6) -- (0,-0.2,0.6) -- (0,-0.15,0.55) -- (0,-0.1,0.59) -- (0,-0.05,0.54) -- (0,0,0.58) -- (0,0.05,0.53) -- (0,0.1,0.57) -- (0,0.15,0.52) -- (0,0.2,0.56) --(0,0.2,-0.64) -- (0,0.15,-0.59) -- (0,0.1,-0.63) -- (0,0.05,-0.58) -- (0,0,-0.62) -- (0,-0.05,-0.57) -- (0,-0.1,-0.61) -- (0,-0.15,-0.56) -- cycle;
\end{scope}

 \begin{scope}[shift={(0,0)}]
    \draw[] node[anchor=west, xshift=-160, yshift=0]{$R_x$};
    \draw[] node[anchor=west, xshift=-45, yshift=0]{$R_x^\dagger$};
 \end{scope}
    
\end{tikzpicture}
    \caption{Schematic of the experimental implementation of an $R_x$ gate and $R_x^\dagger$ gate using transparent adhesive tape for a photon coming out of the page toward the observer.}
    \label{RxGateTape}

\end{figure}

Figure~\ref{RzGate} illustrates the effect of an $R_z$ gate on the Bloch sphere, rotating the state of a qubit around the $z$‑axis by an arbitrary angle $\phi$. Physically, the $R_z$ and $R_z^\dagger$ gates can be constructed as shown in figure~\ref{RzGateTape}.

\begin{figure}[htbp]
\centering

\begin{minipage}[c]{0.35\linewidth}
\centering
\begin{quantikz}
\lstick{$q_0\ \ket{+}$} & \gate{R_z (\phi)} & \qw \\
\end{quantikz}
\end{minipage}
\hfill
\begin{minipage}[c]{0.6\linewidth}
\centering
 
\tdplotsetmaincoords{70}{110} 
\begin{center}
\begin{tikzpicture}[tdplot_main_coords, scale=2]

  \draw[dashed, blue] (0,0,0) circle (1); 
  \draw[thick, tdplot_screen_coords] (0,0) circle (1); 

  \draw[->] (0,0,0) -- (0,0,1) node[anchor=south, yshift=6pt] {$\ket{0}$};
  \draw[->] (0,0,0) -- (0,0,-1) node[anchor=north, yshift=-6pt] {$\ket{1}$};
  \draw[->] (0,0,0) -- (1,0,0) node[anchor=north, xshift=-10] {$\ket{+}$};
  \draw[->] (0,0,0) -- (-1,0,0) node[anchor=south, xshift=10] {$\ket{-}$};
  \draw[->] (0,0,0) -- (0,1,0) node[anchor=west, xshift=6pt, yshift=2] {$\ket{\textit{i}}$};
  \draw[->] (0,0,0) -- (0,-1,0) node[anchor=east, xshift=-6pt, yshift=7] {$\ket{\textit{-i}}$};
\draw [->] (0,0,1.43)  -- (0,0,2) node[anchor=south]{$z$};
  \begin{scope} [canvas is xy plane at z=1.7]
    \draw[blue, <-] (-0.16,0) arc (180:-100:0.2);
  \end{scope}

  \draw[->, thick, blue] (0,0,0) -- (1,0,0);

  \tdplotsetrotatedcoords{0}{0}{0} 
  \begin{scope}[tdplot_rotated_coords]
    \tdplotdrawarc[<-, thick, blue]{(0,0,0)}{1}{150}{1}{anchor=south, xshift=-20, yshift=-25}{$R_z$}
    \end{scope}
  
\end{tikzpicture}
\end{center}
\end{minipage}
 \caption{Rotation of an $R_z$ gate of a quantum
circuit (left) illustrated on the Bloch sphere (right), from the $|+\rangle$
state to a final state through an arbitrary rotation $\phi$.}
     \label{RzGate}
  \end{figure}
  
  \begin{figure}
    \centering
    \begin{center}
        
   \tdplotsetmaincoords{90}{90}
\begin{tikzpicture}[scale=3,tdplot_main_coords]

\begin{scope}[shift={(1,0)}]
\draw[rotate=0, fill=gray,fill opacity=0.6] (0,-0.2,-0.6) -- (0,-0.2,0.6) -- (0,-0.15,0.55) -- (0,-0.1,0.59) -- (0,-0.05,0.54) -- (0,0,0.58) -- (0,0.05,0.53) -- (0,0.1,0.57) -- (0,0.15,0.52) -- (0,0.2,0.56) --(0,0.2,-0.64) -- (0,0.15,-0.59) -- (0,0.1,-0.63) -- (0,0.05,-0.58) -- (0,0,-0.62) -- (0,-0.05,-0.57) -- (0,-0.1,-0.61) -- (0,-0.15,-0.56) -- cycle;
\end{scope}

\begin{scope}[shift={(0,-1,0)}]
\draw[rotate=90, fill=gray,fill opacity=0.6] (0,-0.2,-0.6) -- (0,-0.2,0.6) -- (0,-0.15,0.55) -- (0,-0.1,0.59) -- (0,-0.05,0.54) -- (0,0,0.58) -- (0,0.05,0.53) -- (0,0.1,0.57) -- (0,0.15,0.52) -- (0,0.2,0.56) --(0,0.2,-0.64) -- (0,0.15,-0.59) -- (0,0.1,-0.63) -- (0,0.05,-0.58) -- (0,0,-0.62) -- (0,-0.05,-0.57) -- (0,-0.1,-0.61) -- (0,-0.15,-0.56) -- cycle;
\end{scope}

 \begin{scope}[shift={(0,0)}]
    \draw[] node[anchor=west, xshift=-9, yshift=0]{$R_z^\dagger$};
    \draw[] node[anchor=west, xshift=-90, yshift=0]{$R_z$};
 \end{scope}
    
\end{tikzpicture}
    \end{center}

    \caption{Schematic of the experimental implementation of an $R_z$ gate and an $R_z^\dagger$ gate using transparent adhesive tape for a photon coming out of the page toward the observer.}
    \label{RzGateTape}

\end{figure}

\subsection{\texorpdfstring{$R_x$ and $R_z$}{R\_x and R\_z} implementation}
\begin{figure}
    \centering
    \begin{center}
\includegraphics[width=0.4\textwidth]{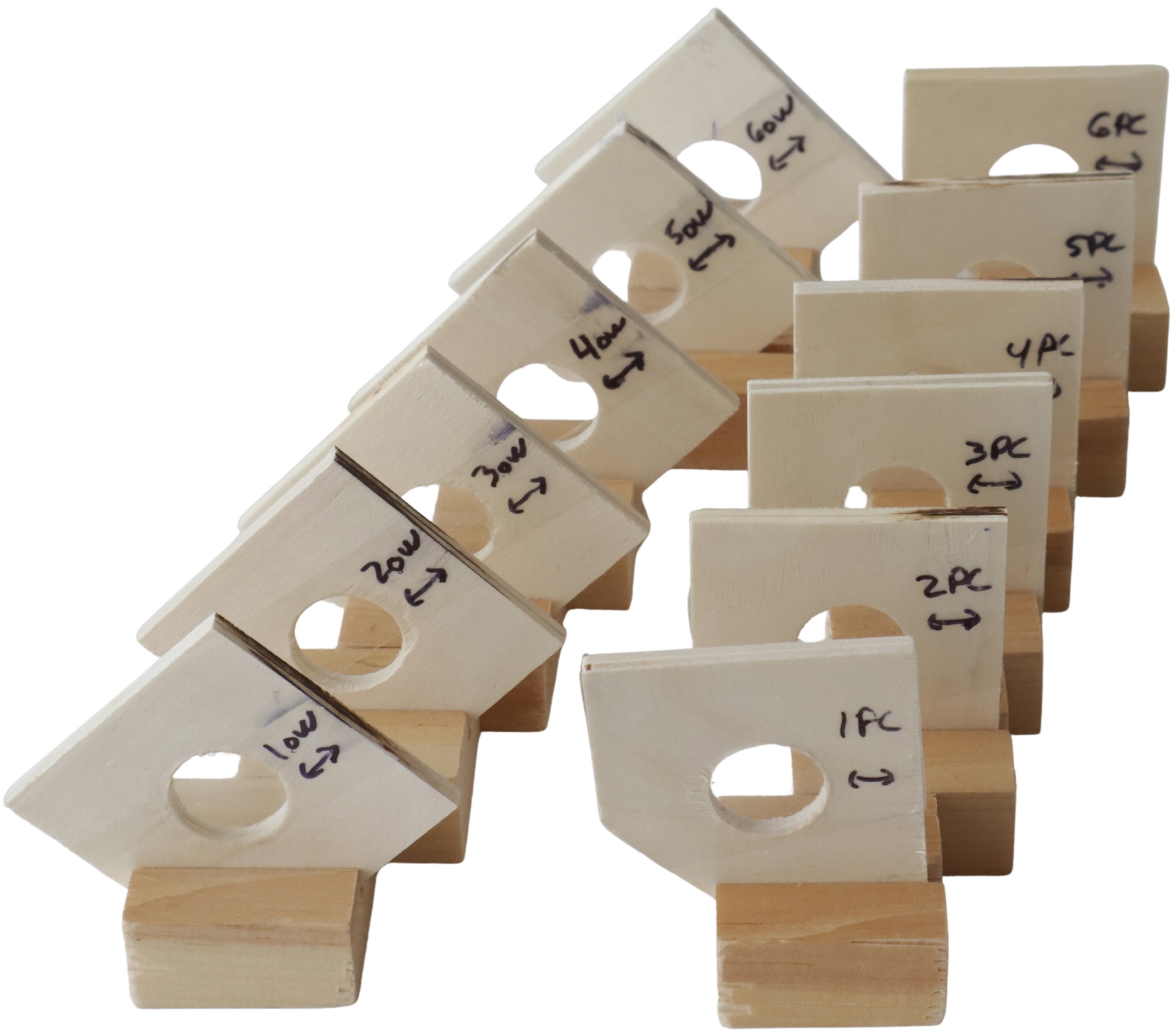}

\end{center}
    \caption{Collection of $R_x^\dagger$ gates (on the left) made of up to 6 layers of Office Works tape and of $R_z$ gates (on the right) made of up to six layers of Canada Post tape.}
    \label{Photo6RxGates}
\end{figure}

Figure~\ref{Photo6RxGates} shows how 5cm × 5cm wooden plates with central apertures can be used to assemble $R_x$ and $R_z$ gates. In our setup, we employed two brands of transparent tape: Office Works tape and Canada Post tape. For each brand of tape, we fabricated six distinct rotation angles using up to six layers of tape. As the two brands of tape have different rotation angles per layer, this results in twelve different rotation angles. One corner of each wooden support is cut at a $45^\circ$ angle, allowing an easy switch between an $R_x$ and an $R_z$ configuration. The $R_x^\dagger$ and $R_z^\dagger$ gates are obtained, respectively, by rotating the $R_x$ gate by $180^\circ$ around the vertical axis or by orienting the optical axis of the tape vertically.

\subsection{Calibration}
\begin{figure}
    \centering
   \begin{center}
\includegraphics[width=0.4\textwidth]{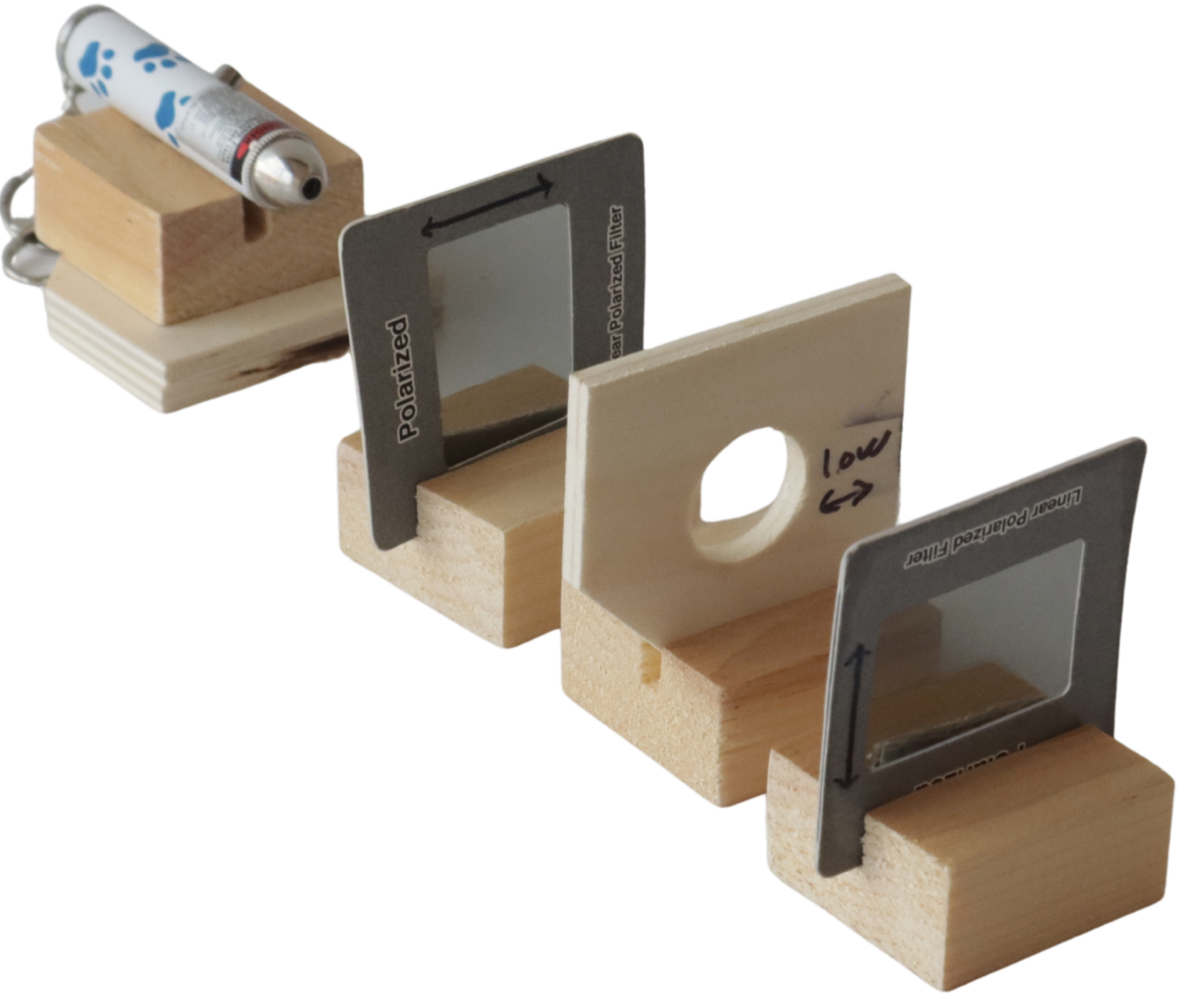}

\end{center}
    \caption{Setup for calibrating the making of the $R_x$ and $R_z$ gates. The transparent tape is placed while in the horizontal position. Since the fast axis of the transparent tape may lie at an angle from the sides of the tape, each layer is added on top of the previous one between a horizontal filter and a vertical one. If the fast axis is perfectly horizontal, even though the tape may not be, no light exits the second filter.}
    \label{PhotoCalibration}
\end{figure}

One of the most delicate aspects of assembling this experiment is the precise alignment of the transparent tape, since any tilt in an $R_x$ or $R_z$ gate will reduce the accuracy of the results. It is also important to note that, although the manufacturing techniques for transparent tape are very stable and yield highly consistent products, the optical axis is not necessarily perfectly parallel to the edge of the tape.
The most reliable method to ensure that the initial $R_z$ gate is properly aligned is to place the tape between two polarizing filters, with one oriented horizontally and the other vertically (figure~\ref{PhotoCalibration}). The tape is correctly aligned for an $R_z$ gate if no light passes through the second filter.

In this configuration, the incident light is horizontally polarized and thus aligned with the optical axis of the tape; as a result, no phase shift is introduced and the polarization remains unchanged upon exiting the tape. The second filter, being vertically oriented, completely blocks the transmitted light.
If, however, the tape is slightly rotated such that the incoming light polarization is not parallel to the optical axis, a relative phase shift is induced between the fast and slow components. This dephasing produces an elliptical polarization, allowing some fraction of the light to pass through the second (vertical) polarizer.
In our setup, the optical axis of the tape was observed to lie approximately $12^\circ$ from the edge of the tape.

\subsection{Birefringence of transparent tape}
\label{SectionBiref1}
Sections \ref{SectionBiref1} to \ref{BirefringenceIndexDetermination} aim at determining the birefringence $\Delta n$ of the two types of transparent tape used in this experiment. While Section \ref{SectionMichelLevyChar} gives an approximate method using the Michel-L{\'e}vy chart, Section \ref{BirefringenceIndexDetermination} leads to a more precise measurement.

Because the birefringence of a material depends on the wavelength of the photon passing through it, determining its value is important for finding the rotation induced by each layer of tape for a given laser. This later gives the ability to predict the rotation angle for any number of layers of a specific tape.

\begin{figure}
      \centering
      \begin{center}
      \includegraphics[width=0.3\textwidth]{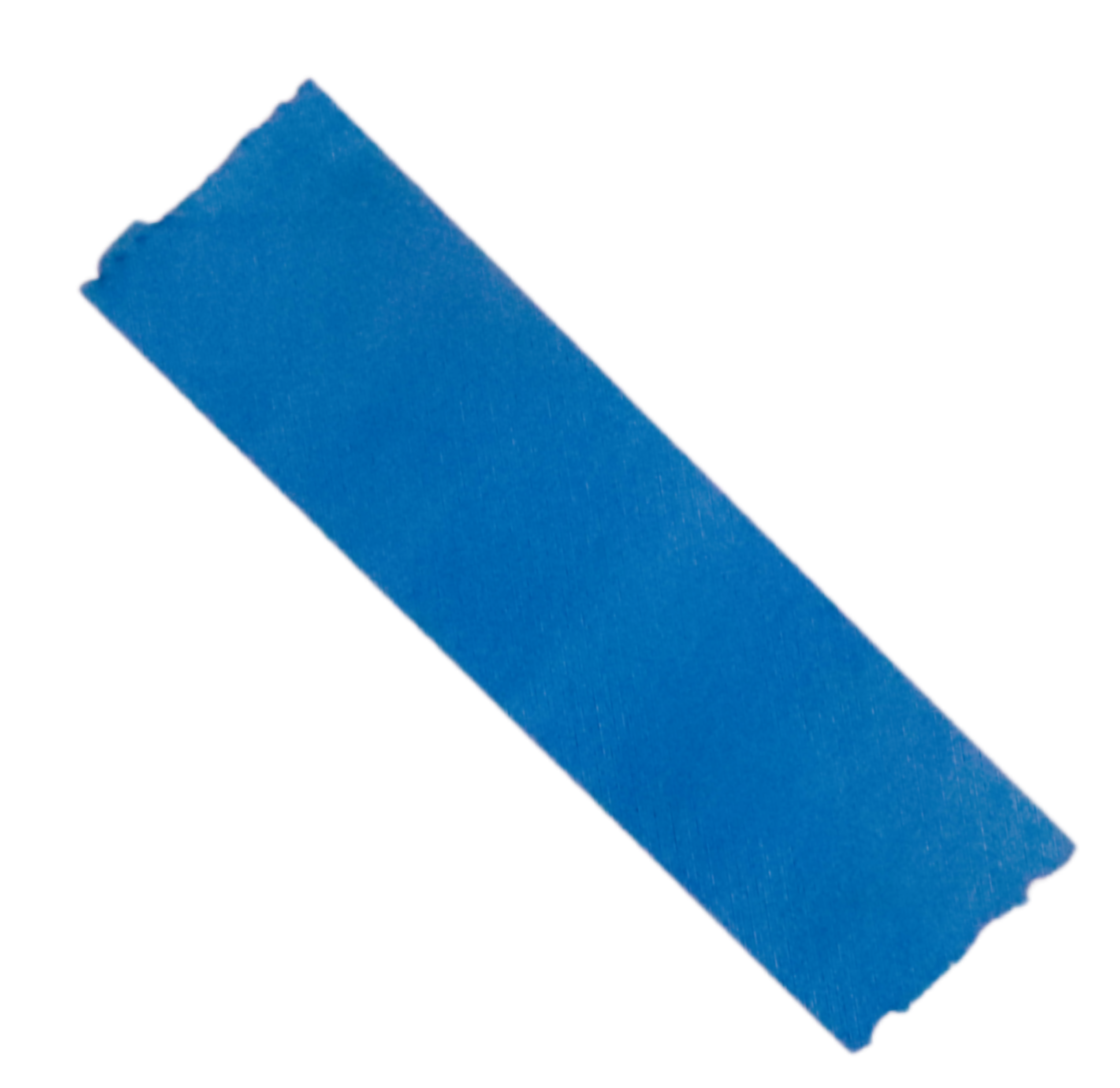}
      \caption{Blue is the perceived colour of horizontally polarized white light passing an anti-diagonal $R_x$ gate made of one layer of Office Works tape, followed by a horizontal polarizer (not shown).}
      \label{TapeBleu}
       \end{center}
    \end{figure}

Before detailing the protocols for calculating these values, it is worth noting that when a single layer of Office Works tape is placed at a $45^\circ$ angle between two horizontal polarizers and illuminated with white light, the perceived colour is blue, as shown in figure~\ref{TapeBleu}.

\begin{figure}[htbp]
  \centering

\tdplotsetmaincoords{70}{110} 
\begin{center}
\begin{tikzpicture}[tdplot_main_coords, scale=2]

  \draw[thick] (0,0,0) circle (1); 
  \draw[thick, tdplot_screen_coords] (0,0) circle (1); 

  \draw[->] (0,0,0) -- (0,0,1) node[anchor=south, yshift=6pt] {$\ket{0}$};
  \draw[->] (0,0,0) -- (0,0,-1) node[anchor=north, yshift=-6pt] {$\ket{1}$};
  \draw[->] (0,0,0) -- (1,0,0) node[anchor=north, xshift=-10] {$\ket{+}$};
  \draw[->] (0,0,0) -- (-1,0,0) node[anchor=south, xshift=10] {$\ket{-}$};
  \draw[->] (0,0,0) -- (0,1,0) node[anchor=west, xshift=6pt, yshift=2] {$\ket{\textit{i}}$};
  \draw[->] (0,0,0) -- (0,-1,0) node[anchor=east, xshift=-6pt, yshift=7] {$\ket{\textit{-i}}$};

  \begin{scope}[canvas is yz plane at x=0]
    \draw[dashed, blue] (0,1) arc (90:-270:1);
  \end{scope}

  \begin{scope}[canvas is yz plane at x=0]
    \draw[blue, thick, ->] (0,1) arc (90:421:1);
  \end{scope}

  \begin{scope}[canvas is yz plane at x=0]
    \draw[green, thick, ->] (0,1) arc (90:336:1);
  \end{scope}

  \begin{scope}[canvas is yz plane at x=0]
    \draw[red, thick, ->] (0,1) arc (90:252:1);
  \end{scope}

\end{tikzpicture}
\end{center}
 \caption{A single layer of transparent tape undergoes a different amount of rotation for each wavelength, depending on the amount of birefringence at each wavelength. Here we show the extent of the rotation for three laser wavelengths (405nm, 532nm, and 650nm) for a single layer of Office Works tape. We see that the blue wavelength retains the largest amount of horizontal polarization, compared to the green and red ones.}
    \label{RxRGB}

  \end{figure}
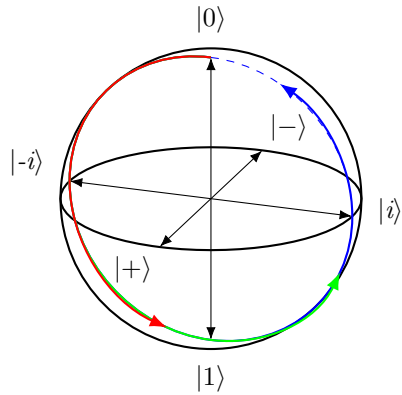

This blue colour arises from the sum of all wavelengths in the visible spectrum that are transformed from horizontal linear polarization into different elliptical polarizations, each corresponding to its own position on the Bloch sphere (figure~\ref{RxRGB}). The second polarizer acts as a measurement device, allowing only a certain percentage of each wavelength to pass through (see appendix~\ref{AppendixMeasurementBasis} through to \ref{SectionPolarizerMeasurement} for more details on measurement). The final perceived colour is therefore the combined contribution of all wavelengths weighted by their transmission probabilities.

Figure~\ref{RxRGB} shows the rotations induced by one layer of Office Works tape for red light (650nm), green light (532nm), and blue light (405nm). The rotation angles are approximately $160^\circ$, $248^\circ$, and $332^\circ$, respectively. Because white light is composed of all visible wavelengths, each mapped to a distinct point on the Bloch sphere, the sum of their transmitted components produces the observed blue colour.

\begin{figure}
  \centering
  \subfloat[]{
    \includegraphics[width=0.48\textwidth]{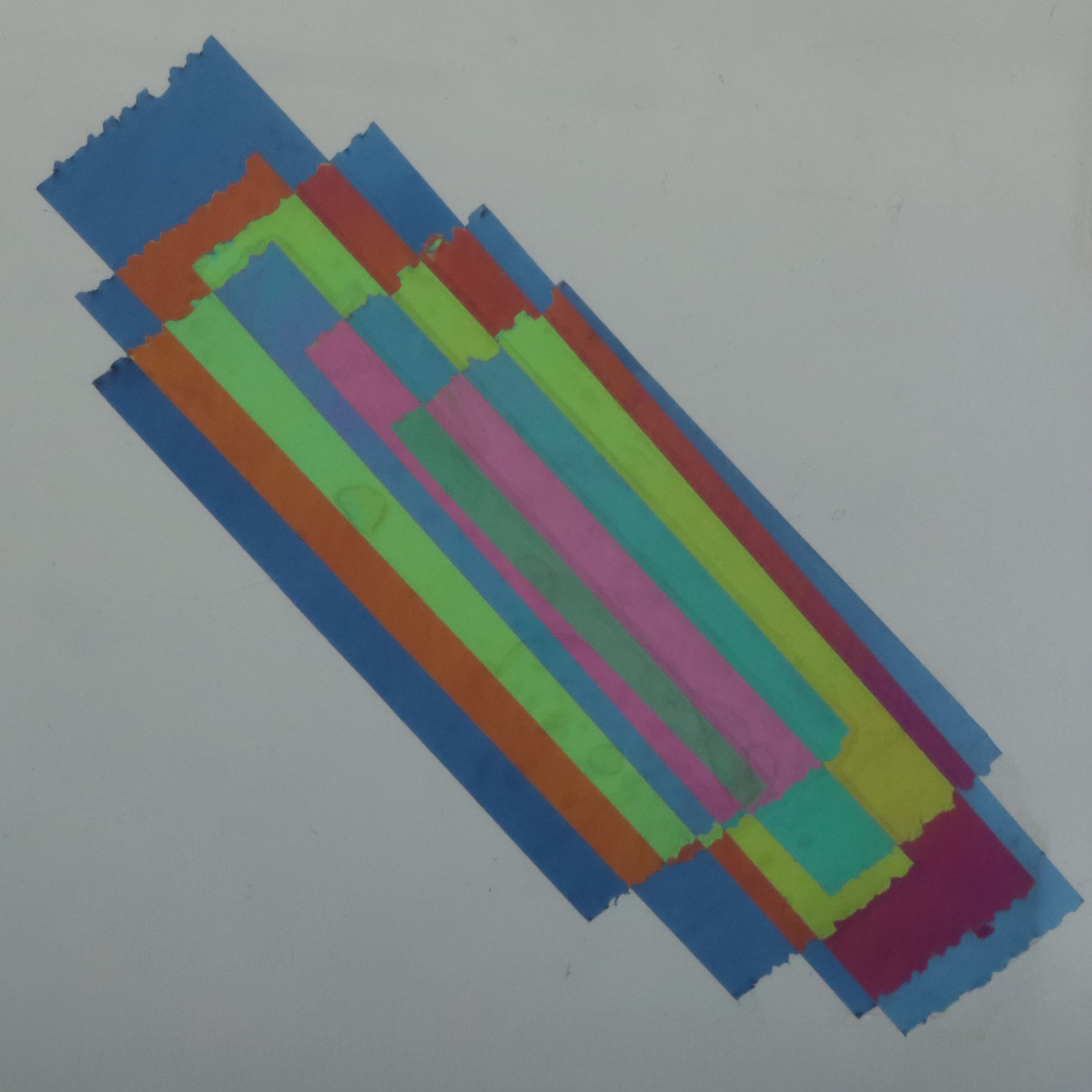}
    \label{fig:6tapesblanc}
  }
  \hfill
  \subfloat[]{
    \includegraphics[width=0.48\textwidth]{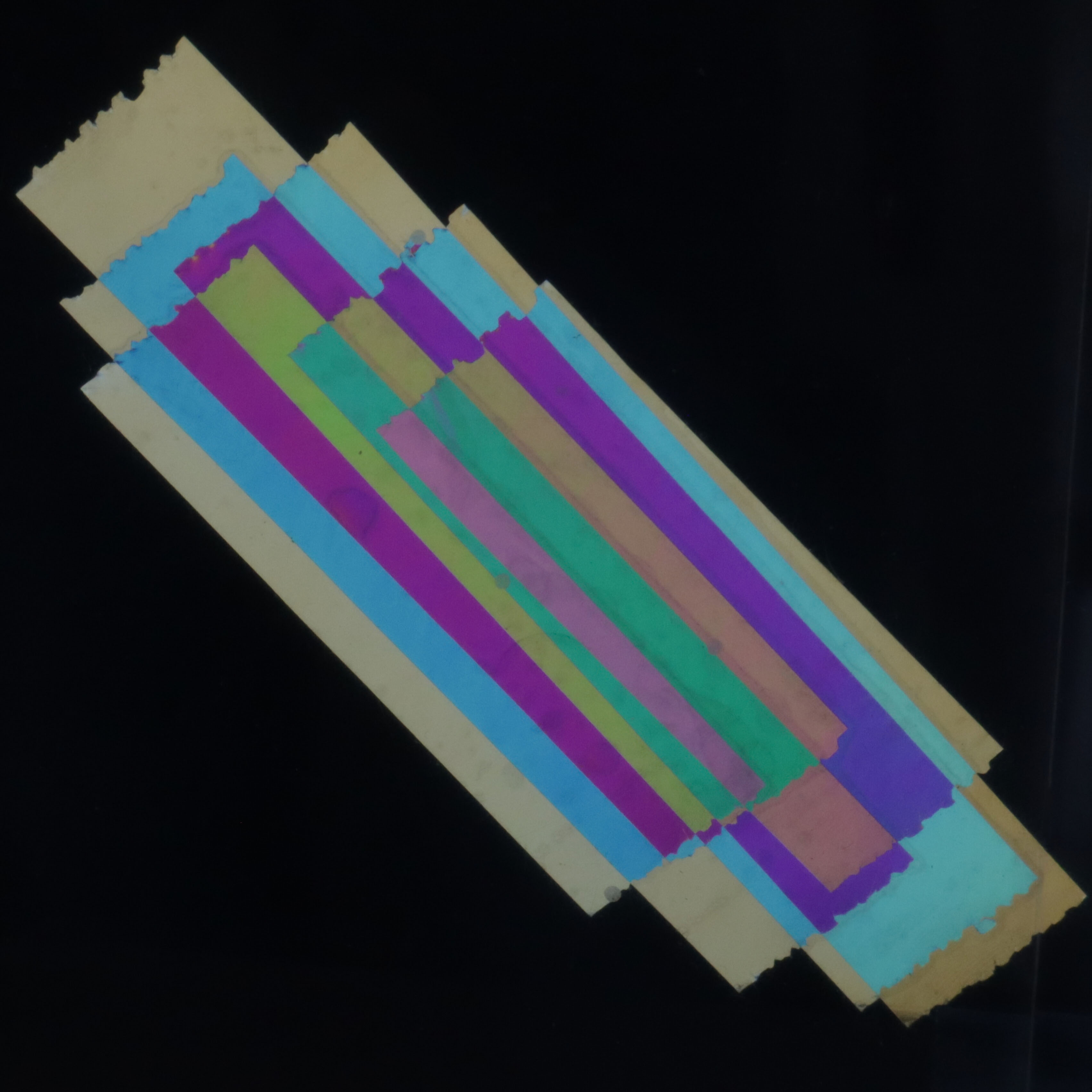}
    \label{fig:6tapesnoir}
  }
  \caption{Perceived colours of up to six layers of Office Works tape when horizontally polarized white light passes through the tape and : (a) through a second polarizing filter (not shown) that is \textit{horizontally} polarized; (b) through a second polarizing filter (not shown) that is \textit{vertically} polarized. Note that the background is black because the second polarizer is blocking all of the light not rotated by the tape.}
  \label{fig:example}
\end{figure}

Figure~\ref{fig:6tapesblanc} shows six layers of transparent tape, arranged to have various amount of overlap and placed at a $45^\circ$ angle between two horizontal polarizers with a background illumination of white light. We see a vibrant array of colours. Figure~\ref{fig:6tapesnoir} presents the same situation with the second polarizer vertically polarized. We again see a vibrant array of colours, though different from the first set as we are now measuring the vertical components of the twisted light. In all, we see that each tape thickness produces its own characteristic perceived colour when measured along the $|0\rangle$ or $|1\rangle$ direction (horizontal or vertical component of polarization, respectively).

\subsection{Michel–L{\'e}vy chart}
\label{SectionMichelLevyChar}
\begin{figure}
    \centering
\begin{center}
\includegraphics[width=1.0\textwidth]{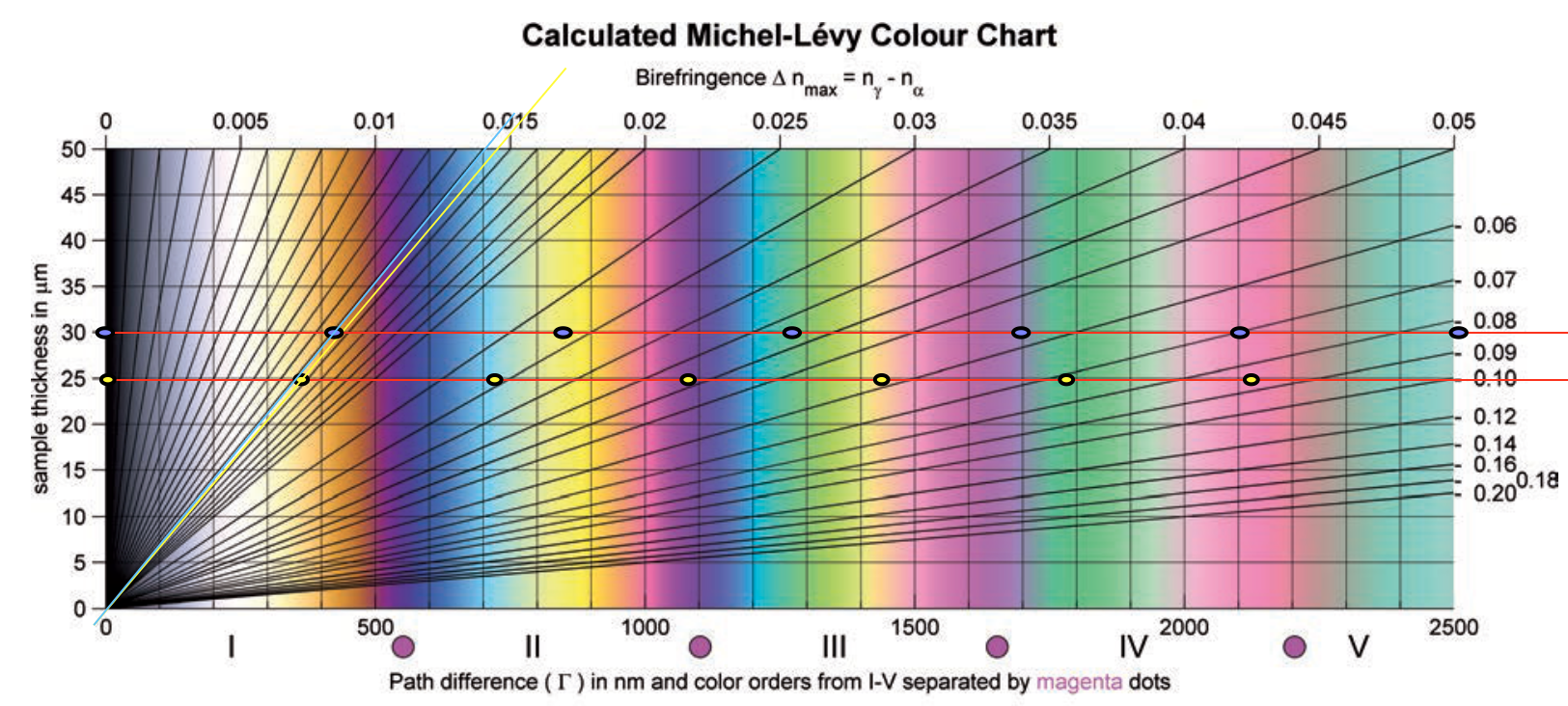}

\end{center}
\caption{Michel-L{\'e}vy chart reproduced from \cite{Sorensen2013_MichelLevy}. The yellow dots correspond to the perceived colours of up to six layers of Office Works tape when put at 45° between two polarizers perpendicular to one another. They should all lie on the yellow line, but points 2 to 6 have been brought down to a thickness of one layer of Office Works tape at 25$\mu$m as they would otherwise lie off the chart. The blue dots correspond to the perceived colours of up to 6 layers of Canada Post tape (30$\mu$m thick) and have been shifted similarly. That tape has a thickness of 30$\mu$m. The measured birefringence values for the two brands of tape are 0.0148 and 0.0140, respectively.}
    \label{MichelLevy}
\end{figure}

We have seen how the rotation angle is wavelength dependent and leads to a characteristic colour when a material is placed between two polarizers. Furthermore, this colour depends on the material thickness, as the amount of rotation of the light depends on its path length through the material. These two phenomena allow us to determine the birefringence of a material using a Michel–L{\'e}vy chart (figure~\ref{MichelLevy}). The chart only provides a single birefringence value \cite{klein_hurlbut_1993}, and so the measurement is by definition approximate for any material whose birefringence is wavelength-dependent \cite{RefSlepkov}. For a known thickness of a birefringent sample, one must identify the perceived colour of white light after it passes through two perpendicular polarizing filters, with the birefringent material placed between them, as in figure~\ref{fig:6tapesnoir}. The maximum colour intensity occurs when the optical axis of the material is oriented at $45^\circ$ with respect to each polarizer \cite{klein_hurlbut_1993}.

Given the thickness of Office Works tape was measured to be $25\mu\text{m}$, the six perceived colours in figure~\ref{fig:6tapesnoir} were placed on the chart (figure~\ref{MichelLevy}) along the horizontal line corresponding to $25\mu\text{m}$. Note that, in principle, we should plot each of the points at the respective thickness of the tape ($25\mu\text{m}$, $50\mu\text{m}$, $75\mu\text{m}$, etc.). However, the vertical limit of the chart is $50\mu\text{m}$.  Plotting the points on the horizontal line allows us to record the perceived colours while still working within the limits of the chart. We can then draw a straight line from the origin through the first point and extend it to the upper boundary of the chart, yielding to an approximate birefringence of $\Delta n \approx 0.0148$.

Previous work \cite{RefSlepkov} has shown that the birefringence of many common transparent adhesive tapes remains nearly constant for parts of the visible spectrum, so the value obtained from the chart is rather reliable, if still technically approximate. This makes the chart useful for validating the more precise results presented in Section~\ref{BirefringenceIndexDetermination}. Using the same method, Canada Post wrapping tape, measured to have a thickness of $30\mu\text{m}$, exhibits a birefringence of approximately $0.0140$ on the Michel–L{\'e}vy chart.

Note that the bottom axis of the chart represents the path difference $\Gamma$ (in nm), defined as
\begin{equation}
\Gamma = \Delta n \cdot t\
\end{equation}
where $\Delta n$ is the birefringence (unitless) and $t$ is the thickness of the material in nm. This path difference is directly related to angle of rotation $\theta$ on the Bloch sphere in the case of an $R_x$ gate and to the angle of rotation $\phi$ in the case of an $R_z$ gate as discussed in Section \ref{SectionRxRzTape}. The phase shift $\phi$ (or $\theta$)(in degrees) is given by
\begin{equation}
\phi = \frac{360^\circ \cdot \Gamma}{\lambda}\
\end{equation}
where $\lambda$ is the wavelength (in nm) of the photon.

\subsection{More precise birefringence determination}
\label{BirefringenceIndexDetermination}
\begin{figure}
    \centering
   \begin{center}
\includegraphics[width=0.2\textwidth]{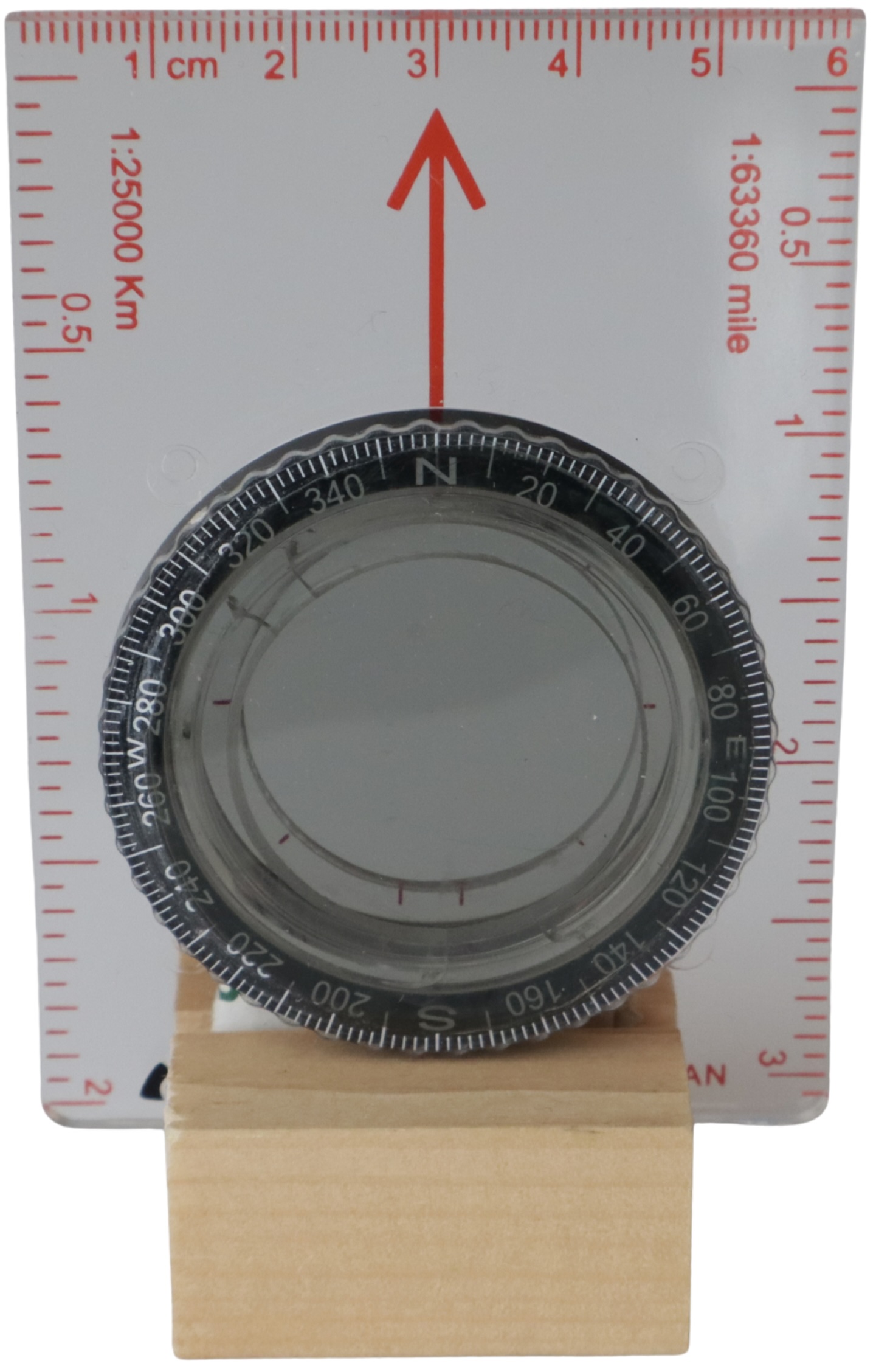}

\end{center}
    \caption{Homemade rotating polarizer made from a general store compass whose centre has been removed and replaced by a one-sided adhesive polarizing filter cut to the right size. Horizontally polarized light corresponds to \textit{North} or 0°.}
    \label{PhotoBoussole}
\end{figure}

One essential tool for experimentally determining the angle of linear polarization is a polarizing filter mounted on a graduated wheel (figure~\ref{PhotoBoussole}). This device enables precise measurement of the polarization angle. In practice, it is easier to identify the angle at which no light passes through the filter, \textit{the extinction angle}, than to determine the angle corresponding to maximum transmitted intensity. Since the extinction angle is exactly $90^\circ$ away from the angle of maximum intensity, the latter can be obtained simply by adding (or subtracting) $90^\circ$.

\begin{figure}
      \centering
      \includegraphics[width=0.48\textwidth]{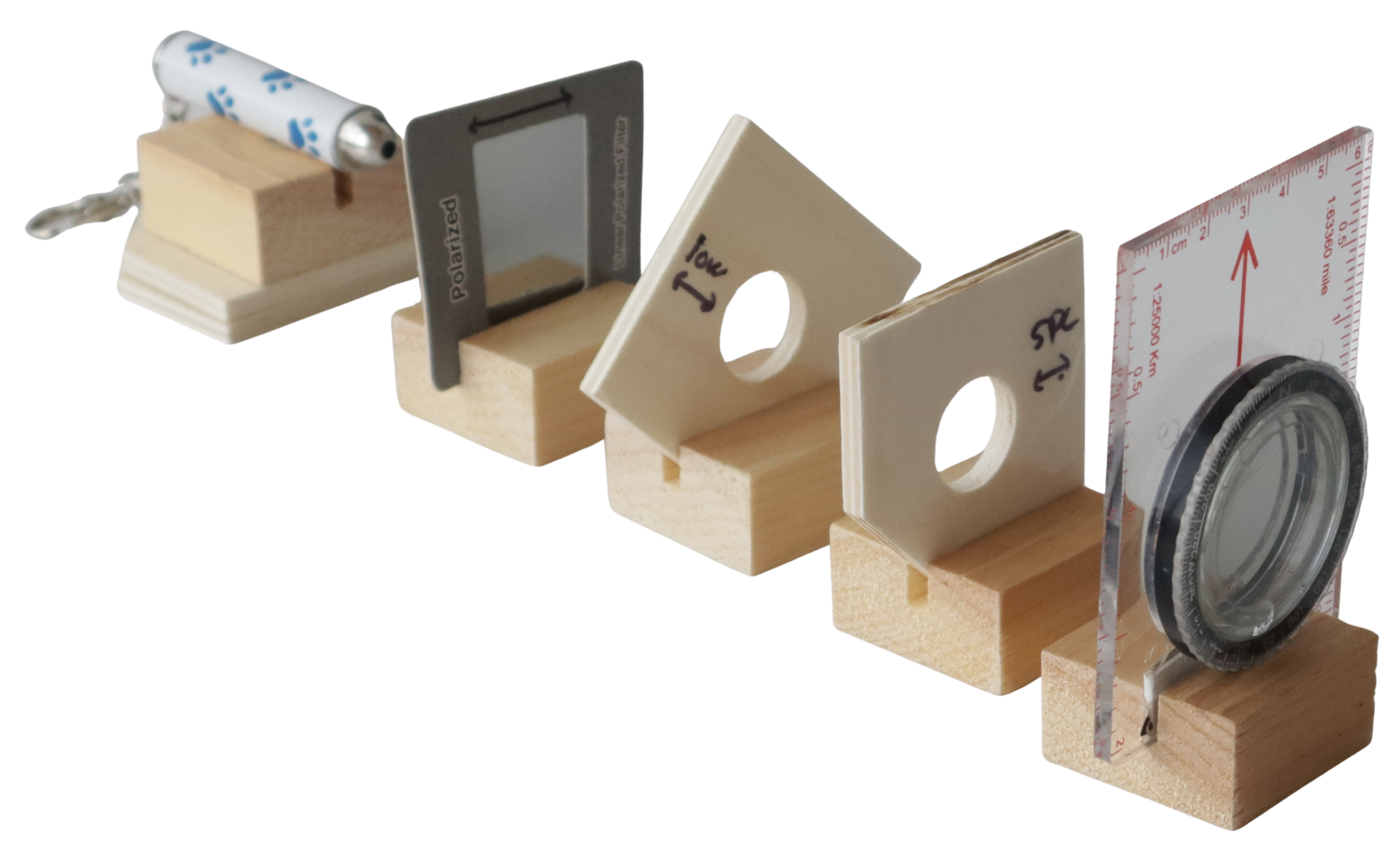}
    \caption{Setup for determining the unknown rotation angle of a specific gate. The laser points toward a horizontal polarizer, followed by the unknown $R_x$ gate, by an $R_z$ gate ($\phi\approx\frac{-\pi}{2}$) and then by the rotating polarizer.}
    \label{PhotoMontageRxRzt}
  \end{figure}

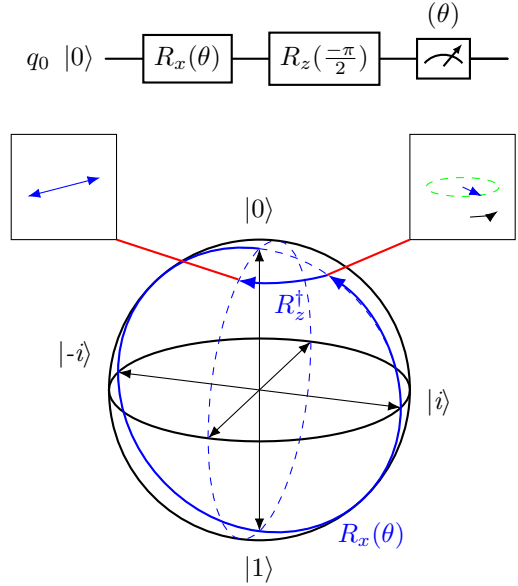
\begin{figure}[htbp]
\centering

\begin{minipage}[c]{0.35\linewidth}
\centering
\begin{quantikz}
\lstick{$q_0\ \ket{0}$} & \gate{R_x (\theta)} & \gate{R_z(\frac{-\pi}{2})} & \meter{(\theta)} & \qw \\
\end{quantikz}
\end{minipage}
\hfill
\begin{minipage}[c]{0.6\linewidth}
\centering

      \tdplotsetmaincoords{70}{110} 

      \begin{tikzpicture}[baseline=(current bounding box.north)]
        \begin{scope}[tdplot_main_coords, scale=2]
          \draw[thick] (0,0,0) circle (1); 
          \draw[thick, tdplot_screen_coords] (0,0) circle (1); 

          \draw[->] (0,0,0) -- (0,0,1) node[anchor=south, yshift=6pt] {$\ket{0}$};
          \draw[->] (0,0,0) -- (0,0,-1) node[anchor=north, yshift=-6pt] {$\ket{1}$};
          \draw[->] (0,0,0) -- (1,0,0);
          \draw[->] (0,0,0) -- (-1,0,0);
          \draw[->] (0,0,0) -- (0,1,0) node[anchor=west, xshift=6pt, yshift=2] {$\ket{\textit{i}}$};
          \draw[->] (0,0,0) -- (0,-1,0) node[anchor=east, xshift=-6pt, yshift=7] {$\ket{\textit{-i}}$};

          \begin{scope}[canvas is yz plane at x=0]
            \draw[dashed, blue] (0,1) arc (90:-270:1);
          \end{scope}

          \def\r{1}
          \draw[dashed, blue,tdplot_main_coords,domain=0:360,smooth,variable=\t]
            plot ({\r*cos(\t)}, 0, {\r*sin(\t)});

          \begin{scope}[canvas is yz plane at x=0]
            \draw[blue, thick, ->] (0,1) arc (90:421:1)
              node[anchor=north west, yshift=-90] {$R_x(\theta)$};
          \end{scope}

          \begin{scope}
            \draw[blue, thick, ->] (0,{cos(61)},{sin(61)}) arc (61:0:0.6)
              node[anchor=north west, xshift=10] {$R_z^\dagger$};
          \end{scope}
        \end{scope}

        \begin{scope}[shift={(0,0)}]
          \coordinate (squareLL) at (2,2);
          \coordinate (squareUR) at ($(squareLL)+(1.4,1.4)$);
          \coordinate (startPT) at ($(squareLL) + (0.7,0.7)$);
          \coordinate (startPTellipse) at ($(startPT)+({cos(29/2)/2-0.04},{sin(-29/2)/2+0.01})$);
          \filldraw[fill=white, draw=black] (squareLL) rectangle (squareUR);
          \draw[rotate=0, dashed, green] (startPT) ellipse [x radius=cos(29/2)/2, y radius=sin(-29/2)/2];
          \begin{scope}[rotate=0]
            \draw[->] ($(startPT) + (0.1,-0.4)$)
              arc[start angle=-80, end angle=-20, x radius=cos(29/2)/2, y radius=sin(29/2)/2];
            \draw[->, blue] (startPT) -- ($(startPT) + (0.26,-0.12)$);
          \end{scope}
          \draw[thick, red] ($(squareLL)+(0,0)$) -- (0.9,1.53);
        \end{scope}

        \def\A{14.5}
        \coordinate (squareLL) at (-3.3,2);
        \coordinate (squareUR) at ($(squareLL)+(1.4,1.4)$);
        \filldraw[fill=white, draw=black] (squareLL) rectangle (squareUR);
        \draw[<->, blue] ($(squareLL)+({0.7-0.5*cos(\A)},{0.7-0.5*sin(\A)})$)
          -- ($(squareUR)-({0.7-0.5*cos(\A)},{0.7-0.5*sin(\A)})$);
        \draw[thick, red] ($(squareLL)+(1.4,0)$) -- (-0.28,1.47);
      \end{tikzpicture}
\end{minipage}
    \caption{The quantum circuit (left) starts in the $|0\rangle$ state followed by the unknown $R_x$ gate, the $R_z(\frac{-\pi}{2})$, and the measurement along the axis tilted at $\theta$. The Bloch sphere (right) shows the rotations for a blue laser (405nm), an $R_x$ gate made of one layer of Office Works tape, and an $R_z$ gate made of three layers of Office Works tape (close to $\frac{-\pi}{2}$). The rotation angle $\theta$ of the $R_x$ gate is measured to be 332$^\circ$.}
    \label{RxRzt}
  \end{figure}

Figure~\ref{PhotoMontageRxRzt} presents the setup used to determine the rotation angle $\theta$ of a given gate plate. The laser first passes through a horizontal polarizing filter, then through the $R_x$ gate to be tested, followed by an $R_z$ gate that applies a rotation as close to $-\frac{\pi}{2}$ on the Bloch sphere as possible, and finally through the rotating polarizer to measure the angle of maximum intensity. The states of the photon polarization are shown in figure~\ref{RxRzt} for a single layer of Office Works tape and a blue laser. First, the light is horizontally polarized ($|0\rangle$ state), then  an unknown rotation angle is applied by an $R_x$ gate, transforming the photon into an elliptically polarized state. In this configuration, rotating the polarizer cannot produce full extinction at any angle since an elliptically polarized photon always has some probability of going through a polarizing filter for all angles of the filter. The presence of the $R_z$ gate with a rotation of $\frac{-\pi}{2}$ brings the qubit’s state vector from the $y$–$z$ plane of the Bloch sphere, where the polarization is elliptical, into the $x$–$z$ plane, where the polarization is linear, while still preserving the latitude of the state vector i.e., the value of $\theta$ remains constant. The rotating polarizer can then be used to measure $\frac{\theta}{2}$.

\begin{figure}
    \centering
   \begin{center}
\includegraphics[width=0.4\textwidth]{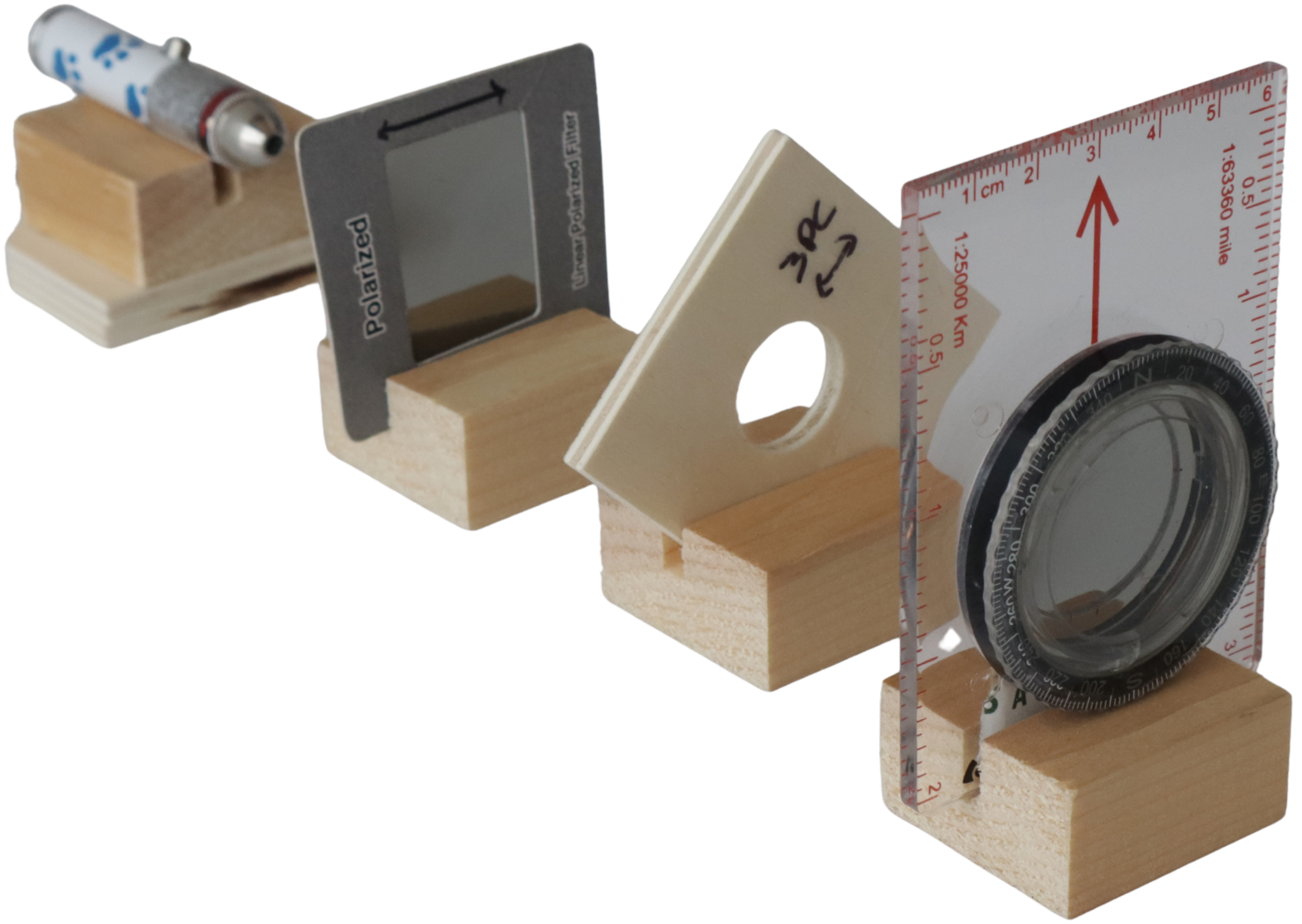}

\end{center}
    \caption{Setup for determining the best $R_z$ gate to use ($\phi\approx\frac{-\pi}{2}$) for a particular wavelength for the setup of figure~\ref{PhotoMontageRxRzt}. The laser points to a horizontal polarizer followed by the gate plate to be tested in the $R_x^\dagger$ position (diagonal). The best gate is the one that shows the least variation of intensity when turning the rotating filter. Intensity is constant when the $R_x^\dagger$ gate has a value of $\theta=\frac{\pi}{2}$ (or $\frac{-\pi}{2}$) and brings the system in the $|i\rangle$ (or $|-i\rangle$) state.}
    \label{PhotoMontageRz90degres}
\end{figure}

To determine which gate plate to use for the $R_z(\frac{-\pi}{2})$ gate, one must identify a number of tape layers for which the transmitted intensity shows no noticeable variation as a horizontally polarized laser passes through the candidate gate and then through the rotating polarizer, as in the setup shown in figure~\ref{PhotoMontageRz90degres}.

When no intensity variation is observed on the final screen, the gate corresponds to an angle close to $\frac{\pi}{2}$ (or $\frac{-\pi}{2}$) on the Bloch sphere, modulo $2\pi$, since full rotations do not affect the outcome and only the remaining fractional rotation is relevant. The sign of the angle is not yet important; only the circular polarization of light ($|i\rangle$ or $|-i\rangle$ state) matters at this stage.

In the example of figure~\ref{RxRzt}, the measured angle is either $28^\circ$ or $-28^\circ$ ($332^\circ$). The correct value is determined once the setup of figure~\ref{PhotoMontageRxRzt} is used with a gate plate made of two layers of the same tape. In this particular case, the $R_x$ gate composed of two layers corresponds to a rotation of $664^\circ$ ($304^\circ$ on the Bloch sphere). From this, we deduce that an $R_x$ gate made of a single layer of Office Works tape has a rotation angle of $332^\circ$.
  
Figure~\ref{graphiqueAnglesMesuresST} and figure~\ref{graphiqueAnglesMesuresPC} present the linear regression for the experimentally measured rotation angles for one to six layers of, respectively, Office Works and Canada Post tapes. All 12 gates were tested with a 405nm blue laser, a 532nm green laser, and a 650nm red laser. The $R^2$ values of all six linear fits are 0.9973 or greater.

\begin{figure}
\centering
\begin{tikzpicture}
\begin{axis}[
  width=7.6cm, height=5.8cm,
  xlabel={Wavelength (nm)},
  ylabel={Birefringence $\Delta n$},
  xmin=380, xmax=680,
  ymin=0, ymax=0.016,
  axis lines=left,
  tick align=outside,
  ytick distance=0.002,
  scaled x ticks=false,
  scaled y ticks=false,
  yticklabel style={/pgf/number format/fixed, /pgf/number format/fixed zerofill, /pgf/number format/precision=3},
  /pgf/number format/1000 sep={\,},
  /pgf/number format/.cd,
  x tick scale label code/.code={},
  y tick scale label code/.code={},
  legend style={draw=none, font=\small, at={(0.02,0.02)}, anchor=south west},
  legend cell align={left},
  every axis plot/.append style={line width=1.2pt}
]

\addplot+[
  only marks,
  mark=o,
  mark size=2.2pt,
  blue!75!black,
] coordinates {
  (405, 0.0149) 
  (532, 0.0147)
  (650, 0.0115)
};
\addlegendentry{Calculated -- Office Works tape}

\addplot+[blue!75!black, dashed, mark size=0pt, domain=380:680, samples=2] {0.0148};
\addlegendentry{Michel-L\'evy -- Office Works tape}

\addplot+[
  only marks,
  mark=o,
  mark size=2.2pt,
  purple!75!black,
] coordinates {
  (405, 0.0144) 
  (532, 0.0142) 
  (650, 0.0076) 
};
\addlegendentry{Calculated -- Canada Post tape}

\addplot+[purple!75!black, dashed, mark size=0pt, domain=380:680, samples=2] {0.0140};

\addlegendentry{Michel-L\'evy -- Canada Post tape}

\end{axis}
\end{tikzpicture}
\caption{Calculated birefringence for Office Works and Canada Post tapes for blue (405nm), green (532nm), and red (650nm) lasers and their Michel-L\'evy constant values.}
\label{graphiqueBirefringenceCalculee}
\end{figure}
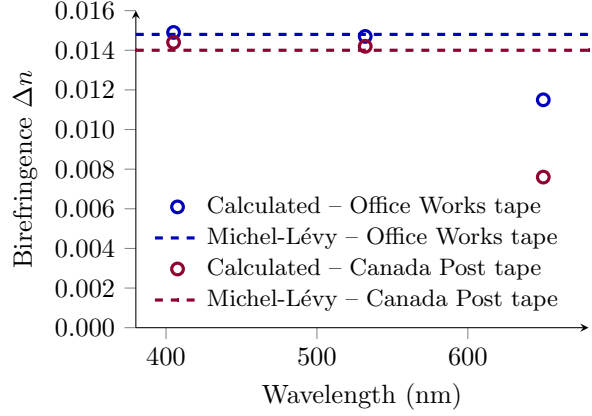

The calculated birefringence for the three laser wavelengths for both types of transparent tape is shown in figure~\ref{graphiqueBirefringenceCalculee}. We observe that the birefringence is almost constant for the 405nm and 532nm for the two types of tape and that there is a drop in the birefringence value for the 650nm light for both of them. The results are consistent with the first approximation obtained using the Michel–L\'evy chart in Section~\ref{SectionMichelLevyChar} where the birefringence of the Office Works tape was estimated to be 0.0148, and that of the Canada Post tape 0.0140. These values agree with the birefringence indices extracted from measurements using the blue and green lasers for both tapes. Only the red laser exhibits significant decreases in birefringence values for both materials. This explains why identifying the overall perceived colours on the Michel–L\'evy chart can be difficult: wavelengths approaching the red region are skewed. A reduction in birefringence for longer wavelengths has also been observed by Slepkov \cite{RefSlepkov} for some transparent tapes, although the values measured in our experiment remain lower than those reported in that study.

Finally, the calculated rotation angles are presented in Table~\ref{tab:AnglesCalcules}. This table provides the necessary information to construct any $R_x$ or $R_z$ gate for the specific configurations used here. It also indicates that the best setups for measuring the rotation angles of figure~\ref{PhotoMontageRxRzt} are: an $R_z$ gate composed of 3 layers of Office Works tape for the blue laser (275°), 4 layers of Office Works tape for the green laser (272°), and 5 layers of Canada Post tape for the red laser (277°). These gates produce $\phi$ values closest to 90° ($\frac{\pi}{2}$) or 270° ($-\frac{\pi}{2}$), as required for this experiment.

\begin{figure}
  \centering
  \subfloat[]{
    
\begin{tikzpicture}
\begin{axis}[
    width=6.6cm, height=6.0cm, 
    xlabel={Number of layers of Office Works tape},
    ylabel={Angle $\phi$ ($^{\circ}$)},
    xmin=0, xmax=6.5,
    ymin=0, ymax=2400,
    axis lines=left,
    tick align=outside,
    legend style={draw=none, at={(0.02,0.98)}, anchor=north west, font=\small},
    legend cell align={left},
    every axis plot/.append style={line width=1.2pt}
]

\addplot[only marks, mark=o, mark size=2.2pt, red]
coordinates {
    (1,164) (2,348) (3,470) (4,630) (5,808) (6,950)
};

\addplot[only marks, mark=o, mark size=2.2pt, green!60!black]
coordinates {
    (1,238) (2,490) (3,744) (4,996) (5,1220) (6,1504)
};

\addplot[only marks, mark=o, mark size=2.2pt, blue]
coordinates {
    (1,324) (2,664) (3,1006) (4,1330) (5,1632) (6,2006)
};

\addplot[red, thick, domain=0:6.5, samples=2] {159.67032967033*x};

\addplot[green!60!black, thick, domain=0:6.5, samples=2] {247.89010989011*x};

\addplot[blue, thick, domain=0:6.5, samples=2] {331.71428571429*x};
\addlegendentry{Red laser}
\addlegendentry{Green laser}
\addlegendentry{Blue laser}

\end{axis}
\end{tikzpicture}

\label{graphiqueAnglesMesuresST}
  }
  \hfill
  \subfloat[]{
   
\begin{tikzpicture}
\begin{axis}[
    width=6.6cm, height=6.2cm, 
    xlabel={Number of layers of Canada Post tape},
    ylabel={Angle $\phi$ ($^{\circ}$)},
    xmin=0, xmax=6.5,
    ymin=0, ymax=2400,
    axis lines=left,
    tick align=outside,
    legend style={draw=none, at={(0.02,0.98)}, anchor=north west, font=\small},
    legend cell align={left},
    every axis plot/.append style={line width=1.2pt}
]

\addplot[only marks, mark=o, mark size=2.2pt, red]
coordinates {
    (1,130) (2,262) (3,386) (4,504) (5,634) (6,766)
};

\addplot[only marks, mark=o, mark size=2.2pt, green!60!black]
coordinates {
    (1,286) (2,568) (3,854) (4,1154) (5,1444) (6,1720)
};

\addplot[only marks, mark=o, mark size=2.2pt, blue]
coordinates {
    (1,386) (2,758) (3,1150) (4,1550) (5,1926) (6,2306)
};

\addplot[red, thick, domain=0:6.5, samples=2] {127.40659340659*x};

\addplot[green!60!black, thick, domain=0:6.5, samples=2] {287.25274725275*x};

\addplot[blue, thick, domain=0:6.5, samples=2] {384.81318681319*x};
\addlegendentry{Red laser}
\addlegendentry{Green laser}
\addlegendentry{Blue laser}

\end{axis}
\end{tikzpicture}

\label{graphiqueAnglesMesuresPC}
}
  \caption{Experimental angles of rotation for up to six layers of (a) Office Works tape and (b) Canada Post tape for blue (405nm), green (532nm), and red (650nm) lasers.}
  \label{fig:example}
\end{figure}
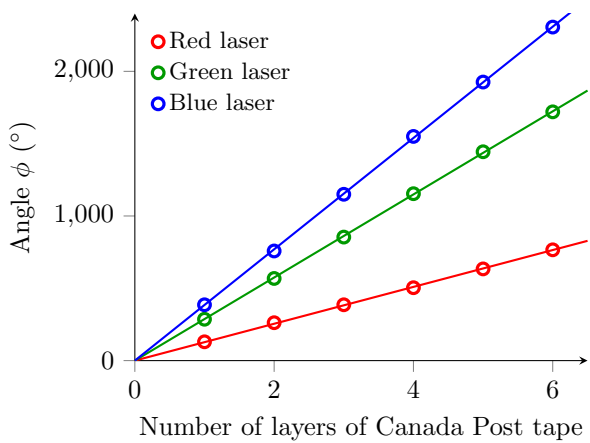

\begin{table}
\caption{Rotation angles $\theta$ ($^\circ$) on the Bloch sphere for $R_x$ and $R_z$ gates.}
\centering
\begin{tabular}{lcccccc}
\hline
Nb of layers & 1 & 2 & 3 & 4 & 5 & 6 \\
\hline
Blue 405nm\\
Office Works & 332 & 303 & 275 & 247 & 218 & 190 \\
Canada Post &  25 &  50 &  74 &  99 & 124 & 149 \\
\hline
Green 532nm\\
Office Works & 248 & 136 &  24 & 272 & 159 &  47 \\
Canada Post & 287 & 214 & 142 &  69 & 356 & 284 \\
\hline
Red 650nm\\
Office Works  & 160 & 319 & 119 & 279 &  78 & 238 \\
Canada Post & 127 & 255 &  22 & 150 & 277 &  44 \\
\hline
\end{tabular}
\label{tab:AnglesCalcules}
\end{table}

\subsection{Hadamard gate: maple syrup, agave syrup, and transparent tape}
\label{SectionHGate}
\begin{figure}
    \centering
\begin{center}
\includegraphics[width=1.0\textwidth]{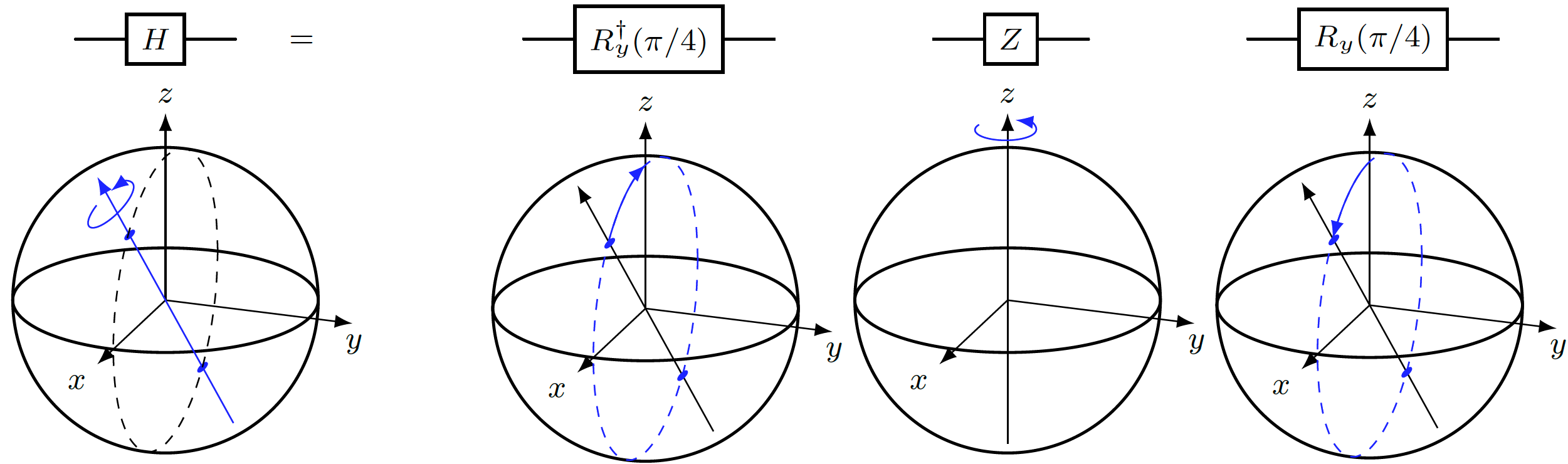}

\end{center}
  \caption{A Hadamard gate is a rotation of 180° around the axis that is located at halfway between the $x-$ and $z-$ axes of the Bloch sphere. It can be decomposed into three native gates of our systems: the $R_y^\dagger (\frac{\pi}{4})$, the $Z$, and the $R_y (\frac{\pi}{4})$ gates.}
  \label{PorteH}
\end{figure}

The Hadamard gate is a very important quantum logic gate and can be described as a rotation of 180° about an axis oriented midway between the $x$- and $z$-axes \cite{nielsen_chuang_2000}, as shown in figure~\ref{PorteH}. The Hadamard gate can be decomposed into a sequence of three native gates in our photonic system. First, an $R_y$ rotation of $-\pi/4$ is applied to the system causing the Hadamard rotation axis to point along the $z$-axis. Then, we apply a $Z$ gate (an $R_z$ rotation of 180°). Finally, an $R_y$ gate of $\pi/4$ returns the axis to its original position, which completes the transformation.

\begin{figure}
    \centering
   \begin{center}
\includegraphics[width=0.4\textwidth]{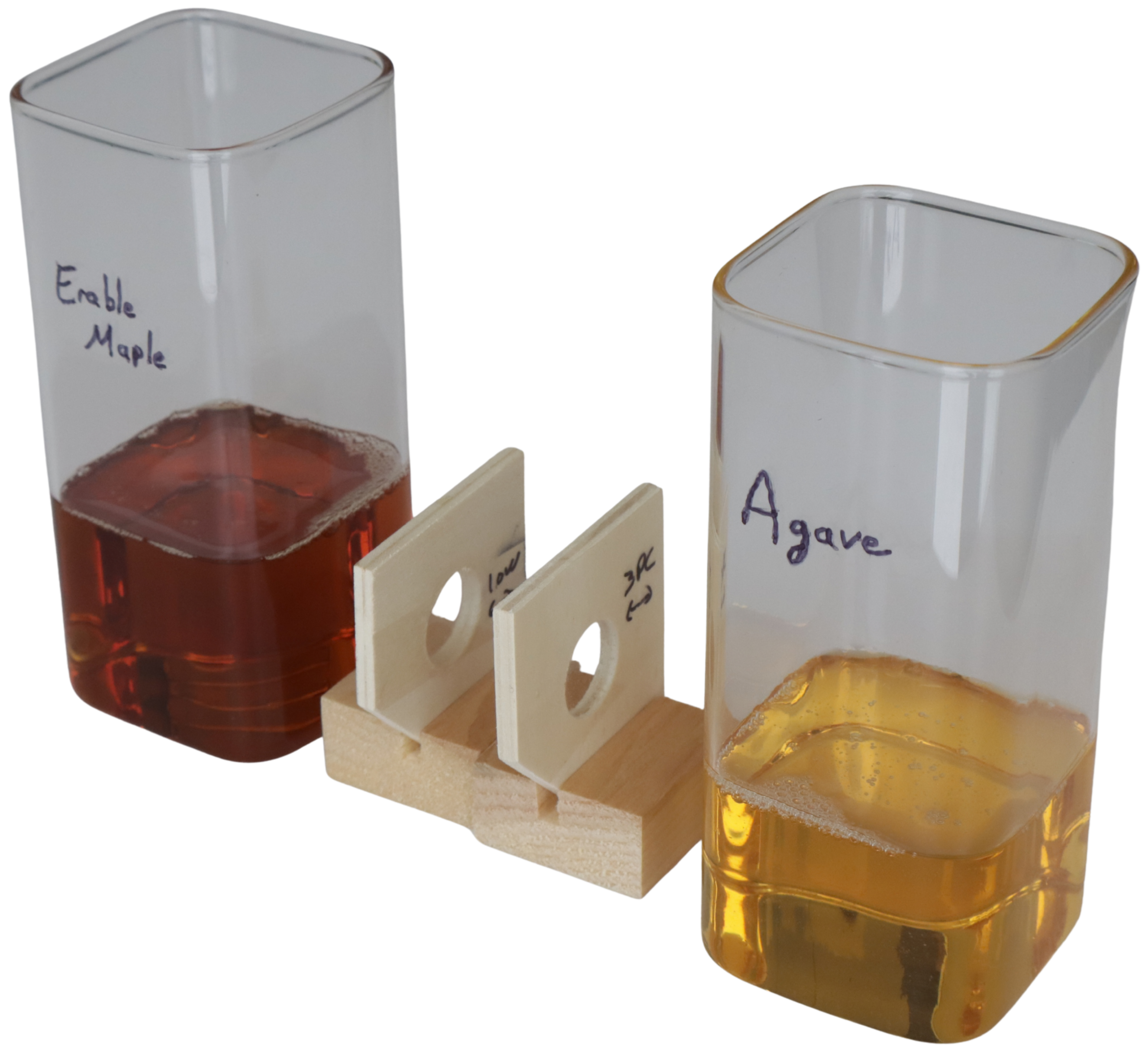}

\end{center}
    \caption{Experimental setup of the Hadamard gate. The maple syrup solution is diluted to create a physical rotation of -22.5° (-45° on the Bloch sphere). The $Z$ gate is an $R_z$ gate with an angle as close to 180° as possible and the agave syrup solution is diluted to create a physical rotation of 22.5° (45° on the Bloch sphere).}
    \label{PhotoHgate}
\end{figure}

For a given laser wavelength, the tools developed in the previous sections can be arranged to implement the Hadamard gate by first placing a maple syrup solution that physically rotates the polarization of light by -22.5° (-45° on the Bloch sphere), as shown in figure~\ref{PhotoHgate}. Then, we use a gate plate in the $R_z$ configuration. For the red laser, we can achieve a rotation angle of 182$^\circ$ with a combination of a single layer of Office Works tape together with three layers of Canada Post tape as calculated from Table~\ref{tab:AnglesCalcules}. This 182° rotation is sufficiently close to the ideal 180° to remain well within the experimental precision of our setup. Finally, we place an agave solution producing a physical rotation of 22.5° (45° on the Bloch sphere) in the circuit.
Exercises involving Hadamard gates can be found in appendix~\ref{AppendixProblemSets}.

\subsection{Calcite crystal as a CNOT gate}
\label{SecCalciteXtal}
 We can use calcite as a CNOT gate for entangling two qubits. The following circuit illustrates the simplest setup for producing entanglement between two qubits. First, $q_0$ is prepared in the $|+\rangle$ state, after which a CNOT gate is applied with $q_0$ as the control qubit and $q_1$ as the target qubit.

\begin{center}
\begin{quantikz}
\lstick{$q_0\ \ket{+}$} & \ctrl{1} & \qw\\
\lstick{$q_1\ \ket{0}$} & \targ{} & \qw       
\end{quantikz}
\end{center}

\begin{figure}
    \centering
   \begin{center}
\includegraphics[width=0.4\textwidth]{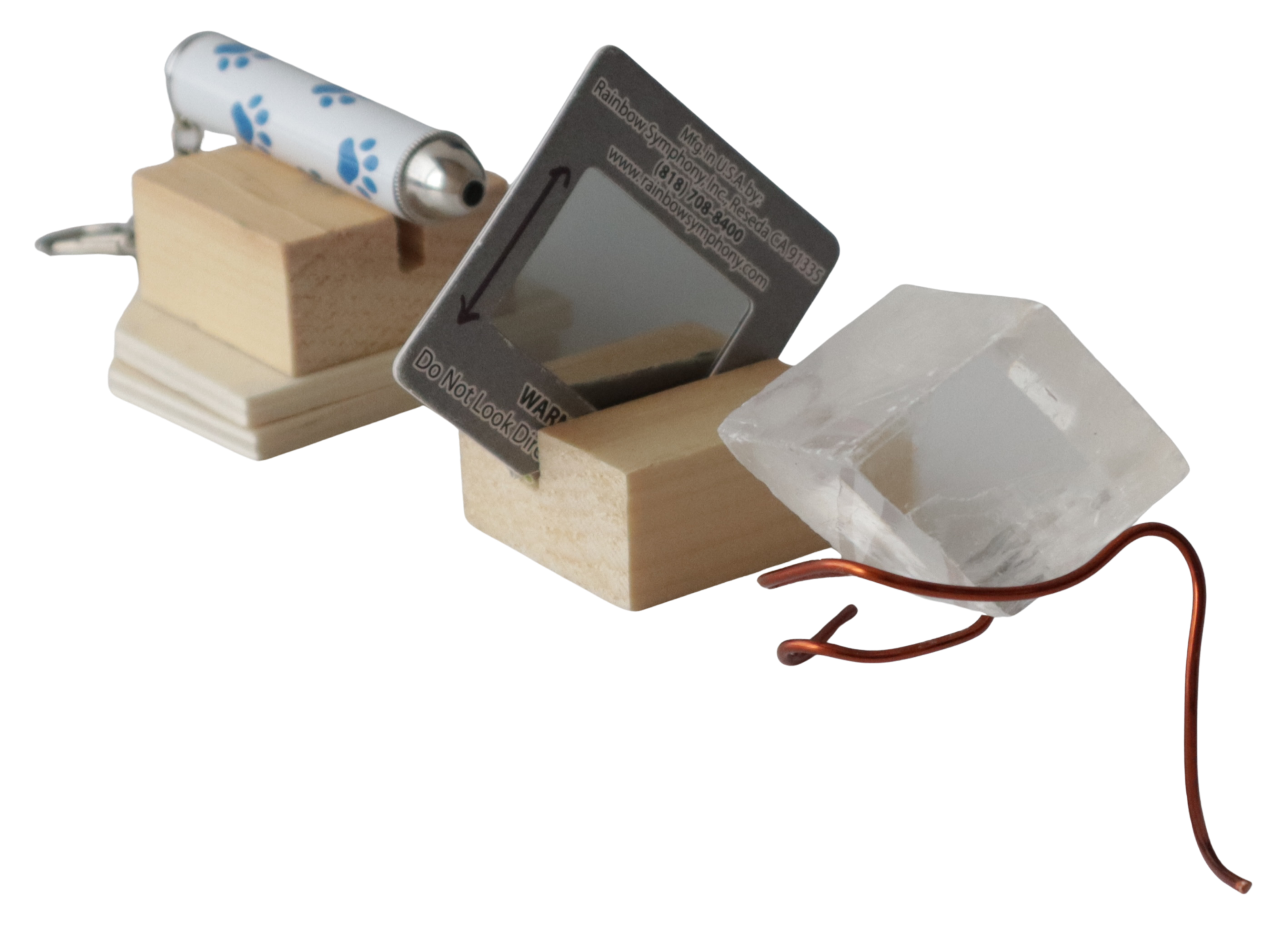}

\end{center}
      \caption{Experimental setup to create entanglement between a qubit $q_0$ encoded in the polarization of the photon and a qubit $q_1$ encoded in the path of the photon. The photon enters the calcite crystal in the $|+\rangle$ state and the calcite acts as a CNOT gate, creating entanglement between the two qubits. In reality, the calcite crystal also adds an unknown phase $\phi$ between the two states.}
      \label{PhotoMontageIntrication}
    \end{figure}

The setup shown in figure~\ref{PhotoMontageIntrication} represents this circuit, where $q_0$ is the qubit encoded in the polarization of the photon and $q_1$ is the qubit encoded in the path of the photon. For the latter, state $|0\rangle$ corresponds to the lower path and state $|1\rangle$ corresponds to the upper path (figure~\ref{PhotoCalcite}). The physical explanation behind this behaviour is : \textit{if the polarization of the photon is vertical ($q_0=|1\rangle$), the photon is deviated upward toward the optical axis ($q_1=|1\rangle$), whereas both qubits stay in the $|0\rangle$ states otherwise}. Since the horizontal polarization is associated to the lower path and the vertical polarization with the upper one, qubits $q_0$ and $q_1$ become entangled.

The photon thus exits the crystal in a superposition of states $|00\rangle$ and $|11\rangle$. In other words, it is either horizontally polarized ($q_0$ in the $|0\rangle$ state) and on the lower path ($q_1$ in the $|0\rangle$ state), or vertically polarized ($q_0$ in the $|1\rangle$ state) and on the upper path ($q_1$ in the $|1\rangle$ state). At the end of the circuit, the photon can be measured in only one of these two states. A more accurate representation of the physical situation includes an $R_z$ gate with an unknown value of $\phi$ acting on $q_1$, since the calcite crystal also introduces an unknown phase between states $|00\rangle$ and $|11\rangle$

\begin{center}
\begin{quantikz}
\lstick{$q_0\ \ket{+}$} & \ctrl{1} & \qw & \qw\\
\lstick{$q_1\ \ket{0}$} & \targ{}  & \gate{R_z (\phi)} & \qw       
\end{quantikz}
\end{center}

and the state vector is 
\begin{equation}
|\psi \rangle=\frac{1}{\sqrt{2}}|00\rangle+\frac{e^{i\phi}}{\sqrt{2}}|11\rangle\
\end{equation}

\subsection{\texorpdfstring{$CR_x$ and $CR_y$ and $CR_z$}{R\_x and R\_y and R\_z} gates}
Controlled gates use a control qubit and a target qubit. If the control qubit is in state $|1\rangle$, a gate is applied on the target qubit. Using the setup of figure~\ref{PhotoMontageIntrication} as the starting point, it is possible to add a $CR_x$, $CR_y$, or $CR_z$ gate by inserting the desired gate on only the upper path. Here is the resulting circuit if a CNOT ($CR_x$ ($180^\circ$)) is used
\begin{center}
\begin{quantikz}
\lstick{$q_0\ \ket{+}$} & \ctrl{1} & \qw        & \targ{}  & \qw        & \qw \\
\lstick{$q_1\ \ket{0}$} & \targ{}  & \gate{R_z (\phi_1)} & \ctrl{-1}& \gate{R_z (\phi_2)} & \qw      
\end{quantikz}
\end{center}

\begin{figure}
    \centering
   \begin{center}
\includegraphics[width=0.4\textwidth]{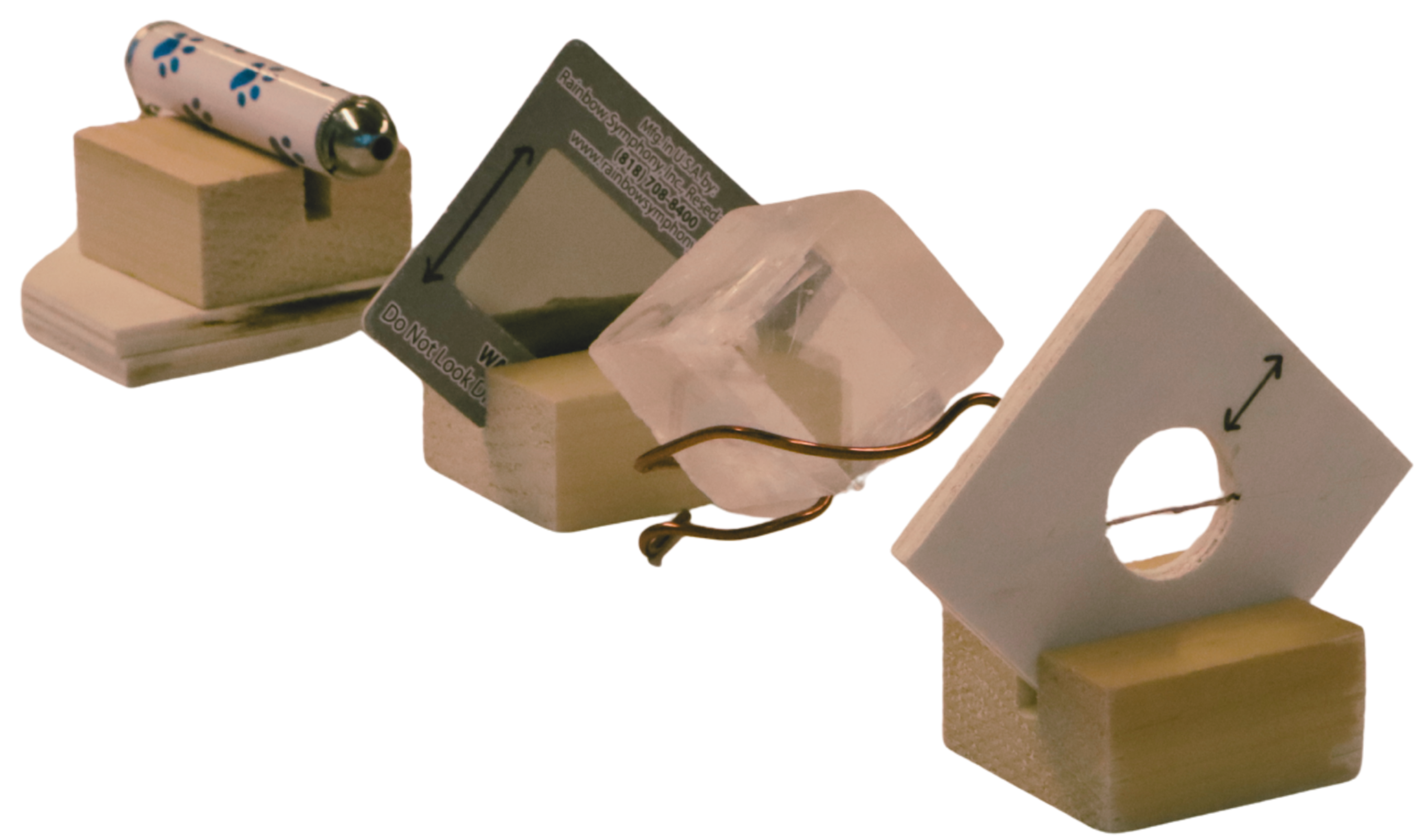}

\end{center}
      \caption{Setup for implementing a CNOT gate using transparent tape. The laser points to a diagonal polarizer followed by a calcite crystal and transparent tape only on the upper path of the photon. This example shows the application of a NOT gate only on the upper path or otherwise said, only if the path qubit is state $|1\rangle$.}
    \label{photoMontageIntricationCNOT}
\end{figure}

Figure~\ref{photoMontageIntricationCNOT} shows the implementation of this last circuit. It is worth noting that the CNOT gate also introduces an unknown $R_z$ rotation into the circuit. In this example, the CNOT is implemented using one layer of Office Works tape and three layers of Canada Post tape with a red laser. The gate is applied only if $q_1$ is in state $|1\rangle$ (i.e., when the photon follows the upper path).
The final state of the circuit is
\begin{equation}
|\psi\rangle = \frac{1}{\sqrt{2}}|00\rangle + \frac{e^{i\phi}}{\sqrt{2}}|10\rangle\
\end{equation}
This shows that $q_0$ (noted on the right of $|00\rangle$ or $|10\rangle$) is always in state $|0\rangle$, and therefore the two qubits are no longer entangled. Additionally, an unknown phase $\phi$ is present, as indicated by the term $e^{i\phi}$. The same idea of applying an $R_y$ or $R_z$ gate on only one of the two paths to create a $CR_y$ or $CR_z$ gate is also valid.
  
\subsection{Measurement apparatus}
\label{sectionMeasurementApparatus}
Two measurement apparatuses are used in these experimental setups. Polarizing filters provide the ability to measure the polarization of a photon along a specific basis. All photons exiting such a measurement device are always linearly polarized in the direction defined by the filter. These filters are used immediately after the lasers to obtain the desired initial state that will later be transformed in the optical setups, and they are also placed at the end of many setups to determine the final state.

The other measurement device is a solid screen (i.e., a piece of paper) placed at the end of all setups, although it is omitted from the various pictures for clarity. The screen measures the position of light. For the 2‑qubit systems described earlier, the laser beams are detected either on the lower path or on the upper path. For the one‑qubit systems, the beam naturally produces a dot at a fixed location, but when used together with a polarizing filter, the screen makes it possible to identify the angles of full extinction and to calculate the angle of maximum intensity. 

\section{Playing with gates}
This section provides sets of simple algorithms to implement using the various gates described above. Additional experiments and exercises are presented in appendix~\ref{AppendixProblemSets}.
\subsection{Inverse gates}
In quantum computing, all gates have an inverse that cancels the effect of the original gate when applied afterward. The usage of $R_y$ and $R_y^\dagger$ gates have already been encountered in the context of fructose and sucrose solutions, which were adjusted to produce the same rotation but in opposite directions. Note that these two gates must be placed back-to-back for the quantum state to return to its initial form after passing through both. A simple experiment demonstrates the final state of the following circuit when the same angle of rotation is used but in opposite directions:

\begin{center}
\begin{quantikz}

\lstick{$q_0\ \ket{0}$} & \gate{R_y^\dagger (\pi/4)} & \gate{R_y (\pi/4)} & {\ket{0}}\\

\end{quantikz}
\end{center}

\begin{figure}
    \centering
    \begin{center}
\includegraphics[width=0.4\textwidth]{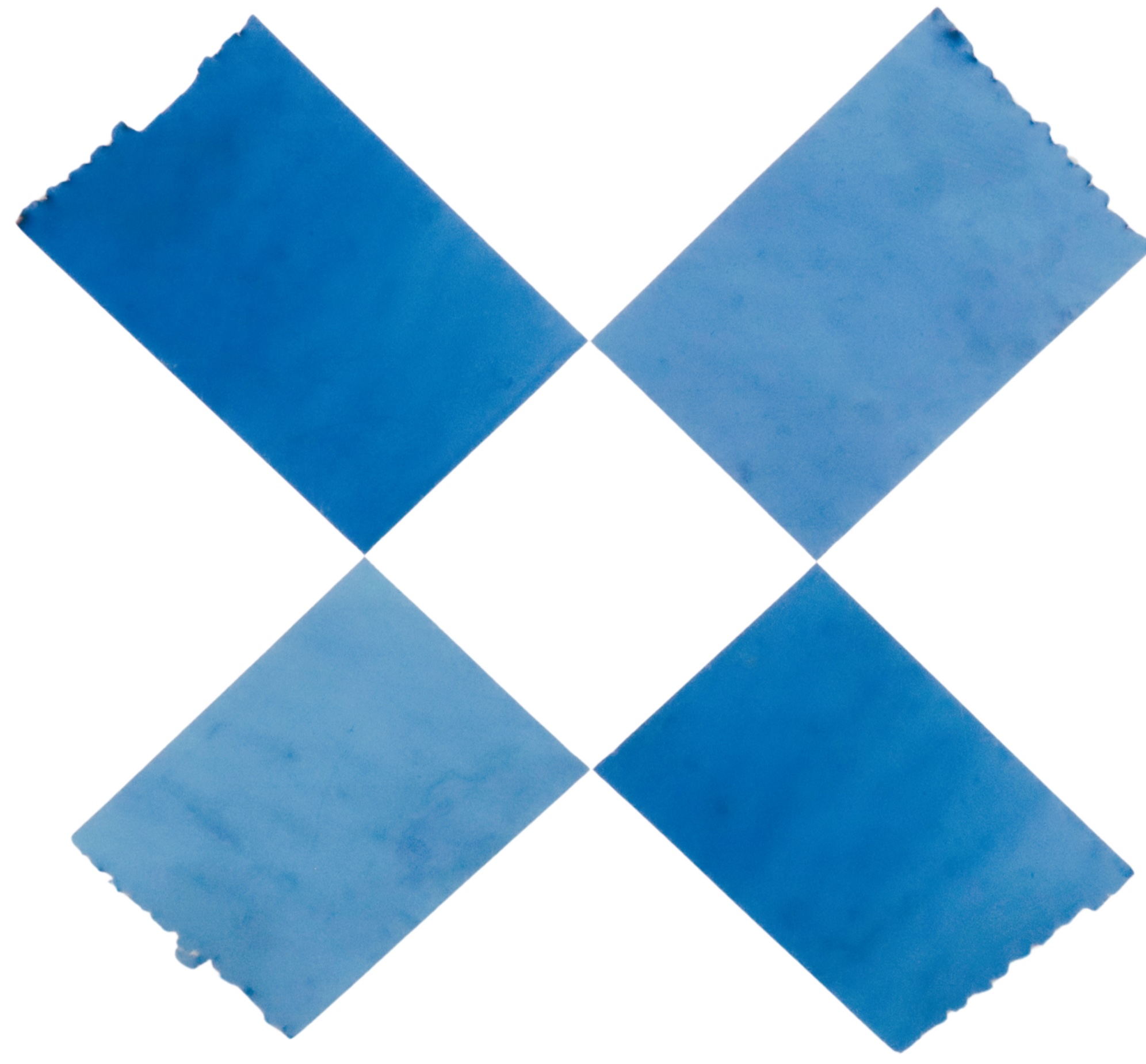}

\end{center}
    \caption{Horizontally polarized white light enters an anti-diagonal single layer of Office Works tape ($R_x$) followed by a diagonal single layer of the same tape ($R_x^\dagger$) and then a horizontal polarizer attached to the camera lens. The white square in the middle indicates that the second gate annihilates the effect of the first one. The blue colour on all four arms is due to a single gate rotation.}
    \label{PhotoRx_Rxt}
\end{figure}

On its side, figure~\ref{PhotoRx_Rxt} shows horizontally polarized white light entering an anti-diagonal single layer of Office Works tape ($R_x$) followed by a diagonal single layer of the same tape ($R_x^\dagger$) and then a horizontal polarizer. The quantum circuit is:

\begin{center}
\begin{quantikz}

\lstick{$q_0\ \ket{0}$} & \gate{R_x (\theta_i)} & \gate{R_x^\dagger (\theta_i)} & {\ket{0}}\\

\end{quantikz}
\end{center}

We see that the central square appears white, matching the incoming light. This is due to the annihilating effect of the second gate. The light that passes through a single layer of tape in each of the four arms appears blue, in accordance with what we discussed in Sections \ref{SectionBiref1} and \ref{SectionMichelLevyChar}.

One may test inverse gates for all the $R_x$, $R_y$, and $R_z$ gates described earlier, as well as for the Hadamard gate, which is its own inverse..

\subsection{Deutsch algorithm implementation}
The Deutsch algorithm is a simple quantum algorithm that demonstrates quantum advantage \cite{nielsen_chuang_2000}. When an object can take two possible states (0 and 1) and a transformation is applied to it, the Deutsch algorithm allows us to use a single query to answer the question:
“\textit{Is the transformation applied to the two possible states the same, or is it different for each state?}” Two possible classes of transformations exist: a transformation either leaves the initial state unchanged, or it flips the state (i.e., transforms 0 to 1 and 1 to 0).

\begin{figure}
    \centering
   \begin{center}
\includegraphics[width=0.48\textwidth]{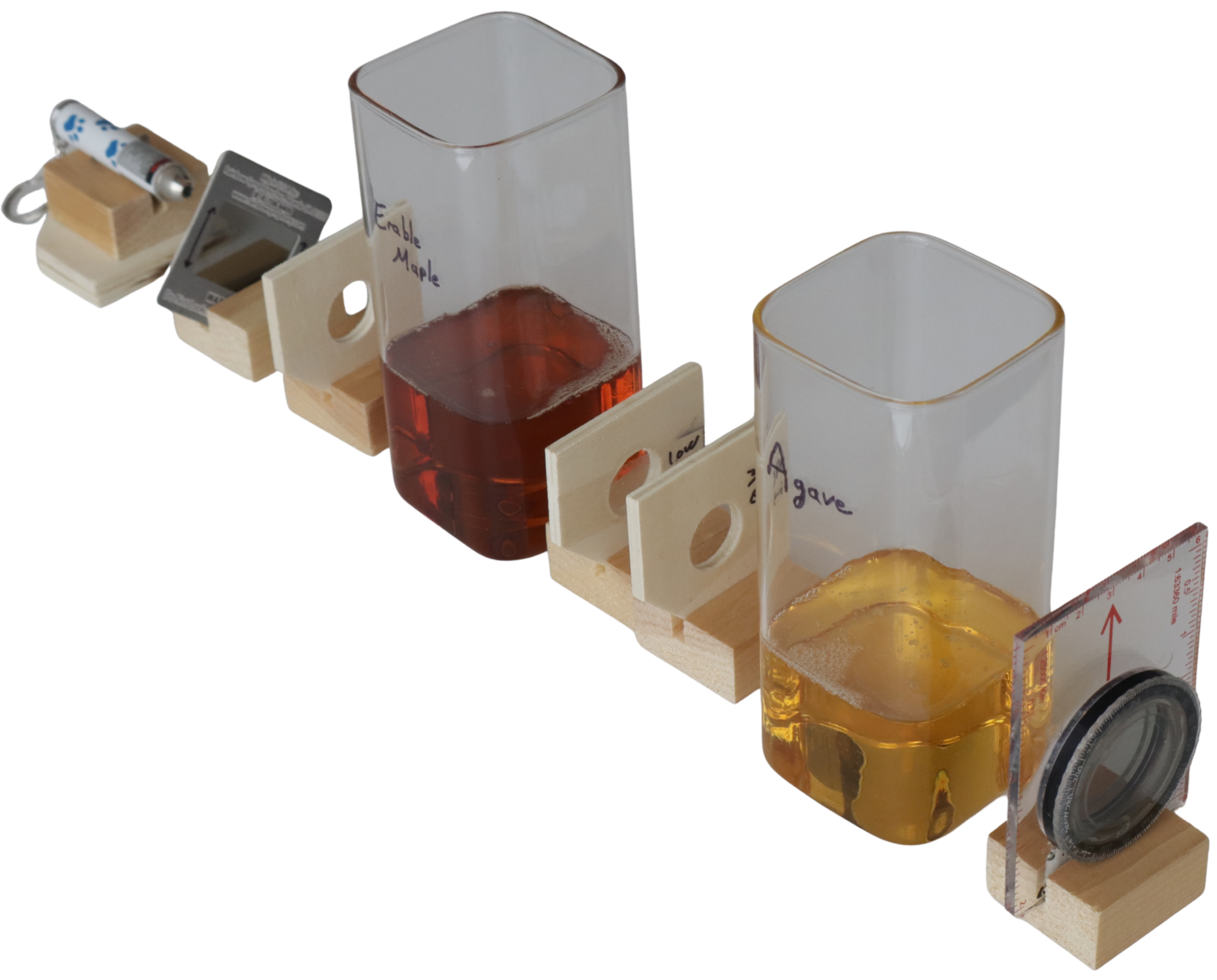}
      \caption{The Deutsch algorithm setup goes as follows: laser, anti-diagonal polarizer, UNKNOWN gate, Hadamard gate, and rotating polarizer. The unknown gate is either the $R_x$, $R_x^\dagger$, $R_z$, or $R_z^\dagger$ gate with an angle of rotation of 180°.}
    \label{PhotoDeutsch}

\end{center}
\end{figure}

If the transformation is the same for both states (either both unchanged or both flipped), the measured qubit yields the result $|0\rangle$. If the transformation is different for the two states (one flipped and the other unchanged), the result is $|1\rangle$. Yoganathan \cite{RefYouTubeYoganathan} presented a YouTube demonstration of a modified version of the Deutsch algorithm that uses only a single qubit instead of two in the original version \cite{nielsen_chuang_2000}. Here, we use the same modified version, but with the final Hadamard gate implemented using the setup described in Section~\ref{SectionHGate}. The circuit for this version of the Deutsch algorithm is:
\begin{center}
\begin{quantikz}
\lstick{$q_0\ \ket{+}$} & \gate{?} & \gate{H} & \meter{}
\end{quantikz}
\end{center}
The box labelled with a question mark represents an operation unknown to the experimentalist and corresponds to the transformation being tested. A gate plate producing a rotation of $\pi$ is placed inside the box in an unknown position producing an $R_x$ ($s_h, s_v$), $R_x^\dagger$ ($f_h, f_v$), $R_z$ ($f_h, s_v$), or $R_z^\dagger$ ($s_h, f_v$) gate. The letters in parentheses indicate how fast the photons travelled through the tape. $s_h$ means the horizontal component travelled (relatively) slowly, while $f_h$ means it travelled (relatively) fast. $s_v$ and $f_v$ are for the vertical components. Recall (Section \ref{SectionBirefringence}) that if both the horizontal and vertical components pass through the tape at the same speed, the polarization state does not change. If however, one is slow and the other fast, the polarization state rotates about an axis in the Bloch sphere.
The experimental setup is shown in figure~\ref{PhotoDeutsch}. In practice, for red‑laser implementation, the unknown gate consists of one layer of Office Works tape and three layers of Canada Post tape ($182^\circ$), oriented in one of four possible directions.

We now turn to why this algorithm works. For a photon in the $|+\rangle$ state, the $R_x$ ($s_h, s_v$) and $R_x^\dagger$ ($f_h, f_v$) gates have no effect on the photon phase, since they do not alter the $|+\rangle$ state. No change occurs because $|+\rangle$ polarization oscillates either parallel or perpendicular to the optical axis, resulting in no change to the relative phase. In contrast, when the inserted gate is $R_z$ ($f_h, s_v$) or $R_z^\dagger$ ($s_h, f_v$), the input photons are transformed from the $|+\rangle$ state to the $|-\rangle$ state.
At the end of the circuit, two outcomes are possible. A photon entering the Hadamard gate in the $|+\rangle$ state is transformed into the $|0\rangle$ state, whereas a photon entering in the $|-\rangle$ state is transformed into the $|1\rangle$ state. Thus, if we measure a $|0\rangle$ state, we know the transformation does not modify the initial state. For $|1\rangle$, we know the transformation does change the initial state. Note that we can determine this with a \textit{single} measurement. Although this modified Deutsch algorithm does not identify which of the four specific gates was used, it reliably determines whether the underlying rotation was about the $x$‑axis or the $z$‑axis.
A more detailed explanation of the Deutsch algorithm suitable for classroom use is provided in a video \cite{RefYouTubeDeutsch} from the \textit{Quantum Enigmas} series.

\subsection{3-qubit circuit for state preparation}
It is also possible to use multiple calcite crystals to produce 3-qubit circuits (or more) for state preparation. If we wished to prepare the following quantum state:

\begin{equation}
|\psi\rangle = \frac{1}{2}|000\rangle + \frac{e^{i\phi_1}}{2}|100\rangle + \frac{e^{i\phi_2}}{\sqrt{2}}|111\rangle\
\label{3qubitalgo}
\end{equation}

we could use the following circuit: 

\begin{center}
\begin{quantikz}
\lstick{$q_0\ \ket{+}$} & \ctrl{1}  & \gate{R_x(90^\circ)} & \octrl{2} & \targ{} & \qw 
& \gategroup[wires=3,steps=2,style={rounded corners,fill=gray!10},label style={anchor=north, yshift=-35}]{$\phi_i$} & \qw & \qw \\
\lstick{$q_1\ \ket{0}$} & \targ{} & \octrl{-1} & \qw & \octrl{-1} & \qw & \qw & \qw & \qw \\
\lstick{$q_2\ \ket{1}$} & \qw & \qw & \targ{} & \ctrl{-1} & \qw & \qw & \qw & \qw 

\end{quantikz}
\end{center}

Note: All two- and three-qubit gates introduce unknown relative phase factors, represented by a final element labeled $\phi_i$.\\ 
As defined previously in Section~\ref{SecCalciteXtal}, $q_0$ is the qubit encoded in the
polarization of the photon ($|0\rangle$ is horizontal polarization and $|1\rangle$ is vertical) and $q_1$ is the qubit encoded in the vertical path of the photon ($|0\rangle$ is bottom and $|1\rangle$ is top). Here, we define $q_2$ as the qubit encoded in the \textit{horizontal} path of the photon ($|0\rangle$ is left and $|1\rangle$ is right). Using the convention where the photon propagates toward the observer, the four possible photon paths are shown in figure~\ref{2pathqubits}.

Figure~\ref{3qubitsetup} shows the evolution of the polarization of the photon depending on the path taken after each gate of the above circuit. For simplicity of the experimental implementation, $q_0$ is initialized in the $|+\rangle$ state and $q_2$ in the $|1\rangle$.

\begin{figure}
    \centering
\begin{center}
    
   \tdplotsetmaincoords{0}{0}
\begin{tikzpicture}[scale=1.7,tdplot_main_coords]

\draw[] (-0.5,0,0) circle (0.04 cm) node[anchor=north west]{$|00\rangle|q_0\rangle$};

\draw[] (0.5,0,0) circle (0.04 cm) node[anchor=north west]{$|10\rangle|q_0\rangle$};

\draw[] (-0.5,1,0) circle (0.04 cm) node[anchor=north west]{$|01\rangle|q_0\rangle$};

\draw[] (0.5,1,0) circle (0.04 cm) node[anchor=north west]{$|11\rangle|q_0\rangle$};

\end{tikzpicture}
    \caption{A setup is constructed in which four optical paths are possible. The lower paths correspond to $q_1=|0\rangle$, while the upper paths correspond to $q_1=|1\rangle$ state. The left-side paths correspond to $q_2=|0\rangle$, and the right-side paths correspond to $q_2=|1\rangle$. In all cases, $q_0$ encodes the polarization of the photon. Note that $|q_2 q_1 q_0\rangle$ can also be expressed as $|q_2 q_1\rangle |q_0\rangle$. The small circle correspond to \textit{no light} in a possible path}
    \label{2pathqubits}
\end{center}
\end{figure}

\begin{figure}
    \centering
\begin{center}

\begin{tikzpicture}[scale=1.7]

\begin{scope}[shift={(0,0)}]
\draw[] (0,0,0) circle (0.03 cm) node[anchor=north west, xshift=-20, yshift=-5]{initial state};
\draw[<->, >=Straight Barb] (0.4-0.0707,-0.0707,0)--(0.4+0.0707,0.0707,0);
\draw[] (0,0.4,0) circle (0.03 cm);
\draw[] (0.4,0.4,0) circle (0.03 cm);

\end{scope}

\begin{scope}[shift={(1,0)}]
\draw[] (0,0.2,0)--(0.2,0.4,0)--(0.4,0.2,0)--(0.2,0,0)--(0,0.2,0) node[anchor=north west, xshift=-7, yshift=-15]{CNOT};
\draw[->] (0.2,0.1,0)--(0.2,0.3,0);
\draw[] (0.15,0.15,0) circle (0.01 cm);
\draw[] (0.25,0.15,0) circle (0.01 cm);
\draw[] (0.15,0.25,0) circle (0.01 cm);
\draw[] (0.25,0.25,0) circle (0.01 cm);
\end{scope}

\begin{scope}[shift={(2,0)}]
\draw[] (0,0,0) circle (0.03 cm);
\draw[<->, >=Straight Barb] (0.3,0,0)--(0.5,0,0);
\draw[] (0,0.4,0) circle (0.03 cm);
\draw[<->, >=Straight Barb] (0.4,0.3,0)--(0.4,0.5,0);
\end{scope}

\begin{scope}[shift={(3,0)}]

  \draw[
  draw=black,
  fill=gray!70,
  fill opacity=0.6,
  variable=\x,
  domain=0:0.4,
  samples=100
]
plot (\x,{0.2-sqrt(0.2^2-(\x-0.2)^2)})
-- (0.4,0.2)
-- (0,0.2)
-- cycle;

\draw[] (0.15,0.15,0) circle (0.01 cm);
\draw[] (0.25,0.15,0) circle (0.01 cm);
\draw[] (0.15,0.25,0) circle (0.01 cm);
\draw[] (0.25,0.25,0) circle (0.01 cm);

\draw[] (0.006,0.15,0) -- (0.144,0.0095,0);
\draw[] (0,0.2,0) -- (0.2,0,0);
\draw[] (0.05,0.2,0) -- (0.243,0.009,0);
\draw[] (0.1,0.2,0) -- (0.277,0.019,0);
\draw[] (0.15,0.2,0) -- (0.31,0.039,0);
\draw[] (0.2,0.2,0) -- (0.34,0.06,0);
\draw[] (0.25,0.2,0) -- (0.365,0.085,0);
\draw[] (0.3,0.2,0) -- (0.38,0.12,0);
\draw[] (0.35,0.2,0) -- (0.395,0.155,0);

\draw[] (0.2,0.2,0) circle (0.2 cm) node[anchor=north west, xshift=-23, yshift=-14]{$CR_x (90^\circ )$};
\end{scope}

\begin{scope}[shift={(4,0)}]
\draw[] (0,0,0) circle (0.03 cm);
\draw[] (0.4+0.0707,0,0) arc[start angle={360}, end angle={0}, x radius={0.0707}, y radius={0.0707}];
\draw[->, >=Straight Barb] (0.4+0.07,0.0018,0)--(0.4+0.0707,0,0);
\draw[] (0,0.4,0) circle (0.03 cm);
\draw[<->, >=Straight Barb] (0.4,0.3,0)--(0.4,0.5,0);
\end{scope}

\begin{scope}[shift={(5,0)}]
\draw[] (0,0.2,0)--(0.2,0.4,0)--(0.4,0.2,0)--(0.2,0,0)--(0,0.2,0) node[anchor=north west, xshift=-7, yshift=-15]{CNOT};
\draw[->] (0.3,0.2,0)--(0.1,0.2,0);
\draw[] (0.15,0.15,0) circle (0.01 cm);
\draw[] (0.25,0.15,0) circle (0.01 cm);
\draw[] (0.15,0.25,0) circle (0.01 cm);
\draw[] (0.25,0.25,0) circle (0.01 cm);
\end{scope}

\begin{scope}[shift={(6,0)}]
\draw[<->, >=Straight Barb] (-0.1,0,0)--(0.1,0,0);
\draw[<->, >=Straight Barb] (0.4,-0.1,0)--(0.4,0.1,0);
\draw[] (0,0.4,0) circle (0.03 cm);
\draw[<->, >=Straight Barb] (0.4,0.3,0)--(0.4,0.5,0);
\end{scope}

\begin{scope}[shift={(7,0)}]

\begin{scope} [rotate=45, shift={(0.083,-0.2)}]

  \draw[
  draw=black,
  fill=gray!70,
  fill opacity=0.6,
  variable=\x,
  domain=0:0.4,
  samples=100
]
plot (\x,{0.2-sqrt(0.2^2-(\x-0.2)^2)})
-- (0.4,0.2)
-- (0,0.2)
-- cycle;

\draw[] (0.007,0.133,0) -- (0.393,0.133,0);
\draw[] (0.025,0.1,0) -- (0.375,0.1,0);
\draw[] (0.05,0.067,0) -- (0.35,0.067,0);
\draw[] (0.09,0.033,0) -- (0.31,0.033,0);
\draw[draw=white, fill=white] (0.005,0.164,0) -- (0.395,0.164,0) -- (0.395,0.236) -- (0.005,0.236) -- (0.005,0.164,0);
\draw[] (0.005,0.164,0) -- (0.395,0.164,0);
\end{scope}

\draw[] (0.15,0.15,0) circle (0.01 cm);
\draw[] (0.25,0.15,0) circle (0.01 cm);
\draw[] (0.15,0.25,0) circle (0.01 cm);
\draw[] (0.25,0.25,0) circle (0.01 cm);

\draw[] (0.2,0.2,0) circle (0.2 cm) node[anchor=north west, xshift=-23, yshift=-14]{$CCNOT$};
\end{scope}

\begin{scope}[shift={(8,0)}]
\draw[<->, >=Straight Barb] (-0.1,0,0)--(0.1,0,0);
\draw[<->, >=Straight Barb] (0.3,0,0)--(0.5,0,0);
\draw[] (0,0.4,0) circle (0.03 cm);
\draw[<->, >=Straight Barb] (0.4,0.3,0)--(0.4,0.5,0);
\end{scope}
\end{tikzpicture}

    \caption{Evolution of the quantum state during the 3-qubit algorithm that prepares the state in Eq.~\ref{3qubitalgo}. The double-sided arrows in the four path representations show the linear polarization state along each path (whose lengths are not proportional to the laser intensity), while a circle with an arrow shows circular polarization, and and empty small circle is the absence of light. For the gates, diamonds represent the displacement of the polarization component parallel to the arrow direction in calcite. Large circles represent the use of transparent tape  (hatched region) to implement two-qubit gates.}
    \label{3qubitsetup}

\end{center}
\end{figure}

\section{Conclusions}
We have demonstrated that simple everyday materials and equipment can be used to simulate an analogue photonic quantum computer. We did this by taking advantage of a mapping between the polarization states of light and the Bloch sphere. That is, manipulating the polarization states is mathematically equivalent to manipulating a qubit. We generated the light with a cat laser pointer and then manipulated these polarization states using transparent adhesive tape, which is birefringent, and syrup solutions, which are optically active. We then introduced the ability to consider two qubits by using the different refractive properties of calcite for horizontally and vertically polarized light to encode the second qubit in a position degree of freedom.

We determined birefringent values via the Michel-L\'evy chart and via direct measurement of polarization rotation angles. We also demonstrated the implementation of various quantum circuits, including a one-qubit version of the Deutsch algorithm.

Our setup costs less than $\$$50 CAD for the single qubit experiments and less than $\$$100 CAD for all of the experiments. Thus, this setup is well suited for at home or in-class implementation, allowing students even at the high school level to gain intuition about quantum computing via a tactile method. To support such experimentation, we have placed multiple exercises in appendix \ref{AppendixProblemSets}.

\section{Acknowledgement}{The author would like to thank the following individuals at the Institut quantique who contributed to the ideas underlying this article and to the revision of the final manuscript: Alexandre Foley, Joshua Tyler Cantin, Maxime Dion, Jean Fr\'ed\'eric Laprade, and Philippe Karan. The author is also grateful to the Institute for Quantum Computing for inspiring one of the exercises in the appendix. The author also wants to acknowledge the financial support of the Ministère de l'\'Economie, de l'Innovation et de l'\'Energie du Qu\'ebec (MEIE), which made this project possible.}

\section{Biography}
Ghislain Lefebvre oversees partnerships at the Institut quantique at the Universit\'e de Sherbrooke. Since 2020, he has developed strategic collaborations in quantum computing with major national and international organizations. Committed to advancing quantum education, he has also led the creation of the animated video series \textit{The Quantum Enigmas} and \textit{The 2-Qubit Dance}, as well as numerous introductory workshops aimed at broadening access to quantum computing.

\begin{appendices}
\section{Quantum Fundamentals}
\label{AppendixFoundamentals}

\subsection{Linearly polarized light}
\label{AppendixLinearPolarization}

The following appendix is based on the work presented in  \cite{Couteau2023,Barz2015,nielsen_chuang_2000,raymer_2017}.

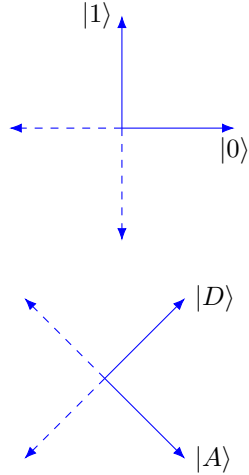
\begin{figure}
    \centering

\tdplotsetmaincoords{0}{0} 
\begin{center}
\begin{tikzpicture}[tdplot_main_coords]
 \begin{scope}
  \draw[->, blue] (0,0,0) -- (1.5,0,0) node[anchor=north, black]{$\ket{0}$};
  \draw[->, blue] (0,0,0) -- (0,1.5,0) node[anchor=east, black]{$\ket{1}$};
  \draw[->, blue, dashed] (0,0,0) -- (-1.5,0,0);
  \draw[->, blue, dashed] (0,0,0) -- (0,-1.5,0);
\end{scope}
\end{tikzpicture}

\end{center}

\begin{center}
\begin{tikzpicture}[tdplot_main_coords]
   \begin{scope} [shift={(0,10)}]     
\def\C{135}
  \draw[->, blue] (0,0,0) -- (1.5/1.4,1.5/1.4,0) node[anchor=west, black]{$\ket{D}$};
  \draw[->, blue, dashed] (0,0,0) -- (-1.5/1.4,1.5/1.4,0);
  \draw[->, blue, dashed] (0,0,0) -- (-1.5/1.4,-1.5/1.4,0);
  \draw[->, blue] (0,0,0) -- (1.5/1.4,-1.5/1.4,0) node[anchor=west, black]{$\ket{A}$};
\end{scope}

\end{tikzpicture}
\end{center}

\caption{By convention, a horizontal polarization vector corresponds to the $|0\rangle$ state and a vertical polarization vector to the $|1\rangle$ state (upper panel). The diagonal and anti-diagonal polarization vectors are shown in the bottom panel.}
    \label{0-1bases}
\end{figure}

For an initial understanding of quantum computing, one can begin with the special case of linearly polarized light. Horizontally polarized light is, by convention, associated to the zero state. As quantum states are written using the ket notation, the zero state is denoted $|0\rangle$. The one state corresponds to vertically polarized light and is written $|1\rangle$. Physically and mathematically, the $|0\rangle$ and $|1\rangle$ states are orthogonal to one another, as illustrated in figure~\ref{0-1bases}.

\begin{figure}
    \centering
\begin{center}
\tdplotsetmaincoords{0}{0}
\begin{tikzpicture}[tdplot_main_coords]
 
  \draw[->] (0,0,0) -- (2,0,0) node[anchor=north]{$\ket{0}$};
  \draw[->] (0,0,0) -- (0,2,0) node[anchor=east]{$\ket{1}$};
  \draw[->, blue] (0,0,0) -- (1.414,1.414,0) node[anchor=west, black] {$\ket{D}=\frac{1}{\sqrt{2}}\ket{0}+\frac{1}{\sqrt{2}}\ket{1}$};
  \draw[dashed] (1.414,1.414,0) -- (1.414,0,0);
  \draw[->] (0,0,0) -- (0,-2,0) node[anchor=east] {$-\ket{1}$};
  \draw[->, blue] (0,0,0) -- (1.414,-1.414,0) node[anchor=west, black] {$\ket{A}=\frac{1}{\sqrt{2}}\ket{0}-\frac{1}{\sqrt{2}}\ket{1}$};
  \draw[dashed] (1.414,-1.414,0) -- (1.414,0,0);

\end{tikzpicture}
\end{center}

\caption{Relationship of the $|D\rangle$ and $|A\rangle$ states to the $|0\rangle$ and $|1\rangle$ states, provided all four vectors are normalized.}
    \label{DAcalc}
\end{figure}
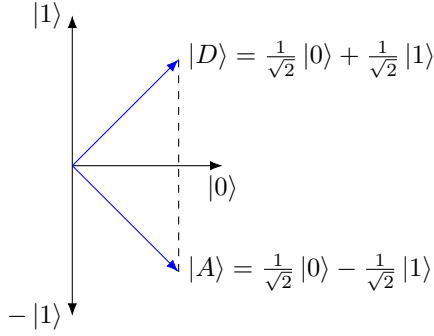

Two other useful states of linearly polarized light are the diagonal $|D\rangle$ and anti-diagonal $|A\rangle$ states, also shown in figure~\ref{0-1bases}. Since the $|D\rangle$ state is oriented at a 45° angle relative to the $|0\rangle$ and $|1\rangle$ states, its state vector can be expressed as
\begin{equation}
\label{EquationPlus}
|D\rangle=\frac{1}{\sqrt{2}}|0\rangle+\frac{1}{\sqrt{2}}|1\rangle\
\end{equation}
assuming that all vectors are normalized (figure~\ref{DAcalc}).

The $|A\rangle$ state, shown in figure~\ref{DAcalc} is defined similarly, but with a minus sign added before the second term
\begin{equation}
\label{EquationMinusState}
|A\rangle=\frac{1}{\sqrt{2}}|0\rangle-\frac{1}{\sqrt{2}}|1\rangle\
\end{equation}

\subsection{Bloch sphere}
\label{AppendixBlochSphere}

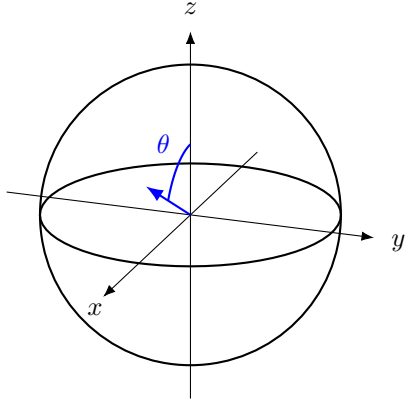
\begin{figure}
    \centering
\tdplotsetmaincoords{70}{110} 
\begin{center}
\begin{tikzpicture}[tdplot_main_coords, scale=2]

  \draw[thick] (0,0,0) circle (1); 
  \draw[thick, tdplot_screen_coords] (0,0) circle (1); 

  \draw[->] (0,0,-1.3) -- (0,0,1.3) node[anchor=south, yshift=3pt] {$z$};
  \draw[->] (-1.3,0,0) -- (1.7,0,0) node[anchor=north, xshift=-3, yshift=1] {$x$};
  \draw[->] (0,-1.3,0) -- (0,1.3,0) node[anchor=west, xshift=3pt, yshift=-2] {$y$};

  \draw[->, thick, blue] (0,0,0) -- (0.87,0,0.5);

  \def\r{0.5}
  \def\startAngle{30}
  \def\endAngle{90}

  \draw[-, thick, blue,tdplot_main_coords,domain=\startAngle:\endAngle,smooth,variable=\t] 
    plot ({\r*cos(\t)}, 0, {\r*sin(\t)}) node[anchor=east, xshift=-4, yshift=0]{$\theta$};

\end{tikzpicture}

\end{center}
\caption{$x$, $y$, and $z$ axes of the Bloch sphere and the (normalized) vector in spherical coordinates; the angle $\theta$ is defined from the $+z$-axis to the blue state vector.}
    \label{BlochGeneral}
\end{figure}

\begin{figure}
   
\begin{center}
\tdplotsetmaincoords{70}{110} 
    \begin{tikzpicture}[tdplot_main_coords, scale=2]

  \draw[thick] (0,0,0) circle (1); 
  \draw[thick, tdplot_screen_coords] (0,0) circle (1); 

  \draw[->, thick] (0,0,0) -- (0,0,1) node[anchor=south, yshift=6pt] {$\ket{0}$};
  \draw[->, thick] (0,0,0) -- (0,0,-1) node[anchor=north, yshift=-6pt] {$\ket{1}$};
  \draw[->, thick] (0,0,0) -- (1,0,0) node[anchor=north, xshift=-10]{$\ket{+}$};
  \draw[->, thick] (0,0,0) -- (-1,0,0) node[anchor=south, xshift=10]{$\ket{-}$};

  \node[right] at (0,1.2,0.15) {$\ket{+}=\frac{1}{\sqrt{2}}\ket{0}+\frac{1}{\sqrt{2}}\ket{1}$};
  \node[right] at (0,1.2,-0.15) {$\ket{-}=\frac{1}{\sqrt{2}}\ket{0}-\frac{1}{\sqrt{2}}\ket{1}$};
  
\end{tikzpicture}
\end{center}
\caption{Positions of the $|0\rangle$, $|1\rangle$, $|+\rangle$, and $|-\rangle$ states on the Bloch sphere. One can see the equation of the $|+\rangle$ state corresponds to the $|D\rangle$ state of figure~\ref{DAcalc} and the $|-\rangle$ state to the $|A\rangle$ state.}
    \label{Bloch+-}
\end{figure}
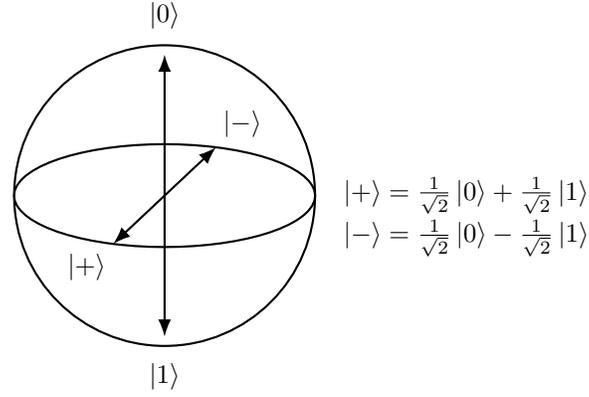

In the quantum realm, information is stored in quantum bits, or qubits. A single qubit can be represented on the Bloch sphere. The $x-$, $y-$, and $z-$ axes are defined as shown in figure~\ref{BlochGeneral}, along with a state vector (in blue) located at an angle $\theta$ from the positive $z-$ axis. By convention, the Bloch sphere has a radius of 1, the $|0\rangle$ state lies at the north pole, and the $|1\rangle$ state lies at the south pole. This representation can be misleading, as the $|0\rangle$ and $|1\rangle$ states do not appear orthogonal on the Bloch sphere (figure~\ref{Bloch+-}). In reality, all vectors on the Bloch sphere directly opposed to each other are orthogonal. That is, the relative angle between the two vectors is doubled in the Bloch sphere representation; see Equation \eqref{GeneralEquationStateVectorxz} of the main text.

The $|D\rangle$ state, being an equal mixture of the $|0\rangle$ and $|1\rangle$ states, is represented along the positive $x$-axis, while the $|A\rangle$ state lies on the negative $x$-axis. The $|D\rangle$ state is commonly referred to as the $|+\rangle$ state (plus-state) in reference to the sign of the second term, and the $|A\rangle$ state is known as the $|-\rangle$ state (minus-state). This notation is used throughout the text.

\subsection{State amplitude}
\label{StateAmplitude}

For a general quantum state $|\psi\rangle$, the amplitude of a contributing state is denoted by the coefficient that precedes the ket of that state. If the system is in state $|0\rangle$, the equation describing the system is
\begin{equation}
|\psi\rangle = 1|0\rangle + 0|1\rangle = |0\rangle\
\end{equation}
and the amplitude of state $|0\rangle$ is 1, whereas the amplitude of state $|1\rangle$ is 0. If the system is placed in the $|+\rangle$ state, as seen in Equation~\ref{EquationPlus}, the amplitudes of both states $|0\rangle$ and $|1\rangle$ are $\frac{1}{\sqrt{2}}$. For the $|-\rangle$ state (Equation~\ref{EquationMinusState}), it is $\frac{+1}{\sqrt{2}}$ and $\frac{-1}{\sqrt{2}}$, respectively, as the sign is considered part of the amplitude. 

\subsection{Quantum phase on the Bloch sphere}
\label{AppendixQuantumPhaseBlochSphere}

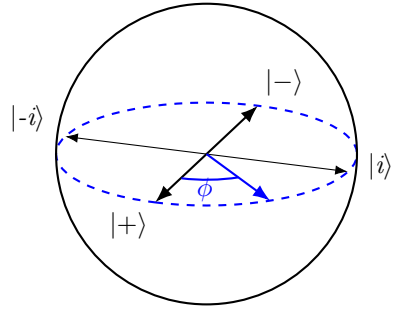
\begin{figure}
   
\begin{center}
\tdplotsetmaincoords{70}{110} 
    \begin{tikzpicture}[tdplot_main_coords, scale=2]

  \draw[thick, dashed, blue] (0,0,0) circle (1); 
  \draw[thick, tdplot_screen_coords] (0,0) circle (1); 

  
  \draw[->, thick] (0,0,0) -- (1,0,0) node[anchor=north, xshift=-10]{$\ket{+}$};
  \draw[->, thick] (0,0,0) -- (-1,0,0) node[anchor=south, xshift=10]{$\ket{-}$};
 \draw[->] (0,0,0) -- (0,1,0) node[anchor=west, xshift=4pt, yshift=2] {$\ket{\textit{i}}$};
 \draw[->] (0,0,0) -- (0,-1,0) node[anchor=east, xshift=-4pt, yshift=7] {$\ket{\textit{-i}}$};
  \draw[->, thick, blue] (0,0,0) -- (1/1.41,1/1.41,0);

  \tdplotsetrotatedcoords{0}{0}{0}; 
  \begin{scope}[tdplot_rotated_coords]
    \tdplotdrawarc[thick, blue]{(0,0,0)}{0.5}{45}{0.5}{anchor=south, xshift=-2, yshift=-12}{$\phi$}
    \end{scope}
    
\end{tikzpicture}
\end{center}
\caption{The phase shift $\phi$ is calculated from the $|+\rangle$ state to the blue state vector.}
    \label{Blochi-i}
\end{figure}

In the case of the minus state, Equation~\ref{EquationMinusState} shows a minus sign as part of the amplitude of state $|1\rangle$, indicating the presence of a phase difference $\phi$ of $\pi$ (figure~\ref{Blochi-i}) between the $|+\rangle$ and $|-\rangle$ states ($e^{i\pi} = -1$). When studying waves, a phase of $\pi$ occurs when two waves of the same frequency have their cycles shifted by one half of a wavelength ($\frac{\lambda}{2}$), which is mathematically expressed as a phase shift of $\pi$. A phase shift of $2\pi$ (a full cycle) corresponds to one wave lagging a complete wavelength behind the other, while their crests and troughs remain perfectly aligned.

\subsection{Measurement basis}
\label{AppendixMeasurementBasis}

The notion of measurement is central to quantum mechanics. When a measurement is made on a quantum state, it collapses into a single state. We call the set of possible states the initial state could collapse into, the measurement basis. The measurement basis depends on the type of measurement made. 

In quantum computing, the so-called "computational" measurement basis for a single qubit includes only the $|0\rangle$ and $|1\rangle$ states. These basis states correspond to a measurement along the $z$-axis. We are not restricted to measurements along the $z$-axis, however, and could measure along any vector on the Bloch sphere (if one has a physical means of doing so). The corresponding measurement basis contains the state the vector points to and the state exactly opposite to it on the Bloch sphere. For a measurement along the $x$-axis, for example, the basis contains the $|+\rangle$ and $|-\rangle$ states.

\subsection{State superposition}
\label{AppendixStateSuperposition}

In quantum mechanics, state superposition occurs when the state vector, which may point anywhere on the Bloch sphere, is not aligned with the measurement axis under consideration. Thus, when measuring along the $z$-axis (the $|0\rangle$ and $|1\rangle$ states) any vector not already pointing toward $|0\rangle$ or $|1\rangle$ is said to be in superposition. Once the measurement is performed, the state of the system collapses to either $|0\rangle$ or $|1\rangle$, and is no longer in a superposition with respect to the $z$-axis. This means that any subsequent measurement along the $z$-axis will leave the system in the same state. For example, a system in state $|0\rangle$ will remain in state $|0\rangle$ regardless of how many times it is measured. The same holds for state $|1\rangle$.

If the system is in either $|0\rangle$ or $|1\rangle$ and a measurement is performed along a different axis, the system is in a superposition state relative to that new measurement basis and subsequently collapses onto one of the two basis states of the new measurement axis.

\subsection{Measurement probability}
\label{AppendixMeasurementProbability}

To calculate the probabilities of obtaining one state or the other after a measurement, we take the modulus of the amplitude squared. For a system in the $|-\rangle$ state and a measurement along the $z$-axis, Equation~\ref{EquationPlus} leads to the probability $P_0$ of obtaining the state $|0\rangle$ as
\begin{equation}
P_0 = \left(\frac{1}{\sqrt{2}}\right)^2 = \frac{1}{2}\
\end{equation}
and the probability $P_1$ of obtaining the state $|1\rangle$ as
\begin{equation}
P_1 = \left(\frac{-1}{\sqrt{2}}\right)^2 = \frac{1}{2}\
\end{equation}
Thus, there is a 50\%–50\% chance of measuring either outcome.

\subsection{Polarizer as a measurement device}
\label{SectionPolarizerMeasurement}
The polarizer we use is a special measurement device and does not correspond exactly to the quantum measurements we have been describing so far. Let us take a horizontal polarizer as an example. If it were a true quantum measurement along the $z$-axis of the Bloch sphere, an incoming photon in the $|-\rangle$ state would always pass through the polarizer, but have a 50$\%$-50$\%$ chance of being in the $|0\rangle$ state or the $|1\rangle$ state. In reality, a single photon either passes through the polarizer or it is absorbed; one will never have a vertically polarized photon pass through a horizontal polarizer. So, a photon in the $|-\rangle$ state that hits the polarizer will have a 50$\%$ chance of passing through (as a horizontally polarized photon) or a 50$\%$ chance of not passing through and being lost.

In our apparatus, the laser is emitting about $10^{15}$ photons per second. If we set up the gates to create a $|-\rangle$ state, we have $10^{15}$ photons in the $|-\rangle$ state per second (ignoring any photon losses up to this point). Then, for a horizontal polarizer, each photon has a 50$\%$ chance of passing through the polarizer. Thus, only $0.5*10^{15}$ photons pass through and the light looks half as bright. The relative intensity of the light we see is proportional to the probability that a photon has the same polarization as the polarizer. Furthermore, we can rotate the polarizer to an arbitrary angle, which allows us to measure any polarization on the x-z plane of the Bloch sphere.

\section{Problem Sets}
\label{AppendixProblemSets}
Here is a set of exercises based on the theory and material presented above:

\textbf{Exercise 1} Using the data in Table~\ref{tab:AnglesCalcules}, and for each laser wavelength, construct a table showing the resulting angles modulo 360° for all combinations of Office Works and Canada Post tapes for up to six layers of tape. Identify the combinations that come closest to 90°, 180°, and 270°.

\textbf{Exercise 2} Construct the experiment for measuring the effect of an $R_z$ gate of 180° on the $|0\rangle$ state, using the tape combination closest to 180° for a chosen wavelength.

\begin{center}
\begin{quantikz}
\lstick{$q_0\ \ket{0}$} & \gate{Z} & \meter{}\\
\end{quantikz}
\end{center}

\textbf{Exercise 3} Show and explain what happens when the $R_z$ gate of 180° is instead applied to the $|1\rangle$, $|+\rangle$, or $|-\rangle$ state.
\begin{center}
\begin{quantikz}
\lstick{$q_0\ \ket{?}$} & \gate{Z} & \meter{}\\
\end{quantikz}
\end{center}

\textbf{Exercise 4} Construct the experiment needed to determine the final state of a photon after the following circuit. Recall that the X gate is the $R_x(180^\circ)$ gate and the Z gate is the $R_z(180^\circ)$ gate.

\begin{center}
\begin{quantikz}
\lstick{$q_0\ \ket{0}$} & \gate{X} & \gate{Z} & \meter{}\\
\end{quantikz}
\end{center}

\textbf{Exercise 5} Construct the Hadamard gate and measure the final state of a photon entering the gate in the $|0\rangle$, $|1\rangle$, $|+\rangle$, or $|-\rangle$ state.

\begin{center}
\begin{quantikz}
\lstick{$q_0\ \ket{?}$} & \gate{H} & \meter{}\\
\end{quantikz}
\end{center}

\textbf{Exercise 6} Construct a Hadamard gate using only transparent tape used as $R_x$ and $R_z$ gates. Write the associate quantum circuit.

\textbf{Exercise 7} Construct an experimental setup using one $R_x$ gate, one $R_y$ gate, and one $R_z$ gate (in any order and for rotation angles of your choice) so that the system starts and ends in state $|0\rangle$. Write the associated quantum circuit.

\textbf{Exercise 8} Starting with a calcite crystal producing the state $|\psi\rangle = \frac{1}{\sqrt{2}}|00\rangle + \frac{e^{i\phi}}{\sqrt{2}}|11\rangle$, create the state $|\psi\rangle = \frac{1}{\sqrt{2}}|01\rangle + \frac{e^{i\phi}}{\sqrt{2}}|10\rangle$. Draw the corresponding quantum circuit.

\textbf{Exercise 9} Using the same initial setup as the previous exercise, describe the effect of placing a piece of horizontal tape on each of the two rays exiting the crystal. Draw the corresponding quantum circuit.

\textbf{Exercise 10} Describe the effect of placing a piece of transparent tape at a 45° angle on only one of the two rays exiting a calcite crystal. Draw the corresponding quantum circuit.

\textbf{Exercise 11} Describe the effect of observing a red computer screen that is vertically polarized through a vertically polarized filter with a 45° angle piece of transparent tape placed between the two.

\textbf{Exercise 12} Calculate the transparent tape thickness required to generate a measurement probability of $1/2$ in the $|1\rangle$ state if the system is initially in $|0\rangle$, a red laser (650 nm) is used, and the birefringence of the tape is 0.0120.

\textbf{Exercise 13} Someone prepares a quantum state at a specific, but unknown, point on the equator of the Bloch sphere. Design and implement a quantum circuit that can determine the exact position of the state.

\end{appendices}
\printbibliography
\end{document}